*Nonlinear Multivariate Analysis of Neurophysiological Signals*

Ernesto Pereda[1], Rodrigo Quian Quiroga[2], Joydeep Bhattacharya[3]

[1] Department of Basic Physics, College of Physics and Mathematics
University of La Laguna, Tenerife, SPAIN
e-mail: eperdepa@ull.es

[2] Department of Engineering, University of Leicester, Leicester, UK
e-mail: rodri@vis.caltech.edu

[3] Commission for Scientific Visualization, Austrian Academy of Sciences, AUSTRIA
e-mail: joydeep@oeaw.ac.at

*Abstract*

Multivariate time series analysis is extensively used in neurophysiology with the aim of studying the relationship between simultaneously recorded signals. Recently, advances on information theory and nonlinear dynamical systems theory have allowed the study of various types of synchronization from time series. In this work, we first describe the multivariate linear methods most commonly used in neurophysiology and show that they can be extended to assess the existence of nonlinear interdependences between signals. We then review the concepts of entropy and mutual information followed by a detailed description of nonlinear methods based on the concepts of phase synchronization, generalized synchronization and event synchronization. In all cases, we show how to apply these methods to study different kinds of neurophysiological data. Finally, we illustrate the use of multivariate surrogate data test for the assessment of the strength (strong or weak) and the type (linear or nonlinear) of interdependence between neurophysiological signals.

**Table of Contents**

*1. Introduction*

One of the most common ways of obtaining information about neurophysiological systems is to study the features of the signal(s) recorded from them by using time series analysis techniques (e.g., (Galka, 2000)). If one is only interested in the features of a single signal, univariate analysis can perfectly carry out this task by itself. But an increasing number of experiments are being carried out in which several neurophysiological signals are simultaneously recorded, and the assessment of the interdependence between signals can give new insights into the functioning of the systems that produce them. Therefore, univariate analysis alone cannot accomplish such a task, as it is necessary to make use of the multivariate analysis.

In spite of their different aims and scopes, univariate and multivariate time series analysis have an important point in common: they have traditionally relied on the use of linear methods in the time and frequency domains (see, e.g., (Bendat and Piersol, 2000)). Unfortunately, these methods cannot give any information about the nonlinear features of the signal. But due to the intrinsic nonlinearity of neuronal activity these nonlinear features might be present in neurophysiological data, which has led researchers to try out other techniques that do not present the aforementioned limitation.

Univariate nonlinear time series analysis methods started to be applied to neurophysiological data about two decades ago (Babloyantz *et al.*, 1985); see, e.g., (Elbert *et al.*, 1994; Faure and Korn, 2001; Galka, 2000; Jansen, 1991; Korn and Faure, 2003; Segundo, 2001; Segundo *et al.*, 1998; Stam, 2005) for surveys. As an example, the EEG has been characterized in terms of its *correlation dimension*, a nonlinear index that has been roughly interpreted as a measure of the irregularity or complexity of a signal[1] (see., e.g., (Kantz and Schreiber, 2004)). This index has been shown useful in sleep-wake research (Pereda *et al.*, 1998; Pradhan *et al.*, 1995), mental load research (Lamberts *et al.*, 2000), monitoring the depth of anesthesia (van den Broek *et al.*, 2005; Widman *et al.*, 2000) and in studies of epilepsy (Pijn, 1990) to name but a few applications (see (Stam, 2005) for a recent review).

---

[1] It might be noted that this index is actually a measure of the randomness of the signal.

Similarly, in the last few years several nonlinear multivariate techniques have started to be used in neurophysiology, mainly as a result of recent advances in information theory (see, e.g., (Kraskov *et al.*, 2004; Schreiber, 2000)) and in the study of the synchronization between chaotic systems (Boccaletti *et al.*, 2002; Pikovsky *et al.*, 2001). Two relevant concepts are: *generalized synchronization* (GS) (Rulkov *et al.*, 1995), a state in which a functional dependence between the systems exist, and *phase synchronization* (PS) (Rosenblum *et al.*, 1996), a state in which the phases of the systems are correlated whereas their amplitudes may not be. In fact, and unlike *complete synchronization* (Fujisaka and Yamada, 1983) (which may exist only between identical systems and entails the exact equality of their variables), GS and PS may exist between nonindentical systems even in the presence of noise. This makes GS and PS methods appealing for the analysis of neurophysiological signals.

The multivariate nonlinear time series methods derived for the study of GS and PS, as well as those based on information theory, are theoretically useful in neurophysiology due to their ability to detect nonlinear interactions, which might not be fully captured by linear techniques. Nevertheless, the application of these methods to neurophysiological signals is not a plain subject. On the one hand, these signals are often noisy, non-stationary and of finite (sometimes quite short) size. On the other hand, the theoretical subtleties underlying the calculation of many nonlinear interdependency indexes from experimental time series must be taken into account before applying them to the data. In this work, we go through all these questions by reviewing the theoretical and practical aspects of the multivariate nonlinear methods more frequently used for the analysis of neurophysiological signals, ranging from recordings of neuronal action potentials (spikes) to the electroencephalogram (EEG) and the magnetoencephalogram (MEG) as typical recordings of integrated neuronal activity.

The paper is organized as follows: we first review the traditional linear tools for the assessment of the interdependence between neurophysiological data in the time and frequency domain; the nonlinear counterparts of the time domain tools are also discussed. Then, we present methods based in information theory as a natural extension of the concept of linear statistical dependence between time series. Next, we explore the idea of PS indexes, which assess the existence of interdependence between the phases of the signals regardless of whether their amplitudes are correlated. Subsequently, state space based methods are introduced, which analyze the interdependence between the amplitudes of the signals in the reconstructed state spaces, and can be used for the assessment of GS. Further, we review the study of the interdependence between signals that present marked events. Finally, we conclude by comparing the performance of the main multivariate nonlinear methods and giving some practical recipes.

Two appendixes are added at the end. The first one deals with the use of multivariate surrogate data for the assessment of the strength (strong or weak) and the character (linear or nonlinear) of the interdependence between neurophysiological signals. The second one is devoted to interesting Internet sites from where it is possible to gather information on how to put into practice the different nonlinear methods reviewed in the text.

## 2. Cross-correlation function

### 2.1. Definition and estimation

This is one of the oldest and most classical measures of interdependence between two time series. The cross-correlation function measures the linear correlation between two variables $X$ and $Y$ as a function of their delay time ($\tau$), which is of interest because such a time delay may reflect a causal relationship between them. In particular, if *X* causes *Y*, one may in principle get a delay from the first signal to the second one. This is, however, not necessarily always the case, since internal delay loops of one of the systems or different distances to the sources may change this interpretation (see e.g. (Quian Quiroga *et al.*, 2000)).

If *x(t)* and *y(t)* are signals normalized to have zero mean and unit variance, their cross-correlation function is:

$$C_{xy}(\tau) = \frac{1}{N-\tau} \sum_{k=1}^{N-\tau} x(k+\tau) y(k) \qquad (1)$$

where *N* is the total number of samples and $\tau$ the time lag between the signals. An example of the cross-correlation function between two EEG signals is shown in Figure 1. This function ranges from -1 (complete linear inverse correlation) to +1 (complete linear direct correlation), with $C_{xy}(\tau) = 0$ suggesting lack of linear interdependence for a given time lag $\tau$. The sign of $C_{xy}$ indicates the direction of correlation: $C_{xy} < 0$ implies inverse correlation, i.e. a tendency of both signals to have similar absolute values but with opposite signs and $C_{xy} > 0$ implies direct correlation, i.e. a tendency of both signals to have similar values with the same sign. The value of $\tau$ that maximizes this function is usually taken as an estimation of the delay between the signals, under the implicit assumption that they are linearly related. It must be stressed again, however, that this delay cannot be directly regarded as a measure of the propagation time of, say, the electrical signal in the cerebral cortex.

In practice, the significance of $C_{xy}(\tau)$ is usually checked, at the desired level of statistical confidence, by calculating the residual cross-correlation from an ensemble of signals with the same autocorrelation than the original ones but completely independent from each other, a procedure that has also become popular in the nonlinear methodology (see *Appendix A*). It must be mentioned that the cross-correlation function at zero time lag is the well-known Pearson's product moment correlation correlation coefficient, ($r_{xy}$ or simply $r$), an index frequently used to measure the linear correlation between two variables.

*2.2. Applications to neurophysiology*

The first approaches to correlation measurements between two simultaneously measured EEG signals were made more than fifty years ago (Brazier and Barlow, 1956; Brazier and Casby, 1952). Before the possibilities of the computation of coherence spectra (see next section) in early 1960s, most of the studies investigated the similarity and the time delay between two EEG signals recorded from two separate regions of the brain by the linear cross-correlation (see e.g., (Gevins and Schaffer, 1980; Shaw and Ongley, 1972) for reviews). After the availability of the fast Fourier transform (FFT) algorithm in 1965, frequency based measures like coherence and phase spectra of neurophysiological signals such as EEG/MEG became increasingly popular. However, the cross-correlation function and its variant, the cross-correlogram histogram (Perkel *et al.*, 1967) remains one of the mostly used measures to reveal the temporal coherence in the firing of cortical neurons from their spike trains (see (Brody, 1999; Nowak and Bullier, 2000) for reviews on the applicability of this method to this kind of signals).

*3. Coherence*

*3.1. Definition and estimation*

The coherence function gives the linear correlation between two signals as a function of the frequency. Coherence, also termed as magnitude squared coherence or coherence spectrum, between two signals is their cross-spectral density function -which is in fact the Fourier transform of Eqn. (1)- normalized by their individual auto-spectral density functions. These spectral quantities are usually derived via the FFT algorithm

(Cooley and Tukey, 1965). However, due to finite size of the neural data, one can only have an estimate of the true spectrum (the *periodogram*). Smoothing techniques are often used to improve the performance of the spectral estimators. Thus, in practice, EEG/MEG signals are usually subdivided into *M* epochs of equal length, and the spectra are estimated by averaging the periodogram over these epochs (Welch's method), so that coherence is normally calculated as:

$$\kappa_{xy}^2(f) = \frac{|\langle S_{xy}(f) \rangle|^2}{|\langle S_{xx}(f) \rangle||\langle S_{yy}(f) \rangle|} \qquad (2)$$

where ⟨•⟩ indicates average over the *M* segments.

For event-related data, spectra are estimated by averaging the periodogram over trials (Andrew and Pfurtscheller, 1996). In this case, however, what we have is not the true coherence, but an estimation of this value, whose confidence interval must be estimated, as detailed below. For this kind of non-parametric spectral estimation, a trade-off has to be made regarding the length of the data segment for analysis, which must be short enough to satisfy the condition of stationarity, and long enough to provide good frequency resolution. Instead, a parametric approach can be used to obtain spectral quantities of signals (Hannan, 1970), which is based on the assumption that a signal can be described as the output of a stochastic process, i.e. autoregressive (AR) or autoregressive-moving-average (ARMA) process (Marple Jr., 1987). This idea has been used to represent EEG signals (Gersch, 1970), to enhance spectral resolution (Franaszczuk *et al.*, 1985) or to classify EEGs (Gersch *et al.*, 1980). See (Blinowska *et al.*, 1981; Davis and Lutchen, 1991; Guler *et al.*, 1995; Spyers-Ashby *et al.*, 1998) for comparative performances of spectral estimators based on FFT and parametric methods.

*3.2. Properties of coherence*

The estimated coherence ranges between 0 and 1. For a given frequency $f_o$, $\kappa_{xy}(f_o)=0$ indicates that the activities of the signals in this frequency are linearly independent, whereas a value of $\kappa_{xy}(f_o)=1$ gives the maximum linear correlation for this frequency. The confidence limit for coherence at the 100% $\alpha$ ($\alpha$ is defined by the confidence probability), is given by $1-(1-\alpha)^{1/(M-1)}$ (Bendat and Piersol, 2000). For a recent theoretical discussion on the estimation of this limit, see (Wang and Tang, 2004). Additionally, specific methods have been derived for the reliable estimation of the coherence function as well as its confident limits in point processes such as sequences of neural action potentials (Jarvis and Mitra, 2001; Pesaran *et al.*, 2002). Other factors that must be carefully considered before EEG coherence estimation are reference electrodes and volume conduction; as an example, we can mention that the application of this technique to study the relationship between EEG channels recorded with electrode $C_z$ as common reference is problematic, because this active common source may introduce interdependence between the electrodes that is not actually present in the signals. We refer the interested readers to (Essl and Rappelsberger, 1998; Nolte *et al.*, 2004; Nunez *et al.*, 1999; Nunez *et al.*, 2001) for detailed treatments of these ideas.

Another important point to take into account is that coherence is sensitive to both phase and amplitude relationships between the signals. Therefore, the relative importance of amplitude and phase covariance in this index is not altogether clear (Lachaux *et al.*, 1999; Varela *et al.*, 2001). If one is only interested in the relationship between the phases without any influence of the amplitudes then other methodology is necessary for this aim, as described in 8.

*3.3. Applications to neurophysiology*

Coherence was first applied to EEG signals more than forty years ago (Adey *et al.*, 1967a; Brazier, 1968; Walter and Adey, 1963; Walter *et al.*, 1966). One of the pioneering efforts was to demonstrate the continuous coherence spectra of the EEG of an astronaut during the Gemini flight GT-7 (Adey *et al.*, 1967b; c). After the introduction of the FFT, the coherence measure could be calculated within a reasonable time and it has been applied to EEG or MEG signals in several cognitive or clinical conditions. It is beyond the scope of the present paper to list all these applications; however, we mention a few key articles that reviewed the applications of coherence to neural data (Dumermuth and Molinari, 1991; French and Beaumont, 1984; Shaw, 1984; Thatcher *et al.*, 1986; Zaveri *et al.*, 1999).

*4. Nonlinear correlation coefficient*

*4.1. Definition and estimation*

This measure is primarily a nonparametric nonlinear regression coefficient, which describes the dependency of *X* on *Y* in a most general way without any direct specification of the type of relationship between them (Lopes da Silva *et al.*, 1989; Pijn *et al.*, 1990). The underlying idea is that if the value of *X* is considered as a function of the value of *Y*, the value of *Y* given *X* can be predicted according to a nonlinear regression curve. The variance of *Y* according to the regression curve is termed as the explained variance, since it is explained or predicted by the knowledge of *X*. The unexplained variance is estimated by subtracting the explained variance from the original one. The correlation ratio $\eta^2$ describes the reduction of variance of *Y* that can be obtained by predicting the *Y* values from those of *X* according to the regression curve as $\eta^2$ = (total variance – unexplained variance)/total variance.

Since this computation involves a step of nonlinear regression, one can only get an estimate of this correlation ratio between two signals of finite data points. The estimate of the above ratio measure is termed as nonlinear correlation (or regression) coefficient ($h^2$). In practice, a scatter plot of Y versus X is studied. The values of X are subdivided into bins; for each bin, the X value of the midpoint ($p_i$) and the average value of Y ($q_i$) are calculated. The curve of regression is approximated by connecting the resulting points ($p_i$, $q_i$) by segments of straight lines. The nonlinear correlation coefficient between demeaned signals X and Y is then calculated as follows:

$$h^2_{y|x} = \frac{\sum_{k=1}^{N} y(k)^2 - \sum_{k=1}^{N}(y(k) - f(x_i))^2}{\sum_{k=1}^{N} y(k)^2} \quad (3)$$

where $f(x_i)$ is the linear piecewise approximation of the nonlinear regression curve. The measure of association in the opposite direction $h^2_{x|y}$ can be calculated analogously.

*4.2. Asymmetry, time delay and direction in coupling*

The estimator $h^2_{y|x}$ ranges from 0 (Y is completely independent of X) to 1 (Y is fully determined by X). If the relationship between these signals is linear, $h^2_{x|y} = h^2_{y|x}$, and this measure approximates the squared linear regression coefficient $r^2$. For a nonlinear relationship, $h^2_{x|y} \neq h^2_{y|x}$ and the difference $\Delta h^2 = h^2_{x|y} - h^2_{y|x}$ indicates the degree of asymmetry of the nonlinear coupling. The nature of the interdependence can be also traced by using multivariate surrogate data, as detailed in *Appendix A*.

By studying the index $h^2$, it is also possible to estimate the delay in the coupling between the signals. For this purpose, $h^2$ has to be calculated as a function the time delay $\tau$. As we already showed in the linear case, the delay at which the maximum value for $h^2$ is obtained is used as an estimate of the time delay between the signals. Indeed, if

*X* causes *Y*, $\tau_{y/x}$ (corresponding to $h^2_{y|x}$) will be positive and $\tau_{x|y}$ (corresponding to $h^2_{y|x}$) will be negative, so that the difference $\Delta\tau = \tau_{x|y} - \tau_{y|x}$ will be also positive.

On combining the information of asymmetry and of time delay in coupling, the following direction index has been recently proposed (Wendling *et al.*, 2001) to provide a robust measure on the direction of coupling:

$$D_{x|y} = \frac{1}{2}\left[\text{sgn}(\Delta h^2) + \text{sgn}(\Delta\tau)\right] \qquad (4)$$

If $D_{x|y}$ = + 1 (or -1), a strong unidirectional coupling *X*→*Y* (or *Y*→*X*) can be concluded. $D_{x|y}$ = 0 indicates bidirectional (*X*↔*Y*) coupling between the signals.

*4.3. Applications to neurophysiology*

Although the index $h^2$ was proposed almost fifteen years back and offers a very general formulation of coupling analysis with little assumptions, its applications have been confined exclusively to epileptic EEG data analysis (Meeren *et al.*, 2002; Pijn *et al.*, 1990; Wendling *et al.*, 2001). One of the first applications involved the recording of epileptiform after-discharges from both hippocampus of rats (Filipe *et al.*, 1989). Here, values of $h^2$ and $r^2$ were almost identical for most of the initial epochs, indicating a predominantly linear relationship between recording sites. But for the later epochs, $h^2$ was significantly larger than $r^2$, indicating an emergence of strong nonlinearity; this nonlinearity became evident when the complexity of the after-discharges increased in the formation of paroxysmal bursts of multiple spikes. A follow up study (Fernandes de Lima *et al.*, 1990) investigated the interhemispheric transfer of hippocampal after-discharges by estimating time delays through $h^2$. Thus, the authors decipher the nature (linear or nonlinear) of coupling and the delay (lead or lag) in transmission between epileptic foci and neighboring brain sites in rats (Meeren *et al.*, 2002). Importantly, from these results they also suggested that absence seizures have a localized cortical origin. However, it must be pointed out that, as already commented, the interpretation of the delay between the signals in terms of signal transmission time must be done carefully, as in general it is not possible to be sure that the latter one is the cause of the former one.

Recently, this nonlinear regression technique has also been applied to study the bspatiotemporal organization of epileptogenic networks in human (Bartolomei *et al.*, 2001; Chavez *et al.*, 2003; Wendling *et al.*, 2001). In order to detect a causal coupling between distant neural populations, the nonlinear regression coefficient ($h^2$) and the direction index (*D*) were first applied (Bartolomei *et al.*, 2001; Wendling *et al.*, 2001) to a neurophysiologically relevant model of EEG generation (Jansen and Rit, 1995; Lopes da Silva *et al.*, 1976). The advantages of using this simulated model are two-fold: (i) for appropriate choices of parameters, both ictal and interictal EEG can be simulated, and (ii) the coupling parameters are explicitly introduced in the model. These two indices were then measured on stereo-EEG signals of human subjects with temporal lobe epilepsy; the results showed that both the indices described abnormal functional interactions between cerebral structures of the temporal regions during seizures. In the previous studies, $h^2$ and *D* indexes were applied to broadband EEG signals, yet the epileptic EEG signals have been found to exhibit a dynamically varying time-frequency structure (Zaveri *et al.*, 1992). Computing $h^2$ to narrowband EEG signals, a significant change of coupling in the focal area was found several minutes before seizure in the frequency band of 10-25 Hz, results that were corroborated afterwards by phase synchrony analysis (Chavez *et al.*, 2003).

*5. Granger causality*

*5.1. Definition and estimation*

In neurophysiology, a question of great interest is whether there exists a causal relation between two brain regions without any specific information on direction. Both the cross-correlation function and the nonlinear correlation coefficient theoretically are able to indicate the delay in coupling, but inferring causality from the time delay is not always straightforward (Lopes da Silva *et al.*, 1989), which encouraged the researchers to develop new methods explicitly tailored for this aim. One of the first attempts involved the method of structural equation modeling (Asher, 1983), where the direction of coupling was first assumed, which was followed by assessing the coupling strength by linear correlation analysis. Methods with similar ideas have been recently applied to neuroimaging data (Buchel and Friston, 2000; McIntosh and Gonzalez-Lima, 1992; 1994).

The importance of temporal ordering in the events (i.e., past and present may cause the future but not vice versa (Granger, 1980)) to the inference of causal relations was first mentioned by the great mathematician Norbert Wiener, who defined causality in a statistical framework as follows: for two simultaneously measured signals, if one can predict the first signal better by incorporating the past information from the second signal than using only information from the first one, then the second signal can be called causal to the first one (Wiener, 1956). This general definition was later given a mathematical formulation by Nobel Prize winning economist Clive Granger in the context of linear stochastic modeling of time series analysis (Granger, 1969). Like Wiener, Granger argued that if $X$ is influencing $Y$, then adding past values of the first variable to the regression of the second one will improve its prediction performance, which can be assessed by comparing the univariate and bivariate fitting of the AR models to the signals. Thus, for the univariate case, one has:

$$x(n) = \sum_{k=1}^{p} a_{xk} x(n-k) + u_x(n)$$
$$y(n) = \sum_{k=1}^{p} a_{yk} y(n-k) + u_y(n)$$
(5)

where $a_{xk}$ and $b_{yk}$ are the model parameters, $p$ is the model order, and $u_x$ and $u_y$ are the uncertainties or the residual noises associated with the model. Here, the prediction error depends only on the past values of the own signal.

On the other hand, for bivariate AR modeling,

$$x(n) = \sum_{k=1}^{p} a_{xyk} x(n-k) + \sum_{k=1}^{p} b_{xyk} y(n-k) + u_{xy}(n)$$
$$y(n) = \sum_{k=1}^{p} a_{yxk} y(n-k) + \sum_{k=1}^{p} b_{yxk} x(n-k) + u_{yx}(n)$$
(6)

where the prediction error for each individual signal depends on the past values of both signals.

The prediction performance for both models can be assessed by the variances of the prediction errors:

$$V_{X|X\_} = \mathrm{var}(u_x) \text{ and } V_{Y|Y\_} = \mathrm{var}(u_y) \text{ for univariate AR model}$$
$$V_{X|X\_,Y\_} = \mathrm{var}(u_{xy}) \text{ and } V_{Y|Y\_,X\_} = \mathrm{var}(u_{yx}) \text{ for bivariate AR model}$$
(7)

where $var(.)$ indicates variance operator, $X|X\_$ and $X|X\_,Y\_$ indicate predicting $X$ by its past values alone and by past values of $X$ and $Y$, respectively. If $V_{X|X\_,Y\_} < V_{X|X\_}$ then $Y$ causes $X$ in the sense of Granger causality. The Granger causality of $Y$ to $X$ can be quantified as:

$$G_{Y \to X} = \ln\left(\frac{V_{X|X\_}}{V_{X|X\_,Y\_}}\right)$$
(8)

If the past of $Y$ does not improve the prediction of $X$, then $V_{X|X\_,Y\_} \approx V_{X|X\_}$ and the causality measure will be close to zero. Any improvement in prediction of $X$ by the inclusion of $Y$ leads to decrease of $V_{X|X_-,Y_-}$, thereby increasing the causality measure. The Granger causality for opposite direction, from $X$ to $Y$, is defined accordingly. If both $G_{X \to Y}$ and $G_{Y \to X}$ are high, it indicates a bidirectional coupling or a feedback relationship between the signals.

*5.2. Nonlinear Granger causality*

The original formulation of causality by Granger assumes that the interacting systems are linear. Accordingly, if the signals are nonlinear, then any measure based on linear regression such as Granger causality may not be appropriate. In the field of economics, there have been some attempts to modify the Granger causality to incorporate nonlinear properties of the signals (Teräsvirta, 1998; Warne, 2000). One of the factors limiting a nonlinear extension of Granger causality in neurophysiology is that the model selected for implementing this measure must be appropriately matched to the dynamical characteristics of the signals, and yet there is no general framework of nonlinear models that are capable of capturing the broad spectrum of characteristics of neural signals. An immediate but approximate solution is to substitute the globally nonlinear model by a locally linear one (Freiwald *et al.*, 1999), where the AR model parameters depend on the current values and a different linear regression model is used for each point of the state-space (Tong, 1990). Recently, a further extension of nonlinear Granger causality has been proposed, which aims to detect whether the causal relation between two signals is due to direct coupling between them or due to a third system that drives them (Chen *et al.*, 2004). However, this extended nonlinear Granger causality index has only been applied to simulated signals, and its practical usefulness are yet to be demonstrated.

*5.3. Applications to neurophysiology*

Although Granger causality was introduced more than thirty five years back, most of its applications to neural data analysis are within the last six years. One of the first studies investigated for the existence of directional or causal interactions by analyzing local field potentials (LFPs) from the macaque inferotemporal cortex (Freiwald *et al.*, 1999). Directional interactions were surprisingly found within the same cortical regions at the same level of the processing hierarchy; directional interactions were also found between spatially separate neuronal populations. Further, it was found that each state of the system is influenced by its own past of up to 60 ms duration, which was defined by the AR model order. This method was also applied to the LFP data recorded from two separate areas (primary and higher visual areas) of the cat visual cortex, in order to investigate the role of bottom-up and top-down interactions in a go/no-go task (Bernasconi *et al.*, 2000) or in a stimulus expectancy task (Salazar *et al.*, 2004). For both cases, task-specific changes in the directed interareal couplings were reported. Recently, a frequency specific Granger causality measure was utilized in LFP recordings from somatosensory and motor cortices of macaque monkeys as they performed a motor maintenance in a visual discrimination task (Brovelli *et al.*, 2004). Synchronous oscillations in the beta frequency band (~ 20 Hz) formed a large scale cortical network with directed influences from primary somatosensory and inferior posterior parietal cortical areas to motor cortex during the task. In human, a time-variant Granger causality measure was applied to EEG signals from the standard color-word conflict Stroop task (Hesse *et al.*, 2003). In conflict situations, a dense cortical network was formed after 400 ms of stimulus presentation, and there was a strong preference of direction of influence from posterior to anterior cortical regions.

*5.4. Comments on Granger causality*

True causality can only be assessed if the set of two time series contains all possible relevant information and sources of activities for the problem (Granger, 1980). From the neurophysiological point of view, rarely two channels of observations fulfill this requirement of completeness of information. As a result, one has to be careful before emphasizing the aspects of causality obtained from bivariate time series. Multiple pairwise analysis is also unable to circumvent this problem (Franaszczuk *et al.*, 1985). There is no unique way either to determine the size of the information set relevant for a given problem. Thus, any result of causality analysis should be interpreted with caution. However, for practical neural data analysis, the activities are recorded often from multiple spatial positions in the brain, so one can create a multivariate modeling framework containing all available information from different channels.

*6. Multichannel analysis*

Most of the methods discussed so far are defined for two signals only: a functional relationship is obtained by pairwise analysis of bivariate signals. However, as discussed earlier, a bivariate method for each pair of signals from a multichannel set of signals does not account for all the covariance structure information from the full data set. In a simple network consisting of one driver and two responses, pairwise analysis is likely to find some correlation between the two responses due to the common driver component, even when the response signals might be fully independent. As a result, a different and maybe erroneous network pattern can be obtained if pairwise analysis is performed as opposed to a genuine multichannel method of correlation analysis (Kus *et al.*, 2004).

*6.1. Partial coherence*

The first extension of bivariate analysis was made by incorporating a third signal into the estimation of a new coherence measure, termed as partial coherence. For signals *X, Y*, and *Z*, the underlying point is to subtract linear influences from other processes to obtain the partial cross-spectrum between *X* and *Y* given all the linear information of *Z*:

$$S_{xy|z}(f) = S_{xy}(f) - S_{xz}(f)S_{zz}^{-1}(f)S_{yz}(f) \qquad (9)$$

Similarly, one can obtain the partial auto-spectra $S_{xx|z}(f)$ and $S_{yy|z}(f)$. The squared partial coherence is estimated as follows (Bendat and Piersol, 2000):

$$\kappa_{xy|z}^2(f) = \frac{|\langle S_{xy|z}(f) \rangle|^2}{|\langle S_{xx|z}(f) \rangle||\langle S_{yy|z}(f) \rangle|} \qquad (10)$$

where, as in the case of normal coherence, ⟨•⟩ indicates average over *M* segments. The partial coherence $\kappa_{xy|z}(f)$ can be represented as the fraction of coherence between *X* and *Y* that is not shared with *Z*. Thus, if three signals are fully sharing with each other, partialization of the coherent activity between any two signals with the remaining signal as the predictor would lead to a zero coherence. In other words, if *Z* contributes to the linear interdependence between *X* and *Y*, then the partial coherence $\kappa_{xy|z}(f)$ will be smaller than the ordinary coherence $\kappa_{xy}(f)$. However, it must be noted that partial coherence is based on the assumption of linearity, so any failure in its reduction might be also caused by nonlinear interaction between signals.

In neurophysiology, partial coherence was first applied to identify epileptic foci using three electrodes (Gersch and Goddard, 1970). To this date, partial coherence has been applied to investigate the nature of connectivity and causal information in various neural signals from spike trains (Cohen *et al.*, 1995), hippocampal field oscillations (Kocsis *et al.*, 1999), intracortical EEG (Lopes da Silva *et al.*, 1980; Mirski *et al.*, 2003), scalp EEG (Liberati *et al.*, 1997; Tucker *et al.*, 1986), and functional magnetic-resonance image (fMRI) data (Sun *et al.*, 2004). However, a recent study demonstrates that the partial coherence measure is very sensitive to noise contamination (Albo *et al.*, 2004): if different signals have different signal-to-noise ratio, this measure tends to identify the signal with the highest ratio as the most influential or driver irrespective of the genuine pattern of underlying connectivity.

*6.2. Partial directed coherence*

This method was introduced recently (Baccala and Sameshima, 2001b; Sameshima and Baccala, 1999) and provides a frequency domain measure for Granger causality. But unlike the original Granger causality, which was introduced for bivariate time series, partial directed coherence (PDC) is based on modeling time series by multivariate autoregressive (MAR) process. Consider an *m*-dimensional (*m* signals simultaneously measured, $X_1, X_2,..., X_m$ ) MAR process with order *p* as follows:

$$\begin{pmatrix} x_1(k) \\ x_2(k) \\ \vdots \\ x_m(k) \end{pmatrix} = \sum_{r=1}^{p} \mathbf{A}_r \begin{pmatrix} x_1(k-r) \\ x_2(k-r) \\ \vdots \\ x_m(k-r) \end{pmatrix} + \begin{pmatrix} \varepsilon_1(k) \\ \varepsilon_2(k) \\ \vdots \\ \varepsilon_m(k) \end{pmatrix} \quad (11)$$

where $\varepsilon_i(k)$ represents independent Gaussian white noise with covariance matrix $\Sigma$, and $\mathbf{A}_1, \mathbf{A}_2, \ldots, \mathbf{A}_p$ are the coefficient matrices (*m* x *m*). This time domain representation can be translated to frequency domain by computing the power spectral density matrix:

$$\mathbf{S}(f) = \mathbf{H}(f)\Sigma\mathbf{H}^H(f) \quad (12)$$

where the superscript $(.)^H$ indicates the Hermitian transpose. **H** is called the transfer function matrix ($\mathbf{H}(f) = \overline{\mathbf{A}}^{-1}(f) = [\mathbf{I} - \mathbf{A}(f)]^{-1}$) where $\mathbf{A}(f)$ is essentially the Fourier transform of the coefficients. Let $\overline{\mathbf{A}}(f) = [\overline{\mathbf{a}}_1(f)\,\overline{\mathbf{a}}_2(f)\cdots\overline{\mathbf{a}}_m(f)]$ and $\overline{a}_{ij}(f)$ is the *i,j*-th element of $\overline{\mathbf{A}}(f)$. Then, the PDC measure from signal *j* to signal *i* is given by:

$$\pi_{ij}(f) = \frac{\bar{a}_{ij}(f)}{\sqrt{\bar{\mathbf{a}}_j^H(f)\bar{\mathbf{a}}_j(f)}} \qquad (13)$$

The PDC from *j* to *i* represents the relative coupling strength of the interaction of a given source, signal *j*, with regard to some signal *i*, as compared to all of *j*'s connections to other signals (Fig. 2). Thus, PDC ranks the relative strength of interaction with respect to a given signal source while fulfilling the following properties: $0 \leq |\pi_{ij}(f)|^2 \leq 1$ and $\sum_{i=1}^{m}|\pi_{ij}(f)|^2 = 1$, for all $1 \leq j \leq m$. For $i = j$, the PDC $\pi_{ii}(f)$ represents how much of $X_i$'s own past is not explained by other signals.

A very similar measure of causal influence, called directed transfer function (DTF) was introduced as follows (Kaminski and Blinowska, 1991):

$$\vartheta_{ij}(f) = \frac{H_{ij}(f)}{\sqrt{\mathbf{h}_i^H(f)\mathbf{h}_i(f)}} \qquad (14)$$

DTF uses the elements of the transfer function matrix **H**, whereas PDC uses those of $\bar{\mathbf{A}}$. Since, unlike DTF calculation, the computation of PDC does not involve any matrix inversion, it is computationally more efficient and more robust than DTF. Further, PDC is normalized with respect to the total inflow of information, but DTF is normalized with respect to the total outflow of the information. For comparative results between these two methods, see (Baccala and Sameshima, 2001a; b; Kus *et al.*, 2004).

### 6.2.1. Comments on multivariate autoregressive modeling

The successful estimation of PDC or DTF depends primarily on the reliability of the fitted MAR model, since all the necessary information is derived from the estimated model parameters. In practice, this issue boils down to the choice of an optimal model order and an optimal epoch length. If the model order is too low, the model will not capture the essential dynamics of the data set, whereas if the model order is too high, it will also capture the unwanted component (i.e. noise), leading to over-fitting and instability. Although there is no direct cook-book approach, several criteria are available (e.g., Akaike Information Criteria, False Prediction Criteria or Schwartz's Criteria) that can be used as a guideline for the selection a proper model order (Marple Jr., 1987). Most of these criteria were originally proposed for univariate AR modeling, so special care should be taken to look for consistent results by comparing the performances of different criteria on a same data window. Further, if there are several data epochs from the same experiment, it is suggested to fix a common model order for all the available epochs to be analyzed even if the optimal model orders may vary between epochs. One of the intermediate but technically important steps is the use of proper algorithms for the estimation of MAR parameters. There is an overwhelming majority of works using the method of Levinson-Robinson-Wiggins, abbreviated as LWR algorithm, due to its robust performance (Haykin and Kesler, 1983; Morf *et al.*, 1978). The next crucial question is how to choose the proper window size: MAR model assumes that the underlying process is stationary, but neurophysiological and cognitive events are themselves transient and may rapidly change their states, insomuch as the neural signals are often non-stationary (Blanco *et al.*, 1995; Kawabata, 1973; Thakor and Tong, 2004). Theoretically, the span of the chosen window can be as short as *p+1* data points, where *p* is the model order. Practically, such a short window would be impossible to achieve for a single realization of the multivariate data set. As a result, a balance has to be maintained between time resolution (limited by stationarity) and the statistical properties of the fitted model. As a rule of thumb, the window length should possess a few times more data points than the number of estimated model parameters. If the signals are found to be non-stationary, time varying MAR model must be adopted, where the model parameters are estimated on a recursive basis (Arnold *et al.*, 1998; Gath *et al.*, 1992; Moller *et al.*, 2001). An alternative solution was offered recently (Ding *et al.*, 2000), where the collection of neural signals from successive trials is treated as an ensemble of realizations of a non-stationary stochastic process with locally stationary segments. The underlying idea is simple and elegant: if the data from successive trials of an experiment are assumed to be different realizations of the same stochastic process, we can obtain the relevant statistical properties of the signals by ensemble averaging as opposed to temporal averaging. In this way, the window length can be as small as the order of the MAR model (Liang *et al.*, 2002; Liang *et al.*, 2000).

### *6.2.2. Applications to neurophysiology*

Since the concept of causality in terms of directional influence between separate brain regions is neurophysiologically very relevant and appealing, the application of PDC and DTF is becoming gradually popular in the field of neural data analysis. Further, both of these methods make use of the global covariance structure in a single multivariate modeling framework.

First we mention the application of these causality measures to neural spike train data. Since MAR model works primarily for continuous data, usually the spike trains are convolved with a Gaussian kernel (Baccala and Sameshima, 2001b; Kaminski *et al.*, 2001). Such pre-processing is especially appropriate when the spike train contains a large number of spikes and the recordings are at least weakly stationary, but may cause distortion of original phase information when the spike trains are sparse (Zhu *et al.*, 2003). PDC analysis showed a clear directional preference from somatosensory cortex to medial thalamic nucleus during whisker twitching than other behaviors in freely moving rats (Fanselow *et al.*, 2001); this suggests that cortex plays a bigger role in sending signal triggering thalamic bursting priming the thalamocortical loop for enhanced signal detection. Another PDC study investigating exploratory activity of rats also found consistent predominant flow of information from cortex to thalamus (Baccala and Sameshima, 2001b). The application of DTF to spike trains of motor neurons in primary motor cortex of the monkey revealed an increase in inter-neuronal coupling during adaptation (Zhu *et al.*, 2003). Restoration of coupling strength to a pre-adaptation level was observed at the end of adaptation period, but the connecting architecture of the neuronal network tended to change as a result of learning and adaptation. From this latter result, the authors suggested that changing the network topology may be more efficient than changing the coupling strength when the target is to achieve a fast response. The MAR modeling thus shows its usefulness to characterize changing interaction patterns between neurons.

The DTF analysis of the LFPs recorded from rat's hippocampus and other regions of the limbic system showed a propagation of signals from CA1 field of hippocampus to enthorhinal-piriform area via subiculum area during locomotion phase but not during resting phase (Korzeniewska *et al.*, 1997). A short-term DTF analysis revealed causal influences on a millisecond time scale in the visual cortex of monkeys while performing task demanding visual pattern discrimination (Liang *et al.*, 2000). It is noteworthy that, by adopting the short-window based adaptive MAR approach by Ding *et al.* (2000), it was possible to demonstrate the rapidly changing neuronal network associated with feedforward, feedback and lateral influences in the ventral regions of the primary visual cortex. This opens up a new possibility to analyze neural data when the underlying neuronal dynamics is comprised of distinct but short cognitive processing steps. Further applications of causality measures on large scale brain signals recorded from human are found in numerous research findings (Franaszczuk *et al.*, 1994; Ginter *et al.*, 2001; Kaminski *et al.*, 1995; Kus *et al.*, 2004; Supp *et al.*, 2004; Wang *et al.*, 1992).

*6.2.3. Practical remarks*

Here, we mention some remarks that need further attention when applying multivariate linear modeling to determine causal relation in neural datasets. Causality measures such as PDC or DTF are meaningful only in statistical sense, because their computation depends completely on an estimation of the model parameters. The statistical properties of these new measures, unlike those of correlation or coherence, have not been yet properly investigated. For example, the confidence limit for testing for nonzero PDC at fixed frequencies is not analytically available (see (Kaminski *et al.*, 1997; Kaminski *et al.*, 2001) for some suggestive limit by simulations). This problem becomes acute for higher order models due to the variability of its estimated parameters. Moreover, we still do not have a clear idea as to how certain pre-processing steps (such as re-referencing or smoothing) affect the causality measures.

Finally, it should be mentioned that DTF itself cannot distinguish between indirect or direct interaction. PDC can detect direct interaction, so it must be used in combination with DTF to get further information on directional influence. But most importantly, no measures based on MAR models, such as bivariate models, can detect true causality or provide directional information if the common input is not included in the model. Nevertheless, we feel that multivariate methods are a valid alternative to pairwise correlation methods whenever a set of multichannel observations is available.

*7. Information-theory based methods*

It might be said that the different methods presented hitherto have a point in common, as they all try to establish whether there is any common *information* between the time series, as a sign of their relationship. Therefore, it has become usual to investigate directly the existence of such relationship by means of information-theoretic tools.

*7.1. Mutual information*

*7.1.1. Definition*

The bases of information theory were derived almost sixty years back (Shannon and Weaver, 1949). Central to this theory is the concept of *entropy,* which can be defined as the average amount of code necessary to encode the draws of a discrete variable *X* with *M* possible outcomes $X_i$, each of them with probability $p_i$. Thus, the Shannon entropy of this set of probabilities is:

$$I_X = -\sum_{i=1}^{M} p_i \log p_i \qquad (15)$$

This entropy is positive and is measured in bits if the base of the logarithm is 2. If the outcomes of *X* are partitionedt into *M* bins, a first estimation for $p_i$ consists in taking it as the fraction of occurrences of $X_i$ after *N* outcomes. Roughly speaking, entropy can be regarded as a measure of the uncertainty of the variable. Thus, a uniform distribution, in which all the states have the same probability, will have the largest entropy, whereas it will be minimum for a delta-type distribution (see Fig. 3).

Let us now consider a pair of random variables *X* and *Y*. The *mutual information* (MI) between them is defined as:

$$MI_{XY} = \sum p_{ij} \log \frac{p_{ij}}{p_i p_j} \qquad (16)$$

where $p_{ij}$ is the joint probability of $X=X_i$ and $Y=Y_j$. This measure, which has been also referred to as "transinformation" (Eckhorn and Popel, 1974) or "redundancy" (Palus, 1996; Panzeri *et al.*, 1999), essentially tells us how much extra information one gets from one signal by knowing the outcomes of the other one. Thus, if there is no relationship between them, $p_{ij} = p_i p_j$, so that the MI is zero for independent processes. Otherwise, $MI_{XY}$ will be positive, attaining its maximal value of $I_X$ for identical signals, although it can be modified to be equal to 1 for this latter case (Wang *et al.*, 2005). The MI is a symmetric measure (i.e., $MI_{XY}=MI_{YX}$), so that it does not provide any information about the direction of the interaction.

The MI is also very useful to study the relationship between a stimulus and its response (Borst and Theunissen, 1999; London *et al.*, 2002). In this context, if we regard *X* as the stimulus and *Y* as its response, $I_Y$ would be the entropy of the response, whereas the conditional entropy $I_{YX}$ would be this same entropy given a certain stimulus. Here, MI represents the reduction in the uncertainty of the response due to the knowledge of the stimulus, and in fact it can also be obtained as $MI_{XY}=I_Y-I_{YX}$. Figure 4 shows an example of this in the case of neural action potentials.

### 7.1.2. Estimating mutual information from time series

Despite the apparent simplicity of the measure, the practical estimation of MI from experimental time series is not an easy task. Indeed, as commented above, the easiest way of estimating the probabilities $p_i$ and $p_j$ from the corresponding variables consists in obtaining the histograms of the series of outcomes and taking, say, $p_i$ as the ratio between the number of samples in the $i_{th}$ bin of the histogram of *X* and the total number of samples. But one limitation of the MI is that, to get an accurate estimate of this measure by using such histogram-derived probabilities, one should have a large number of samples and small bins (see for instance (Quian Quiroga *et al.*, 2002a)). Moreover, if we take bins of the same size for each individual variable, it might happen that there are several values of $p_{ij}=0$ even if $p_i$ and $p_j$ are not. These null joint probabilities produce an underestimation of $MI_{XY}$ that cannot be easily cope with (Kraskov *et al.*, 2004). Optimized estimators can be defined by using adaptive bin sizes geared to produce homogeneous values of $p_{ij}$ (Darbellay and Vajda, 1999; Fraser and Swinney, 1986). Still, they also suffer from systematic errors, that can be corrected, albeit only partially, by using *ad hoc* asymptotic series (Grassberger, 1988; Roulston, 1999), a procedure often used in practice in neurophysiology (e.g. (David *et al.*, 2004; Netoff and Schiff, 2002; Quian Quiroga *et al.*, 2002a)). Recently, refined methods have been derived for the calculation of Eqn. (16), which are not based on the binning of the histograms but on entropy estimates from *k*-nearest neighbors distances (Kraskov *et al.*, 2004). These new algorithms are expected to further improve the estimation of this measure from experimental time series.

The concept of MI can be combined with the procedure of embedding a time series by means of time delay ((Takens, 1980); see also section 9), in such a way that the probabilities $p_i$ are calculated in the space of delay vectors (see e.g., (Duckrow and Albano, 2003; Kraskov *et al.*, 2004; Netoff and Schiff, 2002; Quian Quiroga *et al.*, 2002a)). The difficulties associated with this approach are similar to those already discussed for the univariate case (Kraskov *et al.*, 2004; Quian Quiroga *et al.*, 2002a), but if either the systems or their coupling are nonlinear, this multivariate extension may be more sensitive to the interdependence between the signals. If instead of time delay embedding, space embedding is used (so that the components of each vector are the values of a multichannel data set at time *t*), then we get the average MI among all these channels (i.e., the common information among all the recorded sites), which is useful to determine the global synchronization in spatially extended systems (Kraskov *et al.*, 2004).

In brief, we can summarize all the above results by saying that reliable estimations of the MI often requires a large amount of data, a constraint that is sometimes in conflict with the requisite of stationarity in the case of experimental data.

*7.2. In search of directionality in the interdependence*

Despite the difficulties inherent to their calculation, MI is a useful tool for the assessment of interdependence between experimental signals. Therefore, it would be interesting to extend this concept to rid it from one of its main limitations: the lack of asymmetry. As we have already seen for other indexes, the easiest try consists in introducing a time lag in either of the variables and calculating Eqn. (16) for different lags (Vastano and Swinney, 1988). If it attains a maximum value for a given lag of $X$ with respect to $Y$, it might be argued that $X$ is the cause and $Y$ the effect. Although this approach might be useful in certain situations, it may also be misleading and give rise to erroneous conclusions (Schreiber, 2000). However, this can be solved by incorporating dynamic structure in the index.

*7.2.1. Transfer entropy*

The concept of transfer entropy extends that of Shannon entropy by taking into account the probability of transitions, i.e., the probability $p(i_{n+1}|i_n,\ldots,i_{n-k+1})$ of obtaining a given value of $X$ at instant $n+1$ provided the $k$ former values of $X$ are known. Thus, one defines the *entropy rate* of a time series as:

$$h_X = -\sum_i p(i_{n+1}|i_n,\ldots,i_{n-k+1}) \log p(i_{n+1}|i_n,\ldots,i_{n-k+1}) \qquad (17)$$

which is in fact the difference between Shannon entropies calculated from delay vectors of dimension $k+1$ and $k$ (i.e., $h_X = I_{X^{(k+1)}} - I_{X^{(k)}}$, (Kantz and Schreiber, 2004)). Following a similar reasoning to that leading from Shannon entropy to MI, from Eqn. (17) and its equivalent for $Y$, ($h_Y$), it is possible to calculate the transfer entropy $T_{X \to Y}$, which is a measure of how the transition probabilities of $X$ (i.e., its dynamics) influence those of $Y$ (Kaiser and Schreiber, 2002; Kraskov, 2004; Schreiber, 2000). By explicitly incorporating dynamical information, transfer entropy is asymmetric (i.e., $T_{X \to Y} \neq T_{Y \to X}$ as a rule), so that it gives information about the direction of the interaction.

Unfortunately, as in the case of MI, reliable estimation of this index is only possible for large data sets, which may explain why it has found very few applications in neurophysiology yet (see (Kraskov, 2004) for an example in the case of the EEG).

*7.3. Applications to neurophysiology*

*7.3.1. Information theory, neural coding and synapse efficiency*

As commented above, measures derived from information theory are especially suited to study stimulus-response relationships. The importance of this methodology in the field of neural computation was early recognized (Eckhorn and Popel, 1974; Stein *et al.*, 1972). Later, information theory-based measures were able to detect weak interdependences and to give a better estimation of synaptic connectivity than traditional linear tools (Yamada *et al.*, 1993). These methods were also useful to estimate the maximum information that can be transmitted by a neuron as a function of its firing rate (Wessel *et al.*, 1996), to validate stimulus-response models in real data (Theunissen and Miller, 1991) and to determine the minimum time scale over which neural responses can convey information (Buracas *et al.*, 1998; de Ruyter van Steveninck *et al.*, 1997; Panzeri *et al.*, 1999). In this same context, MI has proven useful both to evaluate the effectiveness of different stimuli in generating various values of information in the encoding neurons (Machens, 2002) -from which it has been hypothesized that neural processing is optimized to represent natural stimuli (Borst and Theunissen, 1999; Hsu *et al.*, 2004)- and to calculate the absolute amount of information transmitted as a test of the goodness of encoding models. In this latter case, interestingly, the results obtained have led several authors to the conclusion that nonlinear encoding models are superior to traditional linear ones (Borst and Theunissen, 1999; Buracas *et al.*, 1998; de Ruyter van Steveninck *et al.*, 1997; Sharpee *et al.*, 2004).

MI has been also used as an index of synaptic efficacy by measuring the relevance of different input parameters such as synaptic position –distal or proximal–, synaptic charge and time-to-peak of the excitatory post-synaptic potentials, as schematized in Fig. 4 (London *et al.*, 2002). This study demonstrated, using both neural models and real data, that the nonlinearities inherent to neurons turn this index into an appropriate tool to study this question. More recently, an *in vitro* result confirmed the ability of primary afferent synapses to transmit the main temporal structure of chaotic impulse trains (Wan *et al.*, 2004), in a study where the dominant role of brief-burst stimulation over single presynaptic action potentials was also suggested in agreement with previous results (Lisman, 1997).

To conclude this section, we mention three recent theoretical works that have shed light on relevant aspects of the application of information theory in neurophysiology. The first one demonstrates that a newly derived directionality measure based on the concept of entropy can be used to detect causal interdependencies between neurons (Dzakpasu and Zochowski, 2005), although results are restricted so far to neural models. The second one has derived an optimized entropy estimator, which performs very well in small data samples, thus opening new possibilities for the information theoretic analysis of neural responses (Nemenman *et al.*, 2004). And finally, the third one has shown the validity in neurons of the data processing inequality (Tiesinga, 2001), which states that, given three random variables *X, Y* and *Z* that form a chain in the order $X \to Y \to Z$, MI between X and Y is greater than or equal to the MI between X and Z. In other words, information cannot be recovered after it has been degraded.

### 7.3.2. Information-theoretic measures in EEG and MEG data

The MI and other indexes derived from information theory have been comparatively less applied to EEG and MEG data than they have been to study neural code. In any case, and since the classical work of Lopes da Silva *et al.* (1989) in real and simulated EEG data, there have been a number of studies on these signals using this methodology.

A first set of studies was aimed to determine whether there are differences between control groups and groups of patients with various neurological disorders, in terms of the flux of information between different cortical areas. Thus, it has been shown that schizophrenic patients present higher intra- and interhemispheric average MI than the normal control group (Na *et al.*, 2002). A similar result was found in patients with Alzheimer's disease, mainly in the frontal and anterior-temporal regions (Jeong *et al.*, 2001). Regarding epilepsy, it has been reported that the information content of the EEG severely decreases prior to a seizure (Chen *et al.*, 2000; Trabka *et al.*, 1989), so that the MI may be used as a good seizure predictor (Kraskov, 2004; Kreuz, 2004) as well as to identify the site of seizure onset (Kraskov, 2004; Mars *et al.*, 1985), a question that, in this context, has been also addressed by asymmetric information measures similar to transfer entropy (Palus *et al.*, 2001). Finally, a variant of the entropy, in which probabilities are calculated not in the time domain but in the time-frequency representation of the EEG provided by wavelets (Quian Quiroga *et al.*, 2001; Rosso *et al.*, 2001), has been used to trace the response of rat's brain to ischemia (Al-Nashash *et al.*, 2003). A strong correlation is found between this entropy in theta, alpha and beta bands and the injury-recovery cycle (see also (Thakor and Tong, 2004)).

A second set of studies has dealt with the working of brain areas during the performance of different tasks. Magnetic field tomography was used to study brain activity during an object and emotion recognition task (Ioannides, 2001; Ioannides *et*

*al.*, 2000), in which MI was able to identify the spatial extension and time course of the brain activity associated with the process. Additionally, this tool detected changes in cortico-cortical connectivity during odor stimulation in subjects classified by occupation (Min *et al.*, 2003). In fact, it showed a frontal specificity of these changes in professional perfume researchers as compared to perfume salespersons and general workers, a result similar to the increase in synchronization found in a group of artists as compared to non-artists in other cognitive tasks (see sections 8.2.1.2.3 and 9.3).

The forecasting ability of information theory measures has also been tested. Thus, an artificial neural network trained with data stemming from the series of MI values between four EEG electrodes was able to classify correctly a high percentage of patients according to their response to anesthesia (Huang *et al.*, 2003). MI was also applied to distinguish between two groups of epileptic patients with different interictal epileptiform discharges during sleep (Varma *et al.*, 1997).

There is another useful application of MI in neurophysiology, which may seem a little counterintuitive at first sight. It is related to its theoretical meaning, which directly estimates the statistical dependence between two time series. This feature can be used for the purpose of blind separation of different sources of activity that integrate to produce a given signal, an idea similar to that carried out in the framework of independent component analysis (Kraskov, 2004; Kraskov *et al.*, 2004; Kreuz, 2004; Stögbauer *et al.*, 2004). This approach has been also used to discriminate between sources of electrical and magnetic activity in combined EEG-MEG recordings, by minimizing the information shared between the gain matrices of both functions (Baillet *et al.*, 1999).

*8. The concept of phase synchronization*

It is well known at present that the phases of two coupled nonlinear (noisy or chaotic) oscillators may synchronize even if their amplitudes remain uncorrelated, a state referred to as *PS* (Pikovsky *et al.*, 2001). By synchronization, it is meant here that the following *phase locking* condition applies for any time *t*:

$$\varphi_{n,m}(t) = |n\phi_x(t) - m\phi_y(t)| \leq \text{constant} \tag{18}$$

where $\phi_x(t)$ and $\phi_y(t)$ are the phases of the signals associated to each system defined on the real line (unwrapped). An example of this is shown in Fig. 5. However, experimental signals are often noisy, and exhibit random phase slips of $2\pi$, so that the fulfillment of Eqn. (18) is normally analyzed from the so-called *cyclic relative phase* $\varphi'_{n,m}(t) = \varphi_{n,m}(t) \bmod 2\pi$ (i.e., the relative phase difference wrapped to the interval $[0,2\pi)$). Therefore, in the case of neurophysiological signals, the phase locking condition must be understood in a statistical sense, for instance, as the existence of a preferred value in the distribution of $\varphi'_{n,m}(t)$ (Rosenblum et al., 2001).

One of the first experimental observations of synchronization between two systems was reported by the Dutch scientist Christiaan Huygens more than three centuries ago (Huygens, 1673): two pendulum clocks hanging from the same beam got synchronized by attaining their maximal amplitudes at the same time but at opposite extremes thanks to the weakly coupling provided by the vibration of the beam in response to their movement. In this case, both clocks had opposite phases, so that the phase difference in Eqn. (18) was equal to π. It is important to note, however, that these pendulum clocks were actually harmonic linear oscillators; the fact that Eqn. (18) may also hold for coupled chaotic oscillators was not proven until recently (Rosenblum *et al.*, 1996).

In order to study the existence of PS synchronization between experimental signals, it is first necessary to obtain their phases, as explained henceforth.

*8.1. Extracting the phases*

Two closely related approaches are mainly used to obtain the phases of a neurophysiological signal. In both cases, the original real-valued signal *x(t)* is transformed with the help of an auxiliary function and turned into a complex-valued signal, from which an instantaneous value of the phase is easily obtained. The most commonly used of these two approaches involves the use of the Hilbert transform (HT), whereby the analytical signal *ζ(t)* is obtained:

$$\zeta(t) = x(t) + i\, x_H(t) \tag{19}$$

where *xH(t)* is the HT of *x(t)*, defined as:

$$x_H(t) = \frac{1}{\pi} P.V. \int_{-\infty}^{\infty} \frac{x(t')}{t-t'} dt' \tag{20}$$

with P.V. denoting the Cauchy principal value.

The second approach makes use of the wavelet transform, and was recently introduced for EEG signal analysis (Lachaux *et al.*, 2000; Lachaux *et al.*, 1999). Here, the function used is the complex Morlet wavelet:

$$\psi(t) = (e^{i\omega_o t} - e^{-\omega_o^2 \sigma_t^2 /2}) e^{-t^2/2\sigma_t^2} \tag{21}$$

where $\omega_o$ and $\sigma_t$ are the centre frequency of the wavelet and a bandwidth parameter determining its rate of decay, respectively[2] (see Fig. 6). Its width $m=\omega_o \sigma_t^2$ determines how many cycles of the corresponding frequency it comprises. It is interesting to note that the generally used Morlet wavelet $\psi(t) = e^{i\omega_o t} e^{-t^2/2\sigma_t^2}$ satisfies the zero mean admissibility condition of a wavelet only for large $\sigma_t$ (when it comprises several oscillations). The additional negative term in Eqn. (21) is introduced in order to avoid spurious effects especially if the signal to be analyzed has nonzero mean or low frequency components (Quian Quiroga *et al.*, 2002a). The Morlet wavelet has a Gaussian modulation both in the time and in the frequency domains and therefore it has an optimal time and frequency resolution (Mallat, 1999; Sinkkonen *et al.*, 1995), a feature that makes it very suitable for the analysis of EEG signals (see e.g. (Sinkkonen *et al.*, 1995; Tallon-Baudry *et al.*, 1996) as earlier examples in this field).

---

[2] The exact expressions for wavelet duration and spectral bandwidth are $2\sigma_t$ and $1/\pi\sigma_t$, respectively.

If this complex wavelet is then convolved with the original signal, one gets:

$$W(t) = (\psi \circ x)(t) = \int \psi(t')x(t-t')dt' = A^W(t)e^{i\phi_W(t)} \quad (22)$$

thereby obtaining an estimation of the phase, $\phi_W(t)$, for each sample.

It has been recently shown that the application of both approaches (i.e., the HT and the wavelet transform) produces essentially the same result (Quian Quiroga *et al.*, 2002a). The main difference between them is that the HT is actually a filter with unit gain at every frequency (Rosenblum *et al.*, 1996), so that the whole range of frequencies is taken into account to define $\phi_H(t)$. Therefore, if the signal is broadband –as it usually happens with the EEG–, it is necessary to pre-filter it in the frequency band of interest before applying the HT, in order to get a proper value of the phase (e.g. (Angelini *et al.*, 2004; Bhattacharya *et al.*, 2001b; Koskinen *et al.*, 2001)). On the other hand, $W(t)$ is non zero only for those frequencies close to $\omega_o$, so that this approach is equivalent to band-pass filtering $x(t)$ at this frequency, which makes the pre-filtering unnecessary.

A third method for the extraction of the phases from experimental time series is especially suited for signals presenting marked events (Rosenblum *et al.*, 2001). Here, each event is regarded as the completion of a new cycle, so that if $t_k$ and $t_{k+1}$ are the times at which events $k$ and $k+1$ take place, there is an increment of $2\pi$ in the phase of the signal between these times. The phase at every instant in-between is defined by linear interpolation. Nevertheless, usually the two approaches mentioned above are preferred for EEG/MEG signals.

*8.2. The assessment of phase synchronization*

*8.2.1. Bivariate phase synchronization*

The most common scenario for the assessment of PS in neurophysiology entails the analysis of the synchronization between pairs of signals. Thus, in a typical experimental set-up, *q* different channels of, say, EEG signals, are recorded, and one studies the PS between each pair of electrodes (*i,j*), from which the series of phases $\phi_i(t)$ and $\phi_j(t)$ have been previously extracted. The direct analysis of the unwrapped phase difference $\varphi_{n,m}(t)$ has been seldom used in practical applications, although it showed the presence of synchronized regimes in human postural control, as the appearance of PS between body sway in anterior-posterior and lateral directions from the stabilograms of two groups of neurological patients (Rosenblum *et al.*, 1998). But as we have already commented, PS between neurophysiological signals must be understood in a statistical sense, and its assessment must be carried out accordingly. Henceforth we detail the different approaches used for such purpose up to this date.

*8.2.1.1. Indexes of bivariate phase synchronization*

Three different indexes have been mostly employed in practice to assess the degree of PS between two signals. The first one is based on information theory and makes use of the concept of conditional probability (Rosenblum *et al.*, 2001). It works by wrapping the individual phases $\phi_x(t)$ and $\phi_y(t)$ into the intervals [0,2π*m*) and [0,2π*n*), respectively. These intervals are then divided into *L* bins, and the index measures the probability of one of the phases to belong to a certain bin provided the other one has a given value:

$$\lambda_{n,m} = \frac{1}{L}\sum_{l=1}^{L}|r_l(t_j)| \qquad (23)$$

where index $j$ corresponds to time, $r_l(t_j)=(1/M_l)\sum e^{i\phi_Y(t_j)}$ for all $j$, such that $\phi_X(t_j)$ belongs to bin $l$ and $M_l$ is the number of points in this bin[3]. This approach is also termed as the *stroboscopic approach*: the phase of one of the oscillators is observed at those instants where that of the other one attains a certain value, and then averaged over all the possible values.

The other two indexes make use of the relative phase difference, $\varphi'_{n,m}(t)$. The second one is termed as *mean phase coherence* (Hoke *et al.*, 1989; Mormann *et al.*, 2000), although it has been also called the *phase locking value* (Lachaux *et al.*, 1999) or the *synchrony factor* (Tallon-Baudry *et al.*, 2001) in neurophysiological applications. It makes use of the concept of directional statistics (Mardia, 1972), and is defined as:

$$\gamma_{n,m} = \sqrt{\left|\left\langle e^{i\varphi'_{n,m}(t)}\right\rangle\right|} = \sqrt{\left\langle \cos\varphi'_{n,m}(t)\right\rangle^2 + \left\langle \sin\varphi'_{n,m}(t)\right\rangle^2} \qquad (24)$$

where $\langle\cdot\rangle$ denotes average over time. It is the intensity of the first mode of the distribution of $\varphi'_{n,m}(t)$ (Rosenblum *et al.*, 2001), which in simpler words means that it is a measure of how the relative phase is distributed over the unit circle. If the two signals are phase synchronized, the relative phase will occupy a small portion of the circle and mean phase coherence is high. On the contrary, lack of PS gives rise to a relative phase that spreads out over the entire unit circle and mean phase coherence is very low (see left side of Fig. 7).

---

[3] Unfortunately, it is not possible to get any information about the direction of the coupling by calculating the index in the opposite sense (i.e., taking the conditional probability on signal *Y*) because the index is symmetric.

Finally, the third index also makes use of $\varphi'_{n,m}(t)$, although in this case it directly studies its distribution by partitioning the interval [0,2π) into $L$ bins[4] and comparing it with the distribution of the cyclic relative phase obtained from two series of independent phases (Tass *et al.*, 1998). This comparison is carried out by estimating the Shannon entropy (Eqn. (15)) of both distributions (i.e., that of the original phases, $I_\varphi$, and that of the independent phases, $I_{MAX}$). As usually, the probability of a phase difference to belong to the $i_{th}$ bin, $p_i$, is normally estimated by taking the ratio between the number of phases in this bin and the total number of samples $N$. In Fig. 3 we already showed an example of three distributions with their corresponding entropies for a given $N$. In the case of independent phases, no preferred value of $\varphi'_{n,m}(t)$ is expected, so that the distribution should be uniform, and $p_i = L/N$, which would give the maximal value $I_{MAX} = \log L$, whereas in the synchronized state the phase distribution presents a sharp peak and thus has a low entropy. However, due to finite size effects, it has been shown that the distribution of phase differences is not uniform as a rule even for completely uncorrelated series ((Kreuz, 2004); top right plot of Fig. 7). Therefore, $I_{MAX}$ must be estimated ad hoc by constructing a set of two independent phases, which can be done, for instance, by randomly shuffling one of the phases while keeping the other unchanged (see *Appendix A* for details). Then, a normalized PS index can be obtained as (see also right side of Fig. 7):

$$\rho_{n,m} = \frac{I_{MAX} - I_\varphi}{I_{MAX}} \qquad (25)$$

The three indexes presented above have the same range of variation: they are close to 0 for uncorrelated signals, whereas they approach 1 if there is strong PS.

---

[4] The optimal number of bins can be calculated exactly as a function of the length of both time series. Thus, for $N$ samples, one has $L = e^{0.626 + 0.4\ln(N-1)}$ (Otnes and Enochson, 1972).

However, they do present some differences. Comprehensive comparative studies of their performance have been carried out in computer simulated as well as in EEG and MEG data (Kreuz, 2004; Rosenblum *et al.*, 2001; Tass *et al.*, 1998). From these results, we can draw some important conclusions. In assessing 1:1 synchronization (the most typical case), $\rho_{l,l}$ strongly depends on the number of bins used to calculate the histogram of $\varphi'_{n,m}(t)$, so that low values of this index can be obtained even for perfect PS in numerical experiments (Rosenblum *et al.*, 2004). On the other hand, $\rho_{l,l}$ and $\gamma_{l,l}$ present the greatest sensibility to the transition from weak coupling to PS state. In contrast, the conditional probability index $\lambda_{l,l}$ is non-zero even in cases where no PS but only very weak interactions are present. Additionally, it has a straightforward way of determining a significance threshold from the data, which may be very useful to avoid spurious detection of PS in uncoupled signals (Kreuz, 2004). Thus, it can be concluded that this index is the right choice for the assessment of weak phase coupling. On the other hand, if our purpose is to cluster a number of time series into groups according to their degree of coupling with a given reference, the other two indexes are more suitable (Rosenblum *et al.*, 2001).

Interesting subtleties must be taken into account when the signals present *n:m* synchronization, with *n>1* and/or *m>1*. As already mentioned, this is certainly not the most usual case in neurophysiological signals, where 1:1 synchronization is almost always the case under study (Angelini *et al.*, 2004; Bhattacharya *et al.*, 2001a; b; David *et al.*, 2004; Kreuz, 2004; Le van Quyen *et al.*, 2001; Mormann *et al.*, 2000; Quian Quiroga *et al.*, 2002a). However, higher order PS is not unlikely to be present in these data (Bhattacharya and Petsche, 2005b; Palva *et al.*, 2005; Tass *et al.*, 1998) and actually, the existence of locking (or entrainment) between frequencies that are close to rational relation ($n\omega_1 \approx m\omega_2$) is one of the hallmarks of this nonlinear phenomenon. If the signals are investigated for 1:1 synchronization but they present, say, 1:2 synchronization, the distribution of phase differences is multimodal, and presents two marked peaks instead of one, which makes the approach based on this distribution useless (Kreuz, 2004). This case of broad distribution of the relative phase can also occur in systems presenting modulated natural frequency, where even if the correct synchronization regime is detected, both $\rho_{n,m}$ and $\gamma_{n,m}$ are of little use. (Rosenblum *et al.*, 2001). So in these cases, the use of $\lambda_{n,m}$ for the assessment of PS is recommended, although other indexes such as Eqn. (25) can certainly be used as well (Bhattacharya and Petsche, 2005b).

*8.2.1.2. Applications to neurophysiology*

*8.2.1.2.1. First results*

The idea of searching for a relationship between the phases of two neurophysiological signals is certainly not new, and early examples can be found more than thirty years ago (e.g. (Butler and Glass, 1974)). But in those cases, the phases were obtained for the harmonics of the FT. However, the concept of PS is based in the assumption that there is a dominant frequency in the signal that leads to a well defined, unique value of the phase for each interacting oscillator. All the pre-processing steps (such as band pass filtering) are aimed to extract this signal from its mixture with other signal and (possibly) broadband noise (Rosenblum *et al.*, 2004). Therefore, if we restrict ourselves to the PS approach, the work of Tass *et al.* (1998) must be regarded as the first application of this idea in neurophysiology. In this pioneering study, the authors extended the concept of PS of chaotic oscillators derived in an earlier work (Rosenblum *et al.*, 1996) to analyze the relationship between the phases of MEGs and records of the muscle activity in a Parkinsonian patient. It is suggested there that the temporal evolution of the peripheral tremor rhythms directly reflects the time course of the synchronization of abnormal activity between cortical motor areas. This work is not only important for their neurological implications, but also for showing that the existence of PS, understood in a statistical sense, can be traced even in noisy experimental signals from neurophysiological records.

*8.2.1.2.2. Phase synchronization and the gamma band*

Another key work in the field demonstrated shortly afterwards the existence of long-range PS in the gamma band (~ 20-60 Hz) of the EEG (Rodriguez *et al.*, 1999). Synchronization in the gamma band of the EEG is thought to reflect the appearing of an integrative mechanism bringing together widely distributed sets of neurons to effectively carry out different cognitive tasks (Damasio, 1990; Roelfsema *et al.*, 1997; Singer and Gray, 1995; Tallon-Baudry and Bertrand, 1999; Varela, 1995). In their work, Rodriguez *et al.* (1999) found increased PS with a latency of 260 ms after the stimulus in the frequency range between 35 and 45 Hz in a group of adult human subjects during visual perception of faces, as opposed to no-perception situation. It must be noted, however, that a recent re-examination of this result, despite reporting similar qualitative outcomes, has indicated the possible effect of the analysis methods as well as the records used, suggesting that PS also occurs during non-visual perception but in a different frequency band (Trujillo *et al.*, 2005).

Another interesting result reported by Rodriguez *et al.* (1999) is the existence of a period of strong desynchronization with latency between 400 and 650 ms after the stimulus, which allegedly reflects the active uncoupling of the neural ensembles necessary to proceed from one cognitive state (visual perception) to another (motor activation) (Rodriguez *et al.*, 1999; Varela, 1995; Varela *et al.*, 2001). The dynamics of such periods of synchronization-desynchronization in the gamma band, which is schematically portrayed in Fig. 8, has been further studied in subsequent papers. In this regard, PS indexes have been used to get insight about the functioning of the sensory cortices (Freeman and Rogers, 2002), and they have also shown the important role of such periods during visual attention in humans (Gross *et al.*, 2004). The latest findings and perspectives on the concept of long-range neuronal synchronization and desynchronization in motor control and cognition in normal as well as pathological conditions have been recently reviewed elsewhere (Schnitzler and Gross, 2005).

Interestingly, PS and desynchronization in the gamma band is also important for the successful formation of declarative memory, as demonstrated by the analysis of the relationship between human EEGs from the hippocampus and the rhinal cortex (Fell *et al.*, 2001). In a later paper, the interaction between gamma band PS and coherence in the theta band was studied (Fell *et al.*, 2003). The authors concluded that, whereas rhinal-hippocampal gamma EEG PS may be closely related to actual memory processes by enabling fast coupling and decoupling of the two structures, theta coherence might be associated with slowly modulated coupling related to an encoding state.

The dynamic patterns of phase clustering and desynchronization have been also analyzed theoretically by means of a model of sparsely coupled neural cell assemblies and checked also in human EEG data (Breakspear *et al.*, 2004).

*8.2.1.2.3. Phase synchronization and the effect of training*

The significance of PS between high frequency EEG bands has been further investigated in a series of studies, in which the performance of trained artists was compared with that of non-artists during perception of music (Bhattacharya and Petsche, 2001; Bhattacharya *et al.*, 2001a; b) and of paintings (Bhattacharya and Petsche, 2002; Bhattacharya and Petsche, 2005a). Interestingly, the artists presented higher degree of PS than non-artists only during the performance of those tasks for which they were trained. Thus, musicians showed higher gamma band PS than untrained subjects during listening to different pieces of music, but listening to a text of neutral content did not elicit such difference (Bhattacharya and Petsche, 2001; Bhattacharya *et al.*, 2001a). In painters, differences during perception were apparent not only in gamma band, but also in beta band –in which PS has been also found during the rehearsal of an object in visual short-term memory (Tallon-Baudry *et al.*, 2001; Tallon-Baudry *et al.*, 1998). Moreover, PS in low frequency bands (mainly delta) appeared during imagery, where also a desynchronization of the activity in alpha band is reported (Bhattacharya and Petsche, 2002; Bhattacharya and Petsche, 2005a). Another interesting result of these works is the existence of hemispheric dominance in artists as compared to non-artists during artistry perception.

A recent study has confirmed the influence of training on gamma band PS (Lutz *et al.*, 2004). In fact, non-specific "meditation" elicits a sharp increase of this synchronization over a wide range of scalp electrodes in subjects with long-term meditative instruction in Buddhist mental training practices, as compared to normal controls. Contrary to the case of artists, differences were not task-related, but appeared also in baseline. Thus, the increase in PS is here a permanent feature related to mental training, and cannot be associated with either increased attention or concentration during the task – elements that are known to mediate task-related high frequency synchronization in humans (Fries *et al.*, 2001; Lutz *et al.*, 2002; Tallon-Baudry *et al.*, 1997).

*8.2.1.2.4. Phase synchronization during motor tasks*

The performance of different motor tasks is another situation giving rise to PS between brain areas. Thus, an interesting work using fMRI activation data found significant PS between the voxel time series and the reference function of an event-related finger-tapping task (Laird *et al.*, 2002). Certainly, these two signals cannot be regarded as stemming from self-sustained oscillators, so this result must be interpreted with caution, yet the PS approach was still useful here in revealing additional information on the complex nature of the fMRI time series (Lin *et al.*, 2004). Phase locking between primary contralateral and secondary ipsilateral sensorimotor cortex was later found in the gamma band of the MEG of healthy volunteers around 80 ms. after the stimulation of right median nerve at the wrist (Simoes *et al.*, 2003). Interestingly, these authors demonstrated that PS was not due to common synchronization of both areas to the stimulus, but to a true significant PS within the sensorimotor system. Distinct PS patterns in the gamma band of the MEG of somatosensory cortex devoted to hand control have been also found during finger tapping (Tass *et al.*, 2003) and also in respond to thumb and little finger stimulation (Tecchio *et al.*, 2004)

Recently, the synchronous activity of a group of single neurons was analyzed in the caudal supplementary motor area of monkeys during the performance of visually guided hand movements (Lee, 2003). The author reported that oscillatory gamma activity of the average population of neurons showed a strong tendency to synchronize immediately before and after the onset of the movement. Furthermore, the duration of the synchronization increased when the onset of the next target was delayed, suggesting that gamma band synchronization is related to anticipation and attention. Interestingly, the phases of coherent oscillations for a given neuron pair were often similar across different movements, showing that PS is also present between individual neurons even at sub threshold level (Freeman and Rogers, 2002; Makarenko and Llinás, 1998).

*8.2.1.2.5. Phase synchronization and pathological brain rhythms*

The concept of PS can be also used to get insight about the mechanisms underlying certain neurological pathologies, such as Parkinson's disease (Tass *et al.*, 1998), mania (Bhattacharya, 2001), migraine (Angelini *et al.*, 2004) and especially epilepsy (Bhattacharya, 2001; Jerger *et al.*, 2001; Kraskov, 2004; Kraskov *et al.*, 2002; Kreuz, 2004; Kreuz *et al.*, 2004; Le van Quyen *et al.*, 2001; Mormann *et al.*, 2003; Mormann *et al.*, 2000). In all these works, decreased long-range synchronization for subjects with pathologies as compared with controls is the typical result, although in the case of migraine, differences were apparent only during visual stimuli, but not in spontaneous EEG (Angelini *et al.*, 2004). In epilepsy, the decrease in synchronization presents a dynamical character, because a further decrease in baseline PS is normally detected prior to the onset of a seizure (Kreuz *et al.*, 2004; Mormann *et al.*, 2003; Mormann *et al.*, 2000). Additionally, the average synchronization in the focal hemisphere of the brain of epileptic patients is higher than in the non-focal hemisphere during the interictal period (Kraskov, 2004). The existence of PS in this pathology has also been studied *in vitro* by using advanced dual-cell patch-clamp techniques, where seizure-like activity was pharmacologically induced in pyramidal neurons (Netoff and Schiff, 2002). This work also raised an interesting question, because the authors concluded that a linear method (namely, the cross-correlation function) was the most sensitive for detecting PS during periods of highly burst, whereas the nonlinear methodology prevailed in the pre-seizure state.

Although not pathological, special rhythms appear as well in the brain activity of subjects under anesthesia. Also here, specific patterns of PS among different cortical areas and different frequency bands have been found (Koskinen *et al.*, 2001), with highly asymmetric behavior between the induction and the recovery periods in most of the low frequency bands (<20 Hz).

### *8.2.1.2.6. Implications of the results*

Taken together, the above results have a double implication. From the analytical point of view, it is evident that the analysis of bivariate PS from neurophysiological signals yields new information about the dynamical co-operation between neuronal assemblies. From a neurophysiological perspective, it has been claimed that this kind of synchronization in high frequency bands might be one of the mechanisms contributing to the so-called temporal binding model of perception (Engel and Singer, 2001; Treisman, 1996), a model against which, however, important concerns have been also raised (Shadlen and Movshon, 1999). These studies of PS in the gamma band have been also very helpful in proposing the so-called 'match-and-utilization model', which aims at explaining the significance of gamma band activity in human as well as in animals within a common framework (Herrmann *et al.*, 2004). According to this model, early gamma band responses (with latency after stimulus lower than 150 ms) would be the result of a match between the stimulus (bottom-up information) and the memory (top-down information), whereas late gamma band responses (latency greater than 200 ms) would reflect the readout and utilization of the information resulting from this match. If the model proves correct, it would allow not only understanding the generation of gamma band oscillations but also predicting the appearing of these oscillations in different experiments.

Finally, we have also seen that PS analysis represents a very useful tool for the assessment of both dynamical and permanent changes in the connectivity patterns of cortical brain in connection with abnormal brain rhythms.

### 8.2.2. In search for directionality and delay in phase synchronization

Despite the usefulness of the bivariate PS indexes reviewed above, they are in principle not suited to give any information about the directionality (if any) of the coupling that produces the synchronization. However, it is indeed possible to study whether such directionality exists, provided we make some assumptions about the dynamics of the phases. Thus, if we assume that the interaction between the two systems is weak, the influence of the amplitudes on the phase dynamics can be neglected, so that this dynamics can be modeled mathematically with a relatively simple equation that does not include the amplitudes (Rosenblum and Pikovsky, 2001). The basic idea is that, if the dynamics of $\phi_x(t)$ depends more on $\phi_y(t)$ than vice versa, then there should exist a certain direction in the interaction from *Y* to *X*, which can be assessed from the series of instantaneous phases by means of different indexes. We refer the interested reader to the appropriate literature for details (Rosenblum *et al.*, 2002; Rosenblum and Pikovsky, 2001; Smirnov and Andrzejak, 2005; Smirnov and Bezruchko, 2003). Interestingly this approach has been already applied in neuroscience with promising results (Cimponeriu *et al.*, 2003; Gross *et al.*, 2002).

More recently, the question of estimating the delay in the PS from time series has been also addressed (Cimponeriu *et al.*, 2004). The existence of delay is an issue of especial relevance in those cases in which the propagation time of the signal through the pathway connecting the interacting systems cannot be neglected in comparison with the characteristic oscillation period (see (Golomb *et al.*, 2001) for an example in the context of neural populations). The identification of the delay by analyzing the interrelations between the series of instantaneous phases of both signals is indeed possible provided the interacting oscillators are sufficiently noisy (Cimponeriu *et al.*, 2004), because noise plays a constructive role by disrupting the coherence between the current and the past states. These authors also show that, in the case of nonlinear self-sustained oscillators, information about the delay cannot be accurately estimated by using linear techniques such as cross-correlation or coherence. Although there are no applications of this method in neurophysiology yet, the role of noise should turn neurophysiological signals into good candidates for the application of this algorithm.

*8.3. Global synchronization of interacting oscillators*

*8.3.1. Synchronization cluster*

Hitherto we have dwelt on the different aspects of assessing pairwise PS. However, as already commented, in many neurological studies it may be interesting to investigate the degree of overall synchronization in a group of multivariate channels. A picture of the PS of the ensemble can be obtained by estimating the pairwise synchronization between every possible pair of channels and then connecting the corresponding sites with lines of different thickness or color according to the strength of their interaction. Changes in PS between a control and a trial situation can be also represented in this way (Bhattacharya *et al.*, 2001b; Rodriguez *et al.*, 1999). Another possibility consists in averaging the corresponding index for all the possible pairs of electrodes, thereby obtaining a raw estimation of the mean PS of the ensemble (see for instance (van Putten, 2003)). Nevertheless, it has been pointed out that these approaches present some drawbacks (Allefeld and Kurths, 2004a). On the one hand, the pairwise representation might be difficult to interpret if the number of lines is large. Additionally, this picture conveys no information about the common integrating structure among the ensemble of electrodes. On the other hand, the averaging approach, as well as other related ones aimed to obtain overall indexes of PS among all recording sites (Haig *et al.*, 2000) normally are not able to either give topographic details or preserve much of the information present in the data. Instead, a new method was recently proposed (Allefeld and Kurths, 2004a), which combines both perspectives (i.e., global and local one) by conceiving the ensemble of oscillators (the different recording sites) as being part of a *cluster* to which each oscillator contributes in different proportion. The cluster has a common rhythm, which is an average of the oscillations of each oscillator, and its dynamics is described by a cluster phase, where the degree of participation of each individual channel can be checked by assessing the PS between the global and the individual phases. Thus, the bivariate PS index based on the conditional probability (Eqn. (23)) between every possible pair of channels is used to estimate the contribution of the individual oscillator to the cluster, thereby obtaining an estimation of the strength of the synchronization of each site with the overall system. This algorithm uses only a few assumptions of the dynamics underlying the data, which makes it useful as a generic data analysis method. Indeed, it has been successfully used to show differences among different experimental situations in the field of cognitive neuroscience, where it provided information about brain dynamics in a time and frequency-specific way (Allefeld and Kurths, 2004a). Hence, it is expected to be useful in EEG studies where the assessment of overall synchronization of multivariate data is the main goal.

*8.3.2. Synchronization in populations of oscillators: the mean field approach*

Although not directly related to the issue of PS analysis, the study of the synchronization of large populations of oscillators as well as its control is a relevant subject in neuroscience, which has recently received great attention (Golomb *et al.*, 2001; Montbrio *et al.*, 2004; Rosenblum and Pikovsky, 2004a; Rosenblum and Pikovsky, 2004b; Tass, 1999). We cover it here because it might be better understood in connection with the concept of globar cluster we have just reviewed.

Briefly, an appropriate model for many systems composed of a population of noisy chaotic oscillators is one in which the elements of the population are supposed to be globally coupled (i.e., where they are all each-to-each coupled, e.g., (Golomb *et al.*, 2001)). In this framework, each unit is regarded as being driven by the force $\varepsilon \bar{X}$, where $\bar{X} = \sum_{i=1}^{N} x_i$ is the *mean field* of the ensemble, $x_i$ is an observable of the $i_{th}$ unit and $\varepsilon$ is a parameter that quantifies the strength of the interaction between the units. In neurophysiology, the counterpart would be a large neural network where all the neurons are interconnected with each other. If $\varepsilon$ remains below a critical value, $\varepsilon_{CR}$, the variance of $\bar{X}$ is small (negligible if $N\to\infty$). Otherwise, macroscopic oscillations of the mean field appear.

Interestingly, the onset of rhythmical brain activity in Parkinson's disease can be regarded as a transition between the former and the later state in large neural populations (Goldberg *et al.*, 2004). Deep brain stimulation, which refers to the electrical stimulation of subcortical brain structures by means of periodic pulse train delivered via a chronically implanted electrode, has proven successful in the treatment of Parkinsonian patients (Titcombe *et al.*, 2001). But the mechanism by which this kind of stimulation reduces the symptoms in these patients is still unknown. Moreover, its efficiency decreases with time due to adaptation to the stimulation, so that it is necessary to understand the way in which this stimulation suppresses the synchronized behavior in order to improve its therapeutic efficiency. In this line, it has been recently shown that collective synchronization can be controlled by using different time-delayed feedback strategies (Rosenblum and Pikovsky, 2004b), in which the past values of the mean field are "reinjected" in the population via external feedback control. The value of the control signal is proportional to the degree of coupling, but can be of very low magnitude as long as it is delivered with the appropriate delay and frequency. Although there have been no practical applications of this result yet, the authors suggest that it might be useful for neuroscientists working on neuronal oscillations in brain slices. They also indicate the possibility of combining this technique with phase-resetting techniques recently applied to control the oscillations of neuronal populations (Tass, 2003).

*9. Assessment of synchronization in state space*

Neurons are highly nonlinear devices, which in some cases show chaotic behavior (Matsumoto and Tsuda, 1988). Thus, the study of their collective activity, as measured by EEG or MEG, could profit from the use of nonlinear measures derived for the study of chaotic dynamical systems. First encouraging results claimed that macroscopic EEG signals also have chaotic structure (Babloyantz *et al.*, 1985), further studies dit not find any strong evidence of chaos in EEG (Pijn, 1990; Theiler *et al.*, 1992; Theiler and Rapp, 1996). At present, there is wide consensus that EEG signals are, at least in a general sense, not (low-dimensionally) chaotic (Lehnertz *et al.*, 2000). In spite of that, as already mentioned in the Introduction, nonlinear chaotic measures are still used with a more pragmatic approach, as invariant quantities from the representation of the signals in a phase space and even if there is no sign of chaoticity, such phase space representation may reveal nonlinear structures hidden to standard linear approaches (see for instance (Stam, 2005)). The present section details on how measures of synchronization can be defined from a state space reconstruction of the signals and describes some applications to EEG data.

The study of synchronization between chaotic systems has been a topic of increasing interest since the beginnings of the '90s. One important step in this direction was the introduction of the already mentioned concept of GS (Rulkov *et al.*, 1995), extending previous studies of coupled identical systems (*complete synchronization,* (Fujisaka and Yamada, 1983)) to the study of coupled systems with different dynamics. Different observables aimed at detecting interdependencies in realistic cases were introduced by several authors. Following an original idea of Rulkov *et al.* (1995), mutual cross-predictabilities were defined and afterwards studied by different authors (Le van Quyen *et al.*, 1998; Le van Quyen *et al.*, 1999; Schiff *et al.*, 1996). In brief, these measures quantify how well one can predict the trajectory in phase space of one of the systems knowing the trajectory of the other. Such a quantification confounds, however, the true synchronization of the systems with their own dynamics and how easy or difficult is to predict each system alone. Variants of this idea have been proposed by different authors in order to improve predictions (Feldmann and Bhattacharya, 2004; Terry and Breakspear, 2003; Wiesenfeldt *et al.*, 2001). Alternatively, a robust set of measures were proposed in (Arnhold *et al.*, 1999; Quian Quiroga *et al.*, 2000; Quian Quiroga *et al.*, 2002a), where instead of looking for predictions, one quantifies how neighborhoods (i.e. recurrences) in one attractor maps into the other. In the following we describe in detail this later approach, which has turned out to be the most reliable way of assessing the extent of GS in time series.

*9.1. Definition of nonlinear interdependences*

From time series measured in two systems $X$ and $Y$, let us reconstruct delay vectors (Takens, 1980) $x_n = (x_n,\ldots,x_{n-(m-1)\tau})$ and $y_n = (y_n,\ldots,y_{n-(m-1)\tau})$, where $n=1,\ldots,N$; $m$ is the embedding dimension and $\tau$ denotes the delay time. Let $r_{n,j}$ and $s_{n,j}$, $j = 1,\ldots,k$, denote the time indices of the $k$ nearest neighbors of $x_n$ and $y_n$, respectively. For each $x_n$, the squared mean Euclidean distance to its $k$ neighbors is defined as:

$$R_n^{(k)}(X) = \frac{1}{k}\sum_{j=1}^{k}(x_n - x_{r_{n,j}})^2 \qquad (26)$$

and the *Y*-conditioned squared mean Euclidean distance is defined by replacing the nearest neighbors of $x_n$ by the equal time partners of the closest neighbors of $y_n$ (see also Fig. 9 and 10),

$$R_n^{(k)}(X|Y) = \frac{1}{k}\sum_{j=1}^{k}(x_n - x_{s_{n,j}})^2 \qquad (27)$$

If the set of reconstructed vectors bworn has an average squared radius *R(X)* then, for strongly correlated systems, $R_n^{(k)}(X|Y) \approx R_n^{(k)}(X) < R(X)$, whereas $R_n^{(k)}(X|Y) \approx R(X) \gg R_n^{(k)}(X)$ for independent systems. Thus, an interdependence measure can be defined accordingly (Arnhold *et al.*, 1999):

$$S^{(k)}(X|Y) = \frac{1}{N}\sum_{n=1}^{N}\frac{R_n^{(k)}(X)}{R_n^{(k)}(X|Y)} \qquad (28)$$

From the reasoning above, it is clear that this measure ranges between 0 and 1 by construction. Low values of this index indicate independence between *X* and *Y*, whereas the measure gives a maximum of 1 for identical systems. Following (Arnhold *et al.*, 1999; Quian Quiroga *et al.*, 2000), it is possible to define another nonlinear interdependence measure as:

$$H^{(k)}(X|Y) = \frac{1}{N}\sum_{n=1}^{N}\log\frac{R_n(X)}{R_n^{(k)}(X|Y)} \qquad (29)$$

where $R_n(X)$ is the average distance of a vector $x_n$ to all the other vectors. A normalized version of this measure can be also defined (Quian Quiroga *et al.*, 2002a). Expression (29) is close to zero if *X* and *Y* are independent, while it is positive if nearness in *Y* implies also nearness in *X* for equal time partners. Theoretical studies with coupled chaotic systems (Quian Quiroga *et al.*, 2000; Schmitz, 2000) have shown that *H* is more robust against noise and easier to interpret than *S*. The opposite interdependences (*S(Y|X)* and *H(Y|X)*) can be defined in complete analogy and they are in general not equal to *S(X|Y)* and *H(X|Y)*, respectively. This asymmetry is one of the main advantages of these indexes over other nonlinear measures, as we detail in the next section.

*9.2. Driver-response relationships*

The nonlinear interdependence is an asymmetric measure, in the sense that $H(X|Y) \neq H(Y|X)$ (the same holds for *S*). This asymmetry can in principle give information about driver-response relationships (Arnhold *et al.*, 1999; Quian Quiroga *et al.*, 2000; Schiff *et al.*, 1996). Suppose the system *X* drives *Y* via unidirectional coupling. Then, *Y* has information about its own dynamics plus that of *X*, whereas *X* does not have any direct information of *Y*. As a consequence, for relatively small couplings it is possible to predict the state of *X* from *Y* but not the other way around. If the coupling is strong enough, then *Y* will tend to follow *X* and both systems can be predicted from each other. Then it is, in principle, easier to predict the state of the driver from the state of the response than vice versa; i.e. $H(X|Y) > H(Y|X)$, if $X \rightarrow Y$. This asymmetry can indeed show driver response relationships, but can also reflect the different dynamical properties of each system (Quian Quiroga *et al.*, 2000). In fact, it has been shown by simulations with coupled dynamical systems that these asymmetries can be biased by different noise levels and different relative frequencies of the systems (Quian Quiroga *et al.*, 2000). This problem can be addressed with the use of proper surrogate testing (see e.g. (Quian Quiroga *et al.*, 2002a)). In particular, one would like to see if an apparent driver-response relationship as reflected by an asymmetry in the interdependencies is also present in data sets with the same characteristics as the original ones but without any coupling. This question is addressed in *Appendix A*.

To conclude this section, we remark that different measures closely to the ones described above and different strategies to form the neighborhood of the reconstructed vectors have been recently defined by several authors (Bhattacharya *et al.*, 2003; Hu and Nenov, 2004; Kramer *et al.*, 2004; Rulkov and Afraimovich, 2003; Stam and van Dijk, 2002).

*9.3. Applications to neurophysiology*

Although the proposal of nonlinear interdependence measures is relatively new, there are already many promising applications reported in the literature, some of which have been recently reviewed (Breakspear, 2004). Their first application to neurophysiological data was by Schiff *et al.* (1996), who used a measure similar to the one defined above and showed its application to coupled dynamical systems and to the study of data from motoneurons within a spinal cord pool. More recently, nonlinear synchronization measures were used for the analysis of EEG data from epileptic patients (Arnhold *et al.*, 1999; Jerger *et al.*, 2001; Le van Quyen *et al.*, 1998; Le van Quyen *et al.*, 1999; Stam and van Dijk, 2002). In all these works, the main goal was to localize the epileptogenic zone and eventually predict the seizure onset, which has certainly a clear clinical relevance.

As it was the case for MI and PS indexes, the use of nonlinear interdependencies seems particularly suited for this purpose because: 1) epilepsy can be broadly defined as an abnormal synchronization in the brain and 2) the landmark of epileptic activity are spikes, which are highly nonlinear and usually appear across several recording channels. At this respect, the utility of nonlinear interdependencies have been based on these arguments and on examples with chaotic toy models. Very few studies, though, posed the question of whether this holds true for real data. In particular, it has been shown with three typical EEG examples, two of them containing spikes, that nonlinear interdependences can disclose information difficult to obtain by visual inspection (Quian Quiroga *et al.*, 2002a).

Another series of studies, which has been already mentioned in the section devoted to PS, aimed at understanding how music may be differentially processed by musicians and non-musicians as reflected by differences in nonlinear interdependencies (Bhattacharya *et al.*, 2003; Bhattacharya *et al.*, 2001a; b). As commented, these studies showed that musicians had a higher synchronization in the gamma band in comparison to non-musicians. Moreover, in musicians synchronization was higher in the left hemisphere. Such findings may reflect a higher ability of correlating various acoustical attributes, a higher involvement of short-term memory when processing music, or a retrieval of larger number of memory patterns from long-term memory. A later related study focused on the asymmetry of the measured interdependencies and reported different patterns in musicians in comparison to non-musicians, thus suggesting different flows of information in the two groups (Feldmann and Bhattacharya, 2004).

Nonlinear interdependencies have also been reported in a relatively small number of short EEG records of resting human subjects (Breakspear and Terry, 2002). Segments showing nonlinear interdependencies have been correlated to a sharpening of the power spectrum in the alpha band (Breakspear and Terry, 2002). Interestingly, a disturbance in the topographic connectivity, as measured by the nonlinear interdependencies, for subjects with a first episode of schizophrenia has also been reported by the same group (Breakspear *et al.*, 2003b), which is in agreement with the view of schizophrenia as a disturbance of the connectivity between cortical areas (Lee *et al.*, 2003; Spencer *et al.*, 2003). Another group has studied the performance of healthy subjects during a working memory task using nonlinear interdependencies in different frequency bands (Stam *et al.*, 2002a). Although these authors did not find a correlation between these measures and memory performance, they described differential patterns of activation and variability of interdependence for the different frequency bands. Using a similar paradigm with MEG data, the same group reported an overall decrease in synchronization for Alzheimer patients (Stam *et al.*, 2002b). Similar decrease in GS was also reported for patients with photo-sensitive epilepsy (Bhattacharya *et al.*, 2004).

Finally, measures of GS have been used to study the patterns of human EEG synchronization during sleep in both adult (Pereda *et al.*, 2001; Terry *et al.*, 2004) and newborns (Pereda *et al.*, 2003). In these works, changes in the interdependences were found among the different sleep stages. Moreover, the combination of GS indexes with the multivariate surrogate data test (*Appendix A*) allowed the authors to show that these interdependences among the analyzed sites were often nonlinear and therefore cannot be explained by using linear indexes alone.

*10. Event synchronization*

All the measures covered up to now are defined for continuous signals, in which we look for linear or nonlinear correlations between amplitude values, frequencies, phases, or trajectories in phase space. As described, these measures have been very useful for different applications. However, many systems in nature express themselves as point-like processes and in this case, the applicability of such measures may be limited. Examples of point processes in neurophysiological signals are spike trains corresponding to the firing of a neuron or the appearance of epileptic spikes in an EEG recording. In this section we describe a very simple measure that can be used for any time series in which we can define events (Quian Quiroga *et al.*, 2002b). In principle, when dealing with signals of different character, the events could be defined differently in each time series, since their common cause might manifest itself differently in each signal. This event synchronization (ES) measure is very simple conceptually and easy to implement. In fact, it can be used on-line and can show rapid changes of synchronization patterns.

*10.1. Definition of event synchronization and delay asymmetry*

For point-like processes the events and times are already given. On the other hand, for continuous time series $x_n$ and $y_n$, $n = 1,...,N$, the first step is to define suitable events and event times $t_i^x$ and $t_j^y$ ($i = 1,...,m_x$; $j = 1,...,m_y$) by taking, e.g. the local maxima, subject to some further conditions. If the signals are synchronized, many events will appear more or less simultaneously. Essentially, we count the fraction of event pairs matching in time, and we count how often each time series leads in these matches. Similar concepts were used by other authors (Brillinger *et al.*, 1976; Pijn, 1990) and a similar idea is extended to multivariate data (Grün *et al.*, 2002a; b; Grün *et al.*, 1999). These measures are also related to the diagonal trace of the joint-peristimulus-time-histograms (Aertsen *et al.*, 1989), which basically consists of a matrix showing coincidences of the spikes of two neurons. Let us first assume that there is a well defined characteristic event rate in each time series. Counter examples include strong chirps and onsets of epileptic seizures where event rates change rapidly. Such cases will be treated below. Allowing a time lag $\pm\tau$ between two 'synchronous' events (which should be smaller than half the minimum inter-event distance to avoid double counting), let us denote by $c^\tau(x|y)$ the number of times an event appears in *x* shortly after it appears in *y*, i.e:

$$c^\tau(x|y) = \sum_{i=1}^{m_x}\sum_{j=1}^{m_y} J_{ij}^\tau \tag{30}$$

where

$$J_{ij}^\tau = \begin{cases} 1 & \text{if} \quad 0 < t_i^x - t_j^y \leq \tau \\ 1/2 & \text{if} \quad t_i^x = t_j^y \\ 0 & \text{otherwise} \end{cases} \tag{31}$$

and analogously for $c^\tau(y|x)$. Next, it is possible to define the following symmetrical and anti-symmetrical combinations:

$$Q_\tau = \frac{c^\tau(y|x) + c^\tau(x|y)}{\sqrt{m_x m_y}}, \quad q_\tau = \frac{c^\tau(y|x) - c^\tau(x|y)}{\sqrt{m_x m_y}} \tag{32}$$

which measure the synchronization of the events and their delay behavior, respectively. They are normalized to $0 \leq Q_\tau \leq 1$ and $-1 \leq q_\tau \leq 1$. We have $Q_\tau = 1$ if and only if the events of the signals are fully synchronized. In addition, if the events in *X* always precede those in *Y*, then $q_\tau = 1$. Figure 11 gives a sketch of the steps involved in the calculation of ES using two simultaneously recorded EEG channels (see (Quian Quiroga *et al.*, 2002b) for details). In cases where we want to avoid a global time scale $\tau$ since event rates change during the recording, we use a local definition $\tau_{ij}$ for each event pair (*ij*). More precisely, we define

$$\tau_{ij} = \min\{t^x_{i+1} - t^x_i, t^x_i - t^x_{i-1}, t^y_{j+1} - t^y_j, t^y_j - t^y_{j-1}\}/2 \qquad (33)$$

$J_{ij}$ is then defined according to Eqn. (31) by replacing the global $\tau$ with the local $\tau_{ij}$. In either case, time resolved variants of Eqn. (30) can be obtained as:

$$c_n(x\,|\,y) = \sum_i \sum_j J_{ij} \Theta(n - t^x_i) \qquad (34)$$

where *n=1,...,N* and $\Theta(x)$ is the step function (0 for $x \leq 0$; 1 otherwise). Similarly, $c_n(y/x)$ can be calculated by exchanging *X* and *Y*. This time resolved variants may be seen as a random walk that takes one step up every time an event in *X* precedes one in *Y* and one step down if vice versa. If the events occur simultaneously or if it appears only in one of the signals, the random walk does not move. Exchanging *X* and *Y* just reverses the walk. For non-synchronized signals, we expect to obtain a random walk with the typical diffusion behavior. With delayed synchronization we will have a bias going up (down) if *X* precedes (follows) *Y*.

The time course of the strength of ES can be obtained from $Q(n)=c_n(y/x)+c_n(x/y)$. If an event is found both in *x* and *y* within the window $\tau$ (resp. $\tau_{ij}$), *Q(n)* increases one step, otherwise it does not change. Of course, *Q(n)* will also not change if there are no new events at all. The synchronization level at time *n*, averaged over the last $\Delta n$ time steps, is thus obtained as:

$$Q'(n) = \frac{Q(n) - Q(n - \Delta n)}{\sqrt{\Delta n_x \Delta n_y}} \qquad (35)$$

*10.2. Applications to neurophysiology*

ES is particularly well suited when one can define clear events in the signals (e.g. spikes). In particular, it is very sensitive to quasi-synchronous appearances of events, even if these are too sparse. Such behavior may be washed-out by other standard measures of synchronization that look at the whole signal. In particular, a measure similar to the one described above was applied to ultra-sparse neuronal data from different structures in the brain of zebra finches (Hahnloser *et al.*, 2002). In this study a sparse quasi-synchronous pattern of activation between different brain structures was correlated to the generation of song motifs. These results were not that obvious in the cross-correlation function between the different spike trains ((Hahnloser *et al.*, 2002), see supplementary material).

ES has been also applied to different sets of EEG data (Quian Quiroga *et al.*, 2002b). An interesting result was that ES showed similar outcomes to other synchronization measures, such as PS, nonlinear interdependencies, cross-correlation, coherency and MI, for the data sets studied in (Quian Quiroga *et al.*, 2002a). Remarkably, this holds true even in cases where the definition of events was not obvious (i.e. a random looking signal). Moreover, it was possible to obtain a better resolved time profile of the synchronization pattern. The same study also reports the application of ES to the study of EEG pre-ictal and ictal activity. In this case, it was possible to identify the recording electrode closer to the epileptic focus (according to clinical evidence) since their events preceded those in the other channels (Quian Quiroga *et al.*, 2002b).

A measure of coincidence spikes (Grün *et al.*, 2002a; b; Grün *et al.*, 1999) similar to the one we have just described, was used to study multiple neuron recordings from the motor cortex of monkeys (Riehle *et al.*, 1997). This study showed the existence of a precise synchronization (of the order of 5 ms) among neurons in the motor cortex, which was associated to distinct phases of the execution of a motor task.

*11. Comparing the different approaches*

*11.1. The current role of linear methods*

After getting acquainted with the different multivariate nonlinear methods, one might be tempted to favor them in prejudice of the linear methods or, at least, to relegate these to the background. But this would be a serious mistake: the nonlinear tools are not intended to substitute linear ones and neither can they be claimed to be superior as such. Instead, they must be regarded as a complement of the linear approach that allows getting a more comprehensive picture of the analyzed data. In fact, we have seen that the information provided by multivariate nonlinear analysis does not necessarily coincide with that of the linear methods (e.g., (Fell *et al.*, 2003)). Both approaches may assess different parts of the interdependence between the signals, to the point that the linear methodology might be even superior in certain cases (e.g., (Freeman and Rogers, 2002)). Additionally, from the methodological point of view, linear methods sometimes present better properties that their nonlinear counterparts, such as robustness against noise.

In consequence, a rigorous approach to the study of any neurophysiological data set should not be biased towards nonlinear methods. Quite on the contrary, the linear approach should be the initial choice, and it is indeed a healthy practice to try first the traditional approaches before going to the more complicated ones. Only if we have good reasons to think that there is any nonlinear structure either in the data themselves or in the interdependence between them should the nonlinear approach be adopted. And even in this case, the best strategy would consist in using both linear and nonlinear methods alike to be sure that we have gathered all the information available from the signals.

*11.2. The relationships between the nonlinear indexes*

At this point, the question naturally arises as to which of the nonlinear strategies should be chosen to analyze a given neurophysiological data set. To clarify this point, comparative studies have been carried out in either animal (Quian Quiroga *et al.*, 2002a) and human epileptic EEGs as well as in model dynamical systems (Kraskov, 2004; Kreuz, 2004; Smirnov and Andrzejak, 2005). A recent work addressed this issue by making use of a neural model in which the interdependence between two simulated neurophysiological signals could be modified by means of an adjustable parameter, thereby having *a priori* knowledge about the results that should be obtained (David *et al.*, 2004). These works, along with other related, more theoretical ones studying the properties of the different indexes within some specific framework (Pereda *et al.*, 2001; Rosenblum *et al.*, 2004) and the relationship between different kinds of synchronization (Parlitz *et al.*, 1996; Zheng and Hu, 2000) allow us to shed some light upon the abovementioned question.

The first remarkable result is that most of the nonlinear indexes are somewhat correlated with each other (Kraskov, 2004; Kreuz, 2004; Quian Quiroga *et al.*, 2002a). Indeed, both the linear correlation coefficient and one information-theoretic index have shown similar results for MI, PS indexes, GS indexes and linear correlation indexes. The correlation between all of them is quite high even in the worst case, although they all cluster in different groups according to their degree of similarity, which roughly speaking coincides with the type of synchronization they assess (Kraskov, 2004). Additionally, GS indexes do not seem to be superior in general to asymmetric PS bivariate indexes in the assessment of weak directional coupling (Smirnov and

Andrzejak, 2005). The most significant result is, however, that it is rather difficult to assess objectively the performance of the different measures either in dynamical systems or in EEG data (Kreuz, 2004). Instead, this author suggests a pragmatic approach, in which the tool to be used depends on the information that one wishes to extract from the data. In this comparative study, the PS indexes based on the HT along with the MI and the cross-correlation function were the most promising in yielding useful information for diagnostic purposes in epilepsy patients. But the joint use of different synchronization measures that gives the maximum non-redundant information (e.g., those with the lowest correlation) might be an interesting approach.

The study using the neural mass model also clarified some important practical issues (David *et al.*, 2004). Concretely, it assessed the sensitivity of the cross-correlation function and several nonlinear synchronization indexes for narrow and broadband signals. The authors concluded that, despite initial claims of PS being a weaker type of synchronization than GS (Parlitz *et al.*, 1996), GS indexes are more sensitive than either PS indexes or MI at weak couplings, which would agree with a later theoretical result indicating that GS and PS may appear independently (Zheng and Hu, 2000). In must be noted, however, that the apparent independency shown in this latter work may be the consequence of an unfortunate definition of the phase. David *et al.* (2004) also showed that MI presented the greatest variation for a given change of the coupling parameter at higher synchronization levels. Likewise, these authors suggest that, although both MI and GS indexes are useful to study changing interdependences, the latter ones should be preferred for the analysis of broadband signals, whereas PS indexes perform well in the narrowband case. Nevertheless, and despite this result, we would like to stress that PS indexes can be used whenever one expects to get information out of the phases disregarding the amplitudes, so that whether the signals are broadband or narrowband should not be a criterion for making this decision. Finally, this work shows that synchronization can be properly detected in relatively short data segments, in agreement with other studies (e.g., (Bhattacharya *et al.*, 2003)), although the detection of coupling is clearly improved as the length of the time series increases.

In conclusion, which is the answer to the question we raised at the beginning of this section? An accurate response would have to take into account the type of analyzed data (David *et al.*, 2004; Kreuz, 2004). It is of course impossible to consider all the possibilities, but we can dare to make some suggestions for some typical cases. Thus, if one is interested in checking the existence of (possibly nonlinear, either symmetric or asymmetric) synchronization between the amplitudes of integrated neural activity recordings such as EEG or MEG signals, methods based on GS, nonlinear Granger causality and the information-theoretic approach may be preferred. In this case, the latter advances from these approaches (Andrzejak *et al.*, 2003; Bhattacharya *et al.*, 2003; Chen *et al.*, 2004; Hu and Nenov, 2004; Kraskov *et al.*, 2004; Quian Quiroga *et al.*, 2002a; Rulkov and Afraimovich, 2003) must be taken into account for optimal performance. If no interdependence between the amplitudes is found, PS indexes can still be used, in order to study the existence of the interdependence between the phases (as it has been done, for instance, during the dynamical formation and destruction of large scale integrative webs in cognitive and binding processes). Those PS indexes that are able to detect the existence of directionality and delay in the synchronization (Cimponeriu *et al.*, 2003; Cimponeriu *et al.*, 2004; Smirnov and Bezruchko, 2003) might be applied to get further information about this interdependence. We must stress that, when using the PS approach, it is fundamental to reliably estimate a meaningful phase from the signals; otherwise the results might be meaningless. In any case, since GS and PS might appear independently, the possible synchronization between the amplitudes and between the phases can be jointly studied, (see for instance (Bhattacharya *et al.*, 2001b; Kraskov, 2004; Kreuz, 2004; Quian Quiroga *et al.*, 2002a))

In dealing with evoked potentials, which are intrinsically non-stationary, the later GS index tailored to cope with non-stationarity (Kramer *et al.*, 2004) as well as PS indexes can be used. Finally, information theory-based methods can be used to study the stimulus-response relationship as well as the features of the information encoded in neural action potentials. The ES indexes may also play an important role in studying synchronization phenomena in such point processes, as in all those signals where marked events can be properly defined.

In all cases, the use of some kind of statistical test for synchronization (such as the one described in *Appendix A*) is advisable, in order to check whether the indexes are actually reflecting the interdependence between the signals, which will avoid drawing erroneous conclusions about the data based on spurious values of these indexes. As already commented, this test may be also useful to get information about the nature of the relationship, thereby checking the convenience of applying nonlinear methods.

*12. Conclusions*

We have reviewed here the current state of the main nonlinear analysis techniques applied to multivariate neurophysiological data, a subject that is earning growing popularity to the extent that the methods have been applied to almost any kind of neurophysiological signals ranging from fMRI data to spike recordings of a neuron. Certainly, and despite its possible advantages, we have also seen that this new approach is not free of caveats. It might be even argued, as suggested in the *Introduction* section, that the very nature of neurophysiological data (which are often non-stationary, short and noisy) as well as that of the methods -whose complex mathematical background may be sometimes discouraging- preclude the growth of this approach in neurophysiology. But we have seen that actually some nonlinear methods rely much less on stationarity than linear ones (e.g. ES vs. coherence) and can be also far much simpler (e.g. ES vs. Granger causality). Moreover, the calculation of nonlinear methods can sometimes be faster than the linear ones.

In any case, it has been clear throughout this work that nonlinear methods might be useful in giving insight into the interdependence between neural assemblies at both short and large time and spatial scales, as they allow the analysis of complex nonlinear interactions from different perspectives and complement the information provided by traditional linear tools. But only the intensive interplay between theorist and applied scientists that is currently taking place in this multidisciplinary research field will allow elucidating whether multivariate nonlinear methods can be actually successfully integrated as standard analytical tools in neurophysiology.

*Appendix A. The multivariate surrogate data: what can they do?*

The surrogate data method was introduced into practice more than a decade ago (Theiler *et al.*, 1992) and it is nowadays the most popular test for non-linearity in experimental data. It belongs to a more general type of statistical tests known as *hypothesis tests*. In the univariate approach, a certain index (*the statistic*) that characterizes a time series is calculated from it. Then, a set of *p* times series is constructed, which share with the original many of its characteristics, but lack the property whose effect on the statistic we want to test. These new series, called the *surrogates*, are used to repeat the calculation of the index, thereby obtaining *p+1* estimations of it. The test consists in determining the probability that the original value of the statistic belongs to the distribution of the surrogates (*the null hypothesis, $H_0$*), which is equivalent to estimate numerically the probability that $H_0$ is true. Different kind of surrogates data are consistent with different null hypothesis, and different statistics can be used (for recent reviews see (Dolan and Spano, 2001; Schreiber and Schmitz, 2000)). However, what is important in our context is that this idea can be extended to deal with multivariate data. In fact multivariate surrogates can be used to get insight into the interdependence between time series in two ways: by studying the significance of the interdependence, as measured by the different indexes, and by determining its nature, i.e., whether it is nonlinear.

*Testing the reliability of the indexes*

Sometimes, the synchronization indexes may present values, which are not reflecting the existence of synchronization between the time series, but are the result of some feature of the individual signals (such as their complexity, their limited length of their non-stationarity, e.g. (Bhattacharya *et al.*, 2003; Pereda *et al.*, 2001; Quian Quiroga *et al.*, 2000)). In order to check whether an index is actually measuring synchronization, multivariate surrogate data can be constructed to test the hypothesis that the signals are independent.

The simplest way of achieving this goal is by randomly shuffling the samples of both time series, thereby obtaining the so called "shuffled surrogates", with the same distribution of the original data but completely independent from each other (Palus, 1996). The problem of these surrogates is that they change the autocorrelation structures of each dataset. Additionally, it is known that the autocorrelation of the data might affect the values of different synchronization indexes (Pereda *et al.*, 2001). Also, the use of such shuffled, white noise-like version of the data as a control condition is not to be recommended, because the prominent autocorrelations inherent to neurophysiological signals (see, e.g., (Linkenkaer-Hansen *et al.*, 2001)) turn this kind of surrogates into very unlikely realizations of any neurophysiological process. Hence, a more feasible $H_0$ can be tested, namely that the time series are two independent, linear stochastic processes with an arbitrary degree of linear autocorrelation. Surrogate data consistent with this $H_0$ would be "traditional" univariate surrogates constructed by any of the available methods, where the surrogating procedure is carried out independently for each time series, so that any cross-correlation is destroyed (see Fig. 12). This idea can be also used to estimate the significance of the coherence function in the linear framework (Faes *et al.*, 2004). Still, this procedure might be insufficient if any of the time series does present nonlinear structure. In such a case, it is necessary to construct surrogates that preserve all the individual structure while destroying all interdependences between the signals (Andrzejak *et al.*, 2003; Quian Quiroga *et al.*, 2002a).

Two possibilities are at hand for this purpose. The first and simplest one consists in comparing the original version of one of the signals with temporally shifted versions of the other one (Bhattacharya *et al.*, 2003; Kraskov, 2004; Netoff and Schiff, 2002; Quian Quiroga *et al.*, 2002a). In a similar way, if one is dealing with evoked potentials then one has pairs of EEG channels whose interdependence is analyzed in a series of several trials. In order to determine the significance of, say, the PS indexes, one possibility consists in randomizing the order of trials in one of the channels in order to check that PS is not spuriously induced by either the measurement devices or the recording procedure (see e.g. (Gross *et al.*, 2004; Lachaux *et al.*, 1999; Rodriguez *et al.*, 1999; Simoes *et al.*, 2003)). The second possibility can be used for instance in the framework of GS analysis. As GS indexes are calculated in the state space of each individual signals, one of the signals is left unchanged, whereas surrogate versions of the other one are constructed. In this way, we obtain for the unchanged signal different estimations of which would be the value of the index if there were no relationship with the other one, as the surrogate versions of the second signal are independent from the first one (e.g. (Bhattacharya *et al.*, 2003; Bhattacharya *et al.*, 2001b; Pereda *et al.*, 2001)).

*The nature of the interdependence*

In order to determine whether the interdependence between two signals is nonlinear, multivariate surrogate data must be constructed in such a way that they preserve the linear cross-correlation between the original data, which can be achieved by keeping constant the relative phase difference between them (Prichard and Theiler, 1994). Thus, the phases of the signals in the frequency domain are randomized by adding the same random quantity to the phases of each signal at each frequency. Both the phases of the Fourier transform (e.g. (Andrzejak *et al.*, 2003; Dumont *et al.*, 2004; Pereda *et al.*, 2001; Prichard and Theiler, 1994)) or those of the wavelet transform (Breakspear *et al.*, 2003a) can be used for this purpose. Additionally, it is also possible to obtain this kind of surrogates by filtering random Gaussian data using the power spectral density of the original data (Dolan, 2004; Dolan and Neiman, 2002).

It must be noted that, although the above surrogating procedure preserves both the autocorrelation of the signals and their linear cross-correlation, the nonlinear individual structure of the individual signals, if any, is also destroyed. In other words, any nonlinearity not only *between* but also *within* the signals is not present in the surrogates. Therefore, these surrogates only test the hypothesis that the data are bivariate stochastic time series with an arbitrary degree of linear auto and cross-correlation (Andrzejak *et al.*, 2003). Nevertheless, if the two signals studied do have any nonlinear structure, it is not possible to ascribe a rejection of the hypothesis that the interdependence is nonlinear to the nonlinearity of the interdependence, because the nonlinearity of the individual signals may also play a role. The generation of surrogate data preserving all the individual structure but destroying only the nonlinear part of the interdependence is currently one of the most challenging tasks in the field, and it is a subject of ongoing research (Andrzejak *et al.*, 2003; Dolan, 2004).

We would like to note that, whether the surrogates are used to test the existence of interdependence or its nature, the underlying idea is always the same: a significance threshold is obtained with the help of the surrogates, beyond which either the synchronization indexes can be regarded as significant or the interdependence can be regarded as nonlinear at a certain level of statistical confidence. It is noteworthy that the derivation of such threshold is closely linked to the ideas traditionally used to estimate the significance of linear indexes such as the cross-correlation or the coherence function. In the nonlinear case, this significance can be tested in different ways, as detailed henceforth.

*Assessing the significance of the test*

Once the values of the statistic for the original and for the set of surrogate time series are obtained, it is necessary to check if the former one is indeed significantly different from the latter ones. In other words, one has to determine whether the corresponding $H_0$ can be rejected at the desired level of confidence. The "classical" approach for this purpose consisted in estimating the mean and the standard deviation of the distribution of the statistic from the surrogates and then comparing them with its value for the original signals. Thus, a Z-score is calculated as follows:

$$Z = \frac{|\xi_o - \bar{\xi}_S|}{\sigma_S} \qquad (36)$$

Here, $\xi_o$ and $\bar{\xi}_S$ are the value of the statistic for the original data and its mean for the surrogate distribution, respectively, and $\sigma_S$ is the standard deviation of this distribution. There is a direct relationship between the number of generated surrogates, $p$, and the minimal value of $Z$ for the difference to be significant. Typically, for $p=19$, in order to reject $H_0$ at the 95% level of confidence one must have $Z>1.96$ for a one-sided test, which is performed in the multivariate case because synchronization indexes are expected to be greater for the original that for the surrogates.

The above approach has been often used in testing for nonlinearity in univariate data, from which it was easily adapted to the multivariate case (Schreiber and Schmitz, 2000). It has the advantage of providing a *quantitative* measure of significance, which might be further used, in principle, as an index of either the degree of coupling or the degree of nonlinearity in the interdependence. However, by assessing the significance in this way, we are implicitly assuming that the indexes from the surrogates are normally distributed, which is certainly not always the case (Schreiber and Schmitz, 2000). Therefore, unless the normality of this distribution is explicitly checked (see (Dumont *et al.*, 2004) for a recent example), a more accurate, nonparametric rank test must be applied, which provides a *qualitative* measure of significance by rejecting $H_0$ if and only if $\xi_o$ is strictly greater than *all* the values for the surrogates. Only if such more restrictive condition is fulfilled can Eqn. (36) be regarded as a measure of the extent of the interdependence and/or of its non-linearity.

An alternative approach can be used to test the significance of the difference at the group level, when one has a set of *n* records carried out in the same experimental situation (Dumont *et al.*, 2004; Fell *et al.*, 1996). In this case, the *n* original values $\xi_o^j$ (*j=1,..,n*) are compared with the *n* surrogate values $\xi_S^j$ obtained after generating one surrogate pair for each original time series. A nonparametric test for dependent samples (e.g., Wilcoxon signed test) must be performed subsequently to get the level of statistical significance of the difference between original and surrogates.

To conclude, we must indicate that there are a few especial situations (such as the study of PS between two time series when a base level of synchronization is present) where a different test for synchronization should be used (Allefeld and Kurths, 2004b).

*Appendix B. Do it yourself…with a little help.*

It is likely that sooner or later one is tempted to apply some of the nonlinear methods described here to his/her own data. This invariably carries the need of programming his/her own code, a task that can be considerably eased by being aware of the work that others have already carried out in the same line, whose results may be available through the Internet. In fact, the World Wide Web has become a fundamental research resource to the extent that it is nowadays common to find references to its content in almost any scientific document. This is not surprising because Internet can be regarded, among other things, as a scientific forum where we can seek information about a particular matter or simply contribute our own work. Considering this, we would like to take advantage of this resource by pointing the interested reader to different web sites where it is possible to get software code (sometimes even complete programs and/or toolboxes) for the multivariate nonlinear analysis of experimental signals. In doing so, we are certainly taking some risks: Internet is a dynamic environment, in which a given site might suddenly move to a different location or even disappear, thus making the corresponding link useless. However, we also think that the potential benefits of including these links clearly outweigh the drawbacks, and at the same time, we pay a tribute to all those researchers who have shared the product of their efforts with the scientific community for the sake of the advance of science.

In what follows, we will restrict ourselves to resources on nonlinear multivariate indexes that are publicly available and possess in our mind a proven reliability in terms of scientific soundness and potential ability to last in their present locations. Unfortunately, and contrary to univariate nonlinear analysis -where the TISEAN package (Hegger *et al.*, 1999) is still the reference- there is not an equivalent in the multivariate context, so that different algorithms for the different indexes must be obtained separately.

Most of the software presented here is coded in MATLAB®. Even if this program is not free, all the references we give are for free, and those researchers who do not own the program and do not want to get it either, can still make use of the code as a guide to produce his/her own custom-written software in the desired language. In this regard, it must be noted that there are two general-purpose file exchange sites that are certainly good starting points (Mathtools.net, ; MatlabCentral).

In all cases, Internet addresses are indicated either as footnotes or in the *References* section.

*Phase synchronization indexes*

To the best of our knowledge, there is only one set of scripts available for the calculation of PS indexes, which can be downloaded from a personal website[5]. These scripts allow calculating both symmetric bivariate PS indexes and the directional PS indexes described in section 8.2.2.

*Mutual information*

Some good software is available to calculate the MI between time series. For instance, the Mutual Information Least Component Analysis package (MILCA, ; Stögbauer *et al.*, 2004) includes the whole source code along with complementary documentation. It implements a MI estimator that is adaptive, data efficient and optimized for minimal bias (Kraskov *et al.*, 2004). On the other hand, the Cross Recurrence Plot Toolbox (CRPTOOL) also includes an implementation of the MI index (Roulston, 1999). Both MILCA and CRPTOOL calculate the MI from either bivariate or multivariate time series as a function of the time delay. Although the latter toolbox

---

[5] Dr. Michael G. Rosenblum's homepage at the University of Potsdam.

URL: www.agnld.uni-potsdam.de/~mros/publications.html

implements a sub-optimal algorithm as compared to the former one, it presents the advantage of giving error estimates for the MI, which might be useful if one does not have long enough data. The TISEAN package and the Time Series Toolbox (TSTOOL) (see below) also include routines for the calculation of MI, but they are only intended for the univariate approach, in which instead of signal *Y* in Eqn. (16), delayed versions of signal *X* are used to calculate the so-called *auto*-MI. The first minimum of this function is often used as a good estimator of the time delay for embedding purposes, but has little usefulness for the multivariate case.

*Generalized synchronization indexes*

One of the authors of this work has made available different routines for the calculation of GS indexes (as presented in (Quian Quiroga *et al.*, 2002a)) along with the data sets necessary to test the results[6]. These routines also allow estimating the cross-correlation and the coherence function. Moreover, similar software, including the calculation of a recently defined nonlinear interdependence measure (Hu and Nenov, 2004) is available upon request from the corresponding author of this paper.

*Multivariate surrogate data*

The already mentioned TISEAN package includes an implementation to construct multivariate surrogates by using both constrained randomization and annealing methods. Furthermore, it is also possible to get a very useful MATLAB® package for the same purpose from (MatlabCentral).

*Other useful software*

Apart from the TISEAN package, there are some other interesting packages, not directly dealing with multivariate analysis that might be anyway of help in our context.

---

[6] R. Quian Quiroga. Software site. URL:www.vis.caltech.edu/~rodri/software.htm

For instance, TSTOOL is a comprehensive MATLAB® toolbox for the nonlinear analysis of time series. Besides giving the option to calculate a whole set of univariate nonlinear indexes, it includes several useful scripts for general-purpose applications such as time delay embedding or nearest neighbours search. It comes with a complete user manual including a large set of bibliographic references, which makes it very useful for those researchers interested in getting started with nonlinear analysis methods.

There are also another two integrated toolboxes worth mentioning, both of them devoted to MEG and EEG analysis and representation (BrainStorm, ; EEGLab). They include many useful data visualization and processing routines.

*13. References*

Adey, W. R., Elul, R., Walter, R. D., Crandall, P. H. 1967a The cooperative behavior of neuronal populations during sleep and mental tasks. Electroencephalogr. Clin. Neurophysiol. 23, 88.

Adey, W. R., Kado, R. T., Walter, D. O. 1967b Analysis of brain wave records from Gemini flight GT-7 by computations to be used in a thirty day primate flight. Life Sci. Space Res. 5, 65-93.

Adey, W. R., Kado, R. T., Walter, D. O. 1967c Computer analysis of EEG data from Gemini flight GT-7. Aerosp. Med. 38, 345-359.

Aertsen, A. M., Gerstein, G. L., Habib, M. K., Palm, G. 1989 Dynamics of neuronal firing correlation: modulation of "effective connectivity". J. Neurophysiol. 61, 900-917.

Al-Nashash, H. A., Paul, J. S., Ziai, W. C., Hanley, D. F., Thakor, N. V. 2003 Wavelet entropy for subband segmentation of EEG during injury and recovery. Annals of Biomedical Engineering 31, 653-658.

Albo, Z., Di Prisco, G. V., Chen, Y., Rangarajan, G., Truccolo, W., Feng, J., Vertes, R. P., Ding, M. 2004 Is partial coherence a viable technique for identifying generators of neural oscillations? Biol. Cybern. 90, 318-326.

Allefeld, C., Kurths, J. 2004a An approach to multivariate phase synchronization analysis and its application to event-related potentials: Synchronization Cluster Analysis. Int. J. of Bifurcation and Chaos 14, 417-426.

Allefeld, C., Kurths, J. 2004b Testing for phase synchronization. Int. J. of Bifurcation and Chaos 14, 405-416.

Andrew, C., Pfurtscheller, G. 1996 Event-related coherence as a tool for studying dynamic interaction of brain regions. Electroencephalogr. Clin. Neurophysiol. 98, 144-148.

Andrzejak, R. G., Kraskov, A., Stogbauer, H., Mormann, F., Kreuz, T. 2003 Bivariate surrogate techniques: Necessity, strengths, and caveats. Phys. Rev. E 68.

Angelini, L., De Tommaso, M., Guido, M., Hu, K., Ivanov, P. C., Marinazzo, D., Nardulli, G., Nitti, L., Pellicoro, M., Pierro, C., Stramaglia, S. 2004 Steady-state visual evoked potentials and phase synchronization in migraine. Phys. Rev. Lett. 93, 038103.

Arnhold, J., Grassberger, P., Lehnertz, K., Elger, C. E. 1999 A robust method for detecting interdependences: application to intracranially recorded EEG. Physica D 134, 419-430.

Arnold, M., Miltner, W. H., Witte, H., Bauer, R., Braun, C. 1998 Adaptive AR modeling of nonstationary time series by means of Kalman filtering. IEEE Trans. Biomed. Eng. 45, 553-562.

Asher, H. B. 1983 Causal Modeling. Sage Publications: Newbury Park, CA.

Babloyantz, A., Salazar, J. M., Nicolis, C. 1985 Evidence of chaotic dynamics of brain activity during the sleep cycle. Phys. Lett. A 111, 152-156.

Baccala, L. A., Sameshima, K. 2001a Overcoming the limitations of correlation analysis for many simultaneously processed neural structures. Prog. Brain Res. 130, 33-47.

Baccala, L. A., Sameshima, K. 2001b Partial directed coherence: a new concept in neural structure determination. Biol. Cybern. 84, 463-474.

Baillet, S., Garnero, L., Marin, G., Hugonin, J. P. 1999 Combined MEG and EEG source imaging by minimization of mutual information. IEEE Trans. Biomed. Eng. 46, 522-534.

Bartolomei, F., Wendling, F., Bellanger, J. J., Regis, J., Chauvel, P. 2001 Neural networks involving the medial temporal structures in temporal lobe epilepsy. Clin. Neurophysiol. 112, 1746-1760.

Bendat, J. S., Piersol, A. G. 2000 Random Data - Analysis and Measurement Procedure. John Wiley & Sons Inc.: New York.

Bernasconi, C., von Stein, A., Chiang, C., Konig, P. 2000 Bi-directional interactions between visual areas in the awake behaving cat. Neuroreport 11, 689-692.

Bhattacharya, J. 2001 Reduced degree of long-range phase synchrony in pathological human brain. Acta Neurobiol. Exp. (Warsz). 61, 309-318.

Bhattacharya, J., Pereda, E., Petsche, H. 2003 Effective detection of coupling in short and noisy bivariate data. IEEE Trans. Syst. Man Cybern. B 33, 85-95.

Bhattacharya, J., Petsche, H. 2001 Musicians and the gamma band - a secret affair? Neuroreport 12, 371-374.

Bhattacharya, J., Petsche, H. 2002 Shadows of artistry: cortical synchrony during perception and imagery of visual art. Cogn. Brain Res. 13, 179-186.

Bhattacharya, J., Petsche, H. 2005a Drawing on mind's canvas: differences in cortical integration patterns between artists and non-artists. Hum. Brain Mapp. 26, 1-14.

Bhattacharya, J., Petsche, H. 2005b Phase synchrony analysis of EEG during music perception reveals changes in functional connectivity due to musical expertise. Signal Processing 85, 2161-2177.

Bhattacharya, J., Petsche, H., Pereda, E. 2001a Interdependencies in the spontaneous EEG while listening to music. Int. J. Psychophysiol. 42, 287-301.

Bhattacharya, J., Petsche, H., Pereda, E. 2001b Long-range synchrony in the gamma band: Role in music perception. J. Neurosci. 21, 6329-6337.

Bhattacharya, J., Watanabe, K., Shimojo, S. 2004 Nonlinear dynamics of evoked neuromagnetic responses signifies potential defensive mechanisms against photosensitivity. Int. J. of Bifurcation and Chaos 14, 2701-2720.

Blanco, S., Garcia, H., Quian Quiroga, R., Romanelli, L., Rosso, O. A. 1995 Stationarity of the EEG time series. IEEE Eng. Med. Biol. Mag. 14, 395-399.

Blinowska, K. J., Czerwosz, L. T., Drabik, W., Franaszczuk, P. J., Ekiert, H. 1981 EEG data reduction by means of autoregressive representation and discriminant analysis procedures. Electroencephalogr. Clin. Neurophysiol. 51, 650-658.

Boccaletti, S., Kurths, J., Osipov, G., Valladares, D., Zhou, C. 2002 The synchronization of chaotic systems. Phys. Rep. 366, 1-101.

Borst, A., Theunissen, F. E. 1999 Information theory and neural coding. Nat. Neurosci. 2, 947-957.

BrainStorm Matlab Toolbox. Available at http://neuroimage.usc.edu/brainstorm/.

Brazier, M. A., Barlow, J. S. 1956 Some applications of correlation analysis to clinical problems in electroencephalography. Electroencephalogr. Clin. Neurophysiol. Suppl. 8, 325-331.

Brazier, M. A., Casby, J. U. 1952 Cross-correlation and autocorrelation studies of electroencephalographic potentials. Electroencephalogr. Clin. Neurophysiol. Suppl. 4, 201-211.

Brazier, M. A. B. 1968 Studies of EEG activity of limbic structures in man. Electroencephalogr. Clin. Neurophysiol. 25, 309-318.

Breakspear, M. 2004 "Dynamic" connectivity in neural systems: theoretical and empirical considerations. Neuroinformatics 2, 205-226.

Breakspear, M., Brammer, M., Robinson, P. A. 2003a Construction of multivariate surrogate sets from nonlinear data using the wavelet transform. Physica D 182, 1-2.

Breakspear, M., Terry, J. 2002 Detection and description of non-linear interdependence in normal multichannel human EEG data. Clin. Neurophysiol. 113, 735-753.

Breakspear, M., Terry, J. R., Friston, K. J., Harris, A. W., Williams, L. M., Brown, K., Brennan, J., Gordon, E. 2003b A disturbance of nonlinear interdependence in scalp EEG of subjects with first episode schizophrenia. Neuroimage 20, 466-478.

Breakspear, M., Williams, L., Stam, C. J. 2004 A novel method for the topographic analysis of neural activity reveals formation and dissolution of "Dynamic Cell Assemblies". J. Comput. Neurosci. 16, 49-69.

Brillinger, D. R., Bryant, H. L., Jr., Segundo, J. P. 1976 Identification of synaptic interactions. Biol. Cybern. 22, 213-228.

Brody, C. D. 1999 Correlations without synchrony. Neural Comput. 11, 1537-1551.

Brovelli, A., Ding, M., Ledberg, A., Chen, Y., Nakamura, R., Bressler, S. L. 2004 Beta oscillations in a large-scale sensorimotor cortical network: directional influences revealed by Granger causality. Proc. Natl. Acad. Sci. U. S. A. 101, 9849-9854.

Buchel, C., Friston, K. 2000 Assessing interactions among neuronal systems using functional neuroimaging. Neural Netw. 13, 871-882.

Buracas, G. T., Zador, A. M., DeWeese, M. R., Albright, T. D. 1998 Efficient discrimination of temporal patterns by motion-sensitive neurons in primate visual cortex. Neuron 20, 959-969.

Butler, S. R., Glass, A. 1974 Asymmetries in the electroencephalogram associated with cerebral dominance. Electroencephalogr. Clin. Neurophysiol. 1974, 481-491.

Cimponeriu, L., Rosenblum, M. G., Fieseler, T., Dammers, J., Schiek, M., Majtanik, M., Morosan, P., Bezerianos, A., Tass, P. A. 2003 Inferring asymmetric relations between interacting neuronal oscillators. Prog. Theor. Phys. Supp. 150, 22-36.

Cimponeriu, L., Rosenblum, M. G., Pikovsky, A. 2004 Estimation of delay in coupling from time series. Phys. Rev. E 70, 046213.

Cohen, M. I., Yu, Q., Huang, W. X. 1995 Preferential correlations of a medullary neuron's activity to different sympathetic outflows as revealed by partial coherence analysis. J. Neurophysiol. 74, 474-478.

Cooley, J. W., Tukey, J. W. 1965 An algorithm for machine calculation of complex Fourier series. Mathematical Computation 19, 297-301.

CRPTOOL Cross Recurrence Plot Toolbox. Available at http://www.agnld.uni-potsdam.de/~marwan/toolbox.php

Chavez, M., Le Van Quyen, M., Navarro, V., Baulac, M., Martinerie, J. 2003 Spatio-temporal dynamics prior to neocortical seizures: amplitude versus phase couplings. IEEE Trans. Biomed. Eng. 50, 571-583.

Chen, F., Xu, J., Gu, F., Yu, X., Meng, X., Qiu, Z. 2000 Dynamic process of information transmission complexity in human brains. Biol. Cybern. 83, 355-366.

Chen, Y. H., Rangarajan, G., Feng, J. F., Ding, M. Z. 2004 Analyzing multiple nonlinear time series with extended Granger causality. Phys. Lett. A 324, 26-35.

Damasio, A. R. 1990 Synchronous activation in multiple cortical regions: a mechanism for recall. Seminars in Neurology 2, 287-297.

Darbellay, G. A., Vajda, I. 1999 Estimation of the information by an adaptive partitioning of the observation space. IEEE Trans. Inf. Theory 45, 1315-1321.

David, O., Cosmelli, D., Friston, K. J. 2004 Evaluation of different measures of functional connectivity using a neural mass model. Neuroimage 21, 659-673.

Davis, K. A., Lutchen, K. R. 1991 Time series versus Fourier transform methods for estimation of respiratory impedance spectra. Int. J. Biomed. Comput. 27, 261-276.

de Ruyter van Steveninck, R. R., Lewen, G. D., Strong, S. P., Koberle, R., Bialek, W. 1997 Reproducibility and variability in neural spike trains. Science 275, 1805-1808.

Ding, M., Bressler, S. L., Yang, W., Liang, H. 2000 Short-window spectral analysis of cortical event-related potentials by adaptive multivariate autoregressive modeling: data preprocessing, model validation, and variability assessment. Biol. Cybern. 83, 35-45.

Dolan, K. 2004 Surrogate analysis of multichannel data with frequency dependant time lag. Fluct. Noise Lett. 4, L75-L81.

Dolan, K., Neiman, A. 2002 Surrogate analysis of coherent multichannel data. Phys. Rev. E 65, 026108.

Dolan, K. T., Spano, M. L. 2001 Surrogate for nonlinear time series analysis. Phys. Rev. E 64, 046128.

Duckrow, R. B., Albano, A. M. 2003 Comment on ``Performance of different synchronization measures in real data: A case study on electroencephalographic signals''. Phys. Rev. E 67, 63901.

Dumermuth, G., Molinari, I. 1991 Relationships among signals: cross-spectral analysis of the EEG. In: Digital biosignal processing *Vol. 5*, pp. 361-398. Ed. R. Weitkunat. Elsevier Science Publishers Amsterdam.

Dumont, M., Jurysta, F., Lanquart, J., Migeotte, P., van de Borne, P., Linkowski, P. 2004 Interdependency between heart rate variability and sleep EEG: linear/non-linear? Clin. Neurophysiol. 115, 2031-2040.

Dzakpasu, R., Zochowski, M. 2005 Discriminating differing types of synchrony in neural systems. Physica D 208, 115-122.

Eckhorn, R., Popel, B. 1974 Rigorous and extended application of information theory to the afferent visual system of the cat: I. Basic concepts. Biol. Cybern. 16, 191-200.

EEGLab Matlab Toolbox. Available at http://sccn.ucsd.edu/eeglab/.

Elbert, T., Ray, W. J., Kowalik, Z. J., Skinner, J. E., Graf, K. E., Birbaumer, N. 1994 Chaos and physiology: deterministic chaos in excitable cell assemblies. Physiol. Rev. 74, 1-47.

Engel, A. K., Singer, W. 2001 Temporal binding and the neural correlates of sensory awareness Trends Cogn. Sci. 5, 16-25.

Essl, M., Rappelsberger, P. 1998 EEG coherence and reference signals: experimental results and mathematical explanations. Med. Biol. Eng. Comput. 36, 399-406.

Faes, L., Pinna, G., Porta, A., Maestri, R., Nollo, G. 2004 Surrogate data analysis for assessing the significance of the coherence function. IEEE Trans. Biomed. Eng. 51, 1156-1166.

Fanselow, E. E., Sameshima, K., Baccala, L. A., Nicolelis, M. A. 2001 Thalamic bursting in rats during different awake behavioral states. Proc. Natl. Acad. Sci. U. S. A. 98, 15330-15335.

Faure, P., Korn, H. 2001 Is there chaos in the brain? I. Concepts of nonlinear dynamics and methods of investigation. C. R. Acad. Sci. III 324, 773-793.

Feldmann, U., Bhattacharya, J. 2004 Predictability improvement as an asymmetrical measure of interdependence in bivariate time series. Int. J. of Bifurcation and Chaos 14, 505-514.

Fell, J., Klaver, P., Elfadil, H., Schaller, C., Elger, C. E., Fernández, G. 2003 Rhinal-hippocampal theta coherence during declarative memory formation: interaction with gamma synchronization? Eur. J. Neurosci. 17, 1082-1088.

Fell, J., Klaver, P., Lehnertz, K., Grunwald, T., Schaller, C., Elger, C. E., Fernández, G. 2001 Human memory formation is accompanied by rhinal-hippocampal coupling and decoupling. Nat. Neurosci. 4, 1259-1264.

Fell, J., Röschke, J., Schaffner, C. 1996 Surrogate data analysis of sleep electroencephalograms reveals evidence for nonlinearity. Biol. Cybern. 75, 85-92.

Fernandes de Lima, V. M., Pijn, J. P., Nunes Filipe, C., Lopes da Silva, F. 1990 The role of hippocampal commissures in the interhemispheric transfer of epileptiform afterdischarges in the rat: a study using linear and non-linear regression analysis. Electroencephalogr. Clin. Neurophysiol. 76, 520-539.

Filipe, C. N., Pijn, J. P., de Lima, V. M., da Silva, F. H. 1989 Propagation of afterdischarges along the septo-temporal axis of the rat hippocampus: a quantitative analysis. Electroencephalogr. Clin. Neurophysiol. 73, 172-178.

Franaszczuk, P. J., Bergey, G. K., Kaminski, M. J. 1994 Analysis of mesial temporal seizure onset and propagation using the directed transfer function method. Electroencephalogr. Clin. Neurophysiol. 91, 413-427.

Franaszczuk, P. J., Blinowska, K. J., Kowalczyk, M. 1985 The application of parametric multichannel spectral estimates in the study of electrical brain activity. Biol. Cybern. 51, 239-247.

Fraser, A. M., Swinney, H. L. 1986 Independent coordinates for strange attractors from mutual information. Phys. Rev. A 33, 1134-1140.

Freeman, W. J., Rogers, L. 2002 Fine temporal resolution of analytic phase reveals episodic synchronization by state transitions in gamma EEGs. J. Neurophysiol. 87, 937-945.

Freiwald, W. A., Valdes, P., Bosch, J., Biscay, R., Jimenez, J. C., Rodriguez, L. M., Rodriguez, V., Kreiter, A. K., Singer, W. 1999 Testing non-linearity and directedness of interactions between neural groups in the macaque inferotemporal cortex. J Neurosci Methods 94, 105-119.

French, C. C., Beaumont, J. G. 1984 A critical review of EEG coherence studies of hemisphere function. Int. J. Psychophysiol. 1, 241-254.

Fries, P., Reynolds, J. H., Rorie, A. E., Demisome, R. 2001 Modulation of oscillatory neuronal synchronization by selective attention. Science 291, 1560-1563.

Fujisaka, H., Yamada, T. 1983 Stability theory of synchronized motion in coupled-oscillator systems. Prog. Theor. Phys. 69, 32-47.

Galka, A. 2000 Topics in Nonlinear Time Series Analysis with Implications for EEG Analysis. World Scientific: Singapore.

Gath, I., Feuerstein, C., Pham, D. T., Rondouin, G. 1992 On the tracking of rapid dynamic changes in seizure EEG. IEEE Trans. Biomed. Eng. 39, 952-958.

Gersch, W. 1970 Spectral analysis of EEG's by autoregressive decomposition of time series. Math. Biosci. 14, 177-196.

Gersch, W., Goddard, G. V. 1970 Epileptic focus location: spectral analysis method. Science 169, 701-702.

Gersch, W., Martinelli, F., Yonemoto, J., Low, M. D., McEwen, J. A. 1980 A Kullback Leibler-nearest neighbor rule classification of EEGs: the EEG population screening problem, an anesthesia level EEG classification application. Comput. Biomed. Res. 13, 283-296.

Gevins, A. S., Schaffer, R. E. 1980 A critical-review of electroencephalographic (EEG) correlates of higher cortical functions. CRC Crit. Rev. Bioeng. 4, 113-164.

Ginter, J., Jr., Blinowska, K. J., Kaminski, M., Durka, P. J. 2001 Phase and amplitude analysis in time-frequency space--application to voluntary finger movement. J. Neurosci. Methods 110, 113-124.

Goldberg, J., Rokni, U., Boraud, T., Vaadia, E., Bergman, H. 2004 Spike synchronization in the cortex-basal ganglia networks of parkinsonian primates reflects global dynamics of the local field potentials. J. Neurosci. 24, 6003-6010.

Golomb, D., Hansel, D., Mato, G. 2001 Mechanisms of synchrony of neural activity in large networks In: Handbook of Biological Physics *Vol. 4, Neuro-Informatics and Neural Modelling*, pp. 887-968. Eds. F. Moss, S. Gielen. Elsevier: Amsterdam.

Granger, C. W. J. 1969 Investigating causal relations by econometric models and cross-spectral methods. Econometrica 37, 424-438.

Granger, C. W. J. 1980 Testing for causality: a personal viewpoint. J Econ Dyn Control 2, 329-352.

Grassberger, P. 1988 Finite sample corrections to entropy and dimension estimates. Phys. Lett. A 128, 369.

Gross, J., Schmitz, F., Schnitzler, I., Kessler, K., Shapiro, K., Hommel, B., Schnitzler, A. 2004 Modulation of long-range neural synchrony reflects temporal limitations of visual attention in humans. Proc. Natl. Acad. Sci. U. S. A. 101, 13050-13055.

Gross, J., Timmermann, J., Kujala, J., Dirks, M., Schmitz, F., Salmelin, R., Schnitzler, A. 2002 The neural basis of intermittent motor control in humans. Proc. Natl. Acad. Sci. U. S. A. 99, 2299-2302.

Grün, S., Diesmann, M., Aertsen, A. 2002a Unitary events in multiple single-neuron spiking activity: I. Detection and significance. Neural Comput. 14, 43-80.

Grün, S., Diesmann, M., Aertsen, A. 2002b Unitary events in multiple single-neuron spiking activity: II. Nonstationary data. Neural Comput. 14, 81-119.

Grün, S., Diesmann, M., Grammont, F., Riehle, A., Aertsen, A. 1999 Detecting unitary events without discretization of time. J. Neurosci. Methods 94, 67-79.

Guler, N. F., Kiymik, M. K., Guler, I. 1995 Comparison of FFT- and AR-based sonogram outputs of 20 MHz pulsed Doppler data in real time. Comput. Biol. Med. 25, 383-391.

Hahnloser, R. H., Kozhevnikov, A. A., Fee, M. S. 2002 An ultra-sparse code underlies the generation of neural sequences in a songbird. Nature 419, 65-70.

Haig, A. R., Gordon, E., Wright, J. J., Meares, R. A., Bahramali, H. 2000 Synchronous cortical gamma-band activity in task-relevant cognition. Comput. Neurosci. 11, 669-675.

Hannan, E. J. 1970 Multiple Time Series. John Wiley & Sons: New York.

Haykin, S., Kesler, S. 1983 Prediction-error filtering and maximum entropy spectral estimatiuon. In: Nonlinear methods of spectral analysis., pp. 9-72. Ed. S. Haykin. Springer-Verlag: Berlin.

Hegger, R., Kantz, H., Schreiber, T. 1999 Practical implementation of nonlinear time series methods: The TISEAN package. Chaos 9, 413-435. Available at http://www.mpipks-dresden.mpg.de/~tisean/TISEAN_2.1/index.html.

Herrmann, C. S., Munk, M. H. J., Engel, A. K. 2004 Cognitive functions of gamma-band activity: memory match and utilization. Trends Cogn. Sci. 8, 347-355.

Hesse, W., Moller, E., Arnold, M., Schack, B. 2003 The use of time-variant EEG Granger causality for inspecting directed interdependencies of neural assemblies. J. Neurosci. Methods 124, 27-44.

Hoke, M., Lehnertz, K., Pantev, C., Lütkenhöner, B. 1989 Spatiotemporal aspects of synergetic processes in the auditory cortex as revealed by magnetoencephalogram. In: Series in Brain Dynamics *Vol. 2*. Eds. E. Basar, T. H. Bullock. Springer: Berlin.

Hsu, A., Woolley, S. M., Fremouw, T. E., Theunissen, F. E. 2004 Modulation power and phase spectrum of natural sounds enhance neural encoding performed by single auditory neurons. J. Neurosci. 24, 9201-9211.

Hu, X., Nenov, V. 2004 Robust measure for characterizing generalized synchronization. Phys. Rev. E 69, 026206.

Huang, L., Yu, P., Ju, F., Cheng, J. 2003 Prediction of response to incision using the mutual information of electroencephalograms during anaesthesia. Med. Eng. Phys. 25, 321-327.

Huygens, C. 1673 Horoloquium Oscilatorium. Paris.

Ioannides, A. A. 2001 Real time human brain function: observations and inferences from single trial analysis of magnetoencephalographic signals. Clinical Electroencephalography 32, 98-111.

Ioannides, A. A., Liu, L. C., Kwapien, J., Drozdz, S., Streit, M. 2000 Coupling of regional activations in a human brain during an object and face affect recognition task. Hum. Brain Mapp. 11, 77-92.

Jansen, B. H. 1991 Quantitative analysis of electroencephalograms: is there chaos in the future? Int. J. Biomed. Comput. 27, 95-123.

Jansen, B. H., Rit, V. G. 1995 Electroencephalogram and visual evoked potential generation in a mathematical model of coupled cortical columns. Biol. Cybern. 73, 357-366.

Jarvis, M. R., Mitra, P. P. 2001 Sampling properties of the spectrum and coherency of sequences of action potentials. Neural Comput. 13, 717-749.

Jeong, J., Gore, J. C., Peterson, B. S. 2001 Mutual information analysis of the EEG in patients with Alzheimer's disease. Clin. Neurophysiol. 112, 827-835.

Jerger, K. K., Netoff, T. I., Francis, J. T., Sauer, T., Pecora, L., Weinstein, S. L., Schiff, S. J. 2001 Early seizure detection. J. Clin. Neurophysiol. 18, 259-268.

Kaiser, A., Schreiber, T. 2002 Information transfer in continuous processes. Physica D 166, 43-62.

Kaminski, M., Blinowska, K., Szclenberger, W. 1997 Topographic analysis of coherence and propagation of EEG activity during sleep and wakefulness. Electroencephalogr. Clin. Neurophysiol. 102, 216-227.

Kaminski, M., Blinowska, K., Szelenberger, W. 1995 Investigation of coherence structure and EEG activity propagation during sleep. Acta Neurobiol. Exp. (Warsz). 55, 213-219.

Kaminski, M., Ding, M., Truccolo, W. A., Bressler, S. L. 2001 Evaluating causal relations in neural systems: granger causality, directed transfer function and statistical assessment of significance. Biol. Cybern. 85, 145-157.

Kaminski, M. J., Blinowska, K. J. 1991 A new method of the description of the information flow in the brain structures. Biol. Cybern. 65, 203-210.

Kantz, H., Schreiber, T. 2004 Nonlinear Time Series Analysis. Cambridge University Press: Cambridge.

Kawabata, N. 1973 Nonstationary Analysis of Electroencephalogram. IEEE Trans. Biomed. Eng. BM20, 444-452.

Kocsis, B., Bragin, A., Buzsaki, G. 1999 Interdependence of multiple theta generators in the hippocampus: a partial coherence analysis. J. Neurosci. 19, 6200-6212.

Korn, H., Faure, P. 2003 Is there chaos in brain? II. Experimental evidence and related methods. C. R. Biol. 326, 787-840.

Korzeniewska, A., Kasicki, S., Kaminski, M., Blinowska, K. J. 1997 Information flow between hippocampus and related structures during various types of rat's behavior. J. Neurosci. Methods 73, 49-60.

Koskinen, M., Seppanen, T., Tuukkanen, J., Yli-Hankala, A., Jantti, V. 2001 Propofol anesthesia induces phase synchronization changes in EEG. Clin. Neurophysiol. 112, 386-392.

Kramer, M. A., Edwards, E., Soltani, M., Berger, M. S., Knight, R. T., Szeri, A. J. 2004 Synchronization measures of bursting data: application to the electrocorticogram of an auditory event-related experiment. Phys. Rev. E 70, 011914.

Kraskov, A. 2004 Synchronization and Interdependence Measures and their Applications to the Electroencephalogram of Epilepsy Patients and Clustering of Data, Dissertation (PhD Thesis). Document publicly available at http://www.fz-juelich.de/nic-series/NIC-Series-e.html. NIC-Directors: Jülich.

Kraskov, A., Kreuz, T., Quiroga, R. Q., Grassberger, P., Mormann, F., Lehnertz, K., Elger, C. E. 2002 Comparison of two phase synchronization analysis techniques for interictal focus lateralization in mesial temporal lobe epilepsy. Epilepsia 43, 48-54.

Kraskov, A., Stögbauer, H., Grassberger, P. 2004 Estimating mutual information. Phys. Rev. E 69, 066138.

Kreuz, T. 2004 Measuring Synchronization in Model Systems and Electroencephalographic Time Series from Epilepsy Patients, Dissertation (PhD Thesis). Document publicly available at http://www.fz-juelich.de/nic-series/NIC-Series-e.html. NIC-Directors: Jülich. .

Kreuz, T., Andrzejak, R. G., Mormann, F., Kraskov, A., Stogbauer, H., Elger, C. E., Lehnertz, K., Grassberger, P. 2004 Measure profile surrogates: A method to validate the performance of epileptic seizure prediction algorithms. Phys. Rev. E 69, 061915.

Kus, R., Kaminski, M., Blinowska, K. J. 2004 Determination of EEG activity propagation: pair-wise versus multichannel estimate. IEEE Trans. Biomed. Eng. 51, 1501-1510.

Lachaux, J. P., Rodriguez, E., Le van Quyen, M., Lutz, J., Martinerie, J., Varela, F. J. 2000 Studying single-trials of phase-synchronous activity in the brain. Int. J. of Bifurcation and Chaos 10, 2429-2441.

Lachaux, J. P., Rodriguez, E., Martinerie, J., Varela, F. 1999 Measuring phase synchrony in brain signals. Hum. Brain Mapp. 8, 194-208.

Laird, A. R., Rogers, B. P., Carew, J. D., Arfanakis, K., Moritz, C. H., Meyerand, M. E. 2002 Characterizing instantaneous phase relationships in whole-brain fMRI activation data. Hum. Brain Mapp. 16, 71-80.

Lamberts, J., van den Broek, P. L. C., Bener, L., van Egmond, J., Dirksen, R., Coenen, A. M. L. 2000 Correlation dimension of the human electroencephalogram corresponds with cognitive load. Neuropsychobiology 41, 149-153.

Le van Quyen, M., Adam, C., Baulac, M., Martinerie, J., Varela, F. J. 1998 Nonlinear interdependencies of EEG signals in human intracranially recorded temporal lobe seizures. Brain Res. 792, 24-40.

Le van Quyen, M., Foucher, J., Lachaux, J. P., Rodriguez, E., Lutz, A., Martinerie, J., Varela, F. 2001 Comparison of Hilbert transform and wavelet methods for the analysis of neuronal synchrony. J. Neurosci. Methods 111, 83.

Le van Quyen, M., Martinerie, J., Adam, C., Varela, F. J. 1999 Nonlinear analyses of interictal EEG map the brain interdependences in human focal epilepsy. Physica D 127, 250-266.

Lee, D. 2003 Coherent oscillations in neuronal activity of the supplementary motor area during a visuomotor task. J. Neurosci. 23, 6798-6809.

Lee, K.-H., Williams, L. M., Breakspear, M., Gordon, E. 2003 Synchronous gamma activity: a review and contribution to an integrative neuroscience model of schizophrenia. Brain Res. Rev. 41, 57-78.

Lehnertz, K., Arnhold, J., Grassberger, P., Elger, C. E. 2000 Chaos in Brain? World Scientific: Singapore.

Liang, H., Bressler, S. L., Ding, M., Truccolo, W. A., Nakamura, R. 2002 Synchronized activity in prefrontal cortex during anticipation of visuomotor processing. Neuroreport 13, 2011-2015.

Liang, H., Ding, M., Nakamura, R., Bressler, S. L. 2000 Causal influences in primate cerebral cortex during visual pattern discrimination. Neuroreport 11, 2875-2880.

Liberati, D., Cursi, M., Locatelli, T., Comi, G., Cerutti, S. 1997 Total and partial coherence analysis of spontaneous and evoked EEG by means of multi-variable autoregressive processing. Med. Biol. Eng. Comput. 35, 124-130.

Lin, F.-H., Witzel, T., Hämäläinen, M. S., Dale, A. M., Belliveau, J. W., Stufflebeam, S. M. 2004 Spectral spatiotemporal imaging of cortical oscillations and interactions in the human brain. Neuroimage 23, 582-595.

Linkenkaer-Hansen, K., Nikouline, V. V., Palva, J. M., Ilmoniemi, R. J. 2001 Long-range temporal correlations and scaling behavior in human brain oscillations. J. Neurosci. 21, 1370-1377.

Lisman, J. E. 1997 Bursts as a unit of neural information: making unreliable synapses reliable. Trends Neurosci. 20, 38-43.

London, M., Schreibman, A., Hausser, M., Larkum, M. E., Segev, I. 2002 The information efficacy of a synapse. Nat. Neurosci. 5, 332-340.

Lopes da Silva, F., Pijn, J. P., Boeijinga, P. 1989 Interdependence of EEG signals: linear vs. nonlinear associations and the significance of time delays and phase shifts. Brain Topogr. 2, 9-18.

Lopes da Silva, F. H., van Rotterdam, A., Barts, P., van Heusden, E., Burr, W. 1976 Models of neuronal populations: the basic mechanisms of rhythmicity. Prog. Brain Res. 45, 281-308.

Lopes da Silva, F. H., Vos, J. E., Mooibroek, J., Van Rotterdam, A. 1980 Relative contributions of intracortical and thalamo-cortical processes in the generation of alpha rhythms, revealed by partial coherence analysis. Electroencephalogr. Clin. Neurophysiol. 50, 449-456.

Lutz, A., Greischar, L. L., Rawlings, N. B., Ricard, M., Davidson, R. J. 2004 Long-term meditators self-induce high-amplitude gamma synchrony during mental practice. Proc. Natl. Acad. Sci. U. S. A. 101, 16369-16373.

Lutz, A., Lachaux, J. P., Martinerie, J., Varela, F. 2002 Guiding the study of brain dynamics by using first person data: Synchrony patterns correlate with ongoing conscious states during a simple visual task. Proc. Natl. Acad. Sci. U. S. A. 99, 1586-1591.

Machens, C. K. 2002 Adaptive sampling by information maximization. Phys. Rev. Lett. 88, 228104.

Makarenko, V., Llinás, R. 1998 Experimentally determined chaotic phase synchronization in a neuronal system. Proc. Natl. Acad. Sci. U. S. A. 95, 15747-15752.

Mallat, S. A. 1999 A Wavelet Tour of Signal Processing. Academic Press: London.

Mardia, K. V. 1972 Probability and Mathematical Statistics: Statistics of Directional Data. Academic Press: London.

Marple Jr., S. L. 1987 Digital Spectral Analysis with Applications. Prentice-Hall: New Jersey.

Mars, N. J., Thompson, P. M., Wilkus, R. J. 1985 Spread of epileptic seizure activity in humans. Epilepsia 26, 85-94.

Mathtools.net at http://www.mathtools.net/.

MatlabCentral at http://www.mathworks.com/matlabcentral/.

Matsumoto, K., Tsuda, I. 1988 Calculation of information flow rate from mutual information. J. Phys. A 21, 1405-1414.

McIntosh, A. R., Gonzalez-Lima, F. 1992 Structural modeling of functional visual pathways mapped with 2-deoxyglucose: effects of patterned light and footshock. Brain Res. 578, 75-86.

McIntosh, A. R., Gonzalez-Lima, F. 1994 Network interactions among limbic cortices, basal forebrain, and cerebellum differentiate a tone conditioned as a Pavlovian excitor or inhibitor: fluorodeoxyglucose mapping and covariance structural modeling. J. Neurophysiol. 72, 1717-1733.

Meeren, H. K., Pijn, J. P., Van Luijtelaar, E. L., Coenen, A. M., Lopes da Silva, F. H. 2002 Cortical focus drives widespread corticothalamic networks during spontaneous absence seizures in rats. J. Neurosci. 22, 1480-1495.

MILCA Mutual Information Least Component Analysis. Available at http://www.fz-juelich.de/nic/cs/software/.

Min, B. C., Jin, S. H., Kang, I. H., Lee, D. H., Kang, J. K., Lee, S. T., Sakamoto, K. 2003 Analysis of mutual information content for EEG responses to odor stimulation for subjects classified by occupation. Chem. Senses 28, 741-749.

Mirski, M. A., Tsai, Y. C., Rossell, L. A., Thakor, N. V., Sherman, D. L. 2003 Anterior thalamic mediation of experimental seizures: selective EEG spectral coherence. Epilepsia 44, 355-365.

Moller, E., Schack, B., Arnold, M., Witte, H. 2001 Instantaneous multivariate EEG coherence analysis by means of adaptive high-dimensional autoregressive models. J. Neurosci. Methods 105, 143-158.

Montbrio, E., Kurths, J., Blasius, B. 2004 Synchronization of two interacting populations of oscillators. Phys. Rev. E 70, 056125.

Morf, M., Vieira, A., Lee, D., Kailath, T. 1978 Recursive multichannel maximum entropy spectral estimation. IEEE Trans. Geosci. Electronics 16, 85-94.

Mormann, F., Kreuz, T., Andrzejak, R. G., David, P., Lehnertz, K., Elger, C. E. 2003 Epileptic seizures are preceded by a decrease in synchronization. Epilepsy Res. 53, 173-185.

Mormann, F., Lehnertz, K., David, P., Elger, C. E. 2000 Mean phase coherence as a measure for phase synchronization and its application to the EEG of epilepsy patients. Physica D 144, 358-369.

Na, S. H., Jin, S. H., Kim, S. Y., Ham, B. J. 2002 EEG in schizophrenic patients: mutual information analysis. Clin. Neurophysiol. 113, 1954-1960.

Nemenman, I., Bialek, W., de Ruyter van Steveninck, R. 2004 Entropy and information in neural spike trains: progress on the sampling problem. Phys. Rev. E 69, 056111.

Netoff, T. I., Schiff, S. J. 2002 Decreased neuronal synchronization during experimental seizures. J. Neurosci. 22, 7297-7307.

Nolte, G., Bai, O., Wheaton, L., Mari, Z., Vorbach, S., Hallet, M. 2004 Identifying true brain interaction from EEG data using the imaginary part of coherency. Clin. Neurophysiol. 115, 2292-2307.

Nowak, L. G., Bullier, J. 2000 Cross-correlograms for neuronal spike trains. Different types of temporal correlation in neocortex, their origin and significance. In: Time and the brain *Vol. 3*, pp. 53-96. Ed. R. Miller. Harwood Academic Publishers: Amsterdam.

Nunez, P. L., Silberstein, R. B., Shi, Z., Carpenter, M. R., Srinivasan, R., Tucker, D. M., Doran, S. M., Cadusch, P. J., Wijesinghe, R. S. 1999 EEG coherency II: experimental comparisons of multiple measures. Clin. Neurophysiol. 110, 469-486.

Nunez, P. L., Wingeier, B. M., Silberstein, R. B. 2001 Spatial-temporal structures of human alpha rhythms: theory, microcurrent sources, multiscale measurements, and global binding of local networks. Hum. Brain Mapp. 13, 125-164.

Otnes, R. K., Enochson, L. 1972 Digital Time Series Analysis. John Wiley & Sons: New York.

Palus, M. 1996 Detecting nonlinearity in multivariate time series. Phys. Lett. A 213, 138-147.

Palus, M., Komarek, V., Hrncir, Z., Sterbova, K. 2001 Synchronization as adjustment of information rates: Detection from bivariate time series. Phys. Rev. E 63, 046211.

Palva, J. M., Palva, S., Kaila, K. 2005 Phase synchrony among neuronal oscillations in the human cortex. J. Neurosci. 25, 3962-3972.

Panzeri, S., Schultz, S. R., Treves, A., Rolls, E. T. 1999 Correlations and the encoding of information in the nervous system. Proc. R. Soc. Lond. B. Biol. Sci. 266, 1001-1012.

Parlitz, U., Junge, L., Lauterborn, W., Kocarev, L. 1996 Experimental observation of phase synchronization. Phys. Rev. E 54, 2115-2117.

Pereda, E., Gamundi, A., Rial, R., Gonzalez, J. 1998 Non-linear behaviour of human EEG: fractal exponent versus correlation dimension in awake and sleep stages. Neurosci. Lett. 250, 91-94.

Pereda, E., Manas, S., De Vera, L., Garrido, J. M., Lopez, S., Gonzalez, J. 2003 Non-linear asymmetric interdependencies in the electroencephalogram of healthy term neonates during sleep. Neurosci. Lett. 337, 101-105.

Pereda, E., Rial, R., Gamundi, A., Gonzalez, J. 2001 Assessment of changing interdependencies between human electroencephalograms using nonlinear methods. Physica D 148, 147-158.

Perkel, D. H., Gerstein, G. L., Moore, G. P. 1967 Neuronal spike trains and stochastic point processes. II. Simultaneous spike trains. Biophys. J. 7, 419-440.

Pesaran, B., Pezaris, J. S., Sahani, M., Mitra, P. P., Andersen, R. A. 2002 Temporal structure in neuronal activity during working memory in macaque parietal cortex. Nat. Neurosci. 5, 805-811.

Pijn, J. P. 1990 Quantitative Evaluation of EEG Signals in Epilepsy, Ph.D. Thesis, Amsterdam University, Amsterdam

Pijn, J. P., Vijn, P. C., Lopes da Silva, F. H., Van Ende Boas, W., Blanes, W. 1990 Localization of epileptogenic foci using a new signal analytical approach. Neurophysiol. Clin. 20, 1-11.

Pikovsky, A., Rosenblum, M. G., Kurths, J. 2001 Synchronization: a Universal Concept in Nonlinear Science. Cambridge University Press: Cambridge.

Pradhan, N., Sadasivan, P. K., Chatterji, S., Narayana, D. 1995 Patterns of attractor dimensions of sleep EEG. Comput. Biol. Med. 25, 455-462.

Prichard, D., Theiler, J. 1994 Generating surrogate data for time series with several simultaneously measured variables. Phys. Rev. Lett. 73, 951-954.

Quian Quiroga, R., Arnhold, J., Grassberger, P. 2000 Learning driver-response relationships from synchronization patterns. Phys. Rev. E 61, 5142-5148.

Quian Quiroga, R., Kraskov, A., Kreuz, T., Grassberger, P. 2002a Performance of different synchronization measures in real data: a case study on electroencephalographic signals. Phys. Rev. E 65, 041903.

Quian Quiroga, R., Kreuz, T., Grassberger, P. 2002b Event synchronization: a simple and fast method to measure synchronicity and time delay patterns. Phys. Rev. E 66, 041904.

Quian Quiroga, R., Rosso, O., Basar, E., Schurmann, M. 2001 Wavelet entropy in event-related potentials: a new method shows ordering of EEG oscillations. Biol. Cybern. 84, 291-299.

Riehle, A., Grun, S., Diesmann, M., Aertsen, A. 1997 Spike synchronization and rate modulation differentially involved in motor cortical function. Science 278, 1950-1953.

Rodriguez, E., George, N., Lachaux, J. P., Martinerie, J., Renault, B., Varela, F. J. 1999 Perception's shadow: long-distance synchronization of human brain activity. Nature 397, 430-433.

Roelfsema, P. R., Engel, A. K., König, P., Singer, W. 1997 Visuomotor integration is associated with the zero time-lag synchronization among cortical areas. Nature 385, 157-161.

Rosenblum, M., Pikovsky, A. 2004a Delayed feedback control of collective synchrony: An approach to suppression of pathological brain rhythms. Phys. Rev. E 70, 041904.

Rosenblum, M. G., Cimponeriu, L., Bezerianos, A., Patzak, A., Mrowka, R. 2002 Identification of coupling direction: Application to cardiorespiratory interaction. Phys. Rev. E 65, 041909.

Rosenblum, M. G., Firsov, G. I., Kuuz, R., Pompe, B. 1998 Human postural control: force plate experiments and modelling. In: Nonlinear Analysis of Physiological Data, pp. 283-306. Eds. H. Kantz, J. Kurths, G. Mayer-Kress. Springer: Berlin.

Rosenblum, M. G., Pikovsky, A. 2001 Detecting direction of coupling in interacting oscillators. Phys. Rev. E 64, 045202(R).

Rosenblum, M. G., Pikovsky, A. S. 2004b Controlling synchronization in an ensemble of globally coupled oscillators. Phys. Rev. Lett. 92, 114102.

Rosenblum, M. G., Pikovsky, A. S., Kurths, J. 1996 Phase synchronization of chaotic oscillators. Phys. Rev. Lett. 76, 1804-1807.

Rosenblum, M. G., Pikovsky, A. S., Kurths, J. 2004 Synchronization approach to analysis of biological systems. Fluct. Noise Lett. 4, L53-L62.

Rosenblum, M. G., Pikovsky, A. S., Kurths, J., Schäfer, C., Tass, P. 2001 Phase synchronization: from theory to data analysis. In: Handbook of Biological Physics *Vol. 4, Neuro-Informatics and Neural Modelling*, pp. 279-321. Eds. F. Moss, S. Gielen. Elsevier Science: Amsterdam.

Rosso, O. A., Blanco, S., Yordanova, J., Kolev, V., Figliola, A., Schurmann, M., Basar, E. 2001 Wavelet entropy: a new tool for analysis of short duration brain electrical signals. J. Neurosci. Methods 105, 65-75.

Roulston, M. S. 1999 Estimating the errors on measured entropy and mutual information. Physica D 25, 285-294.

Rulkov, N. F., Afraimovich, V. S. 2003 Detectability of nondifferentiable generalized synchrony. Phys. Rev. E 67, 066218.

Rulkov, N. F., Sushchik, M. M., Tsimring, L. S., Abarbanel, H. D. I. 1995 Generalized synchronization of chaos in directionally coupled chaotic systems. Phys. Rev. E 51, 980-994.

Salazar, R. F., Konig, P., Kayser, C. 2004 Directed interactions between visual areas and their role in processing image structure and expectancy. Eur. J. Neurosci. 20, 1391-1401.

Sameshima, K., Baccala, L. A. 1999 Using partial directed coherence to describe neuronal ensemble interactions. J. Neurosci. Methods 94, 93-103.

Schiff, S. J., So, P., Chang, T., Burke, R. E., Sauer, T. 1996 Detecting dynamical interdependence and generalized synchrony through mutual prediction in a neural ensemble. Phys. Rev. E 54, 6708-6715.

Schmitz, A. 2000 Measuring statistical dependence and coupling of subsystems. Phys. Rev. E 62, 7508-7511.

Schnitzler, A., Gross, J. 2005 Normal and pathological oscillatory communication in the brain. Nature Rev. Neurosci. 6, 285-296.

Schreiber, T. 2000 Measuring information transfer. Phys. Rev. Lett. 85, 461-464.

Schreiber, T., Schmitz, A. 2000 Surrogate time series. Physica D 142, 346-382.

Segundo, J. P. 2001 Nonlinear dynamics of point process systems and data. Int. J. of Bifurcation and Chaos 13, 2035-2116.

Segundo, J. P., Sugihara, G., Dixon, P., Stiber, M., Bersier, L. 1998 The spike trains of inhibited pacemaker neurons seen through the magnifying glass of nonlinear analyses. Neuroscience 87, 741-766.

Shadlen, M. N., Movshon, J. A. 1999 Synchrony unbound: A critical evaluation of the temporal binding hypothesis. Neuron 24, 67-77.

Shannon, C. E., Weaver, W. 1949 The Mathematical Theory of Information. University Press: Urbana, Illinois.

Sharpee, T., Rust, N. C., Bialek, W. 2004 Analyzing neural responses to natural signals: maximally informative dimensions. Neural Comput. 16, 223-250.

Shaw, J. C. 1984 Correlation and coherence analysis of the EEG: a selective tutorial review. Int. J. Psychophysiol. 1, 255-266.

Shaw, J. C., Ongley, C. 1972 The measurement of synchronization. In: Synchronization of EEG activity in epilepsies, pp. 204-216. Eds. H. Petsche, M. A. B. Brazier. Springer-Verlag: Wien.

Simoes, C., Jensen, O., Parkkonen, L., Hari, R. 2003 Phase locking between human primary and secondary somatosensory cortices. Proc. Natl. Acad. Sci. U. S. A. 100, 2691-2694.

Singer, W., Gray, C. M. 1995 Visual feature integration and the temporal correlation hypothesis. Annu. Rev. Neurosci. 18, 555-586.

Sinkkonen, J., Tiitinen, H., Näätänen, R. 1995 Gabor filters - An informative way for analyzing event-related brain activity. J. Neurosci. Methods 56, 99-104.

Smirnov, D. A., Andrzejak, R. G. 2005 Detection of weak directional coupling: Phase-dynamics approach versus state-space approach. Phys. Rev. E 71, 036207.

Smirnov, D. A., Bezruchko, B. 2003 Estimation of interaction strength and direction from short and noisy time series. Phys. Rev. E 68, 046209.

Spencer, K. M., Nestor, P. G., Niznikiewicz, M. A., Salisbury, D. F., Shenton, M. E., McCarley, R. W. 2003 Abnormal neural synchrony in schizophrenia. J. Neurosci. 23, 7407-7411.

Spyers-Ashby, J. M., Bain, P. G., Roberts, S. J. 1998 A comparison of fast Fourier transform (FFT) and autoregressive (AR) spectral estimation techniques for the analysis of tremor data. J. Neurosci. Methods 83, 35-43.

Stam, C. J. 2005 Nonlinear dynamical analysis of EEG and MEG: Review of an emerging field. Clin. Neurophysiol. 116, 2266-2301.

Stam, C. J., van Cappellen van Walsum, A. M., Micheloyannis, S. 2002a Variability of EEG synchronization during a working memory task in healthy subjects. Int. J. Psychophysiol. 46, 53-66.

Stam, C. J., van Cappellen van Walsum, A. M., Pijnenburg, Y. A., Berendse, H. W., de Munck, J. C., Scheltens, P., van Dijk, B. W. 2002b Generalized synchronization of MEG recordings in Alzheimer's Disease: evidence for involvement of the gamma band. J. Clin. Neurophysiol. 19, 562-574.

Stam, C. J., van Dijk, B. W. 2002 Synchronization likelihood: an unbiased measure of generalized synchronization in multivariate data sets. Physica D 163, 236-251.

Stein, R. B., French, A. S., Holden, A. V. 1972 The frequency response, coherence, and information capacity of two neuronal models. Biophys. J. 12, 295-322.

Stögbauer, H., Kraskov, A., Astakhov, S. A., Grassberger, P. 2004 Least-dependent-component analysis based on mutual information. Phys. Rev. E 70, 066123.

Sun, F. T., Miller, L. M., D'Esposito, M. 2004 Measuring interregional functional connectivity using coherence and partial coherence analyses of fMRI data. Neuroimage 21, 647-658.

Supp, G. G., Schlogl, A., Gunter, T. C., Bernard, M., Pfurtscheller, G., Petsche, H. 2004 Lexical memory search during N400: cortical couplings in auditory comprehension. Neuroreport 15, 1209-1213.

Takens, F. 1980 Detecting strange attractors in turbulence. In: Dynamical Systems and Turbulence *Vol. 898*, pp. 366-381. Eds. D. A. Rand ,L. S. Young Springer-Verlag: Warwick.

Tallon-Baudry, C., Bertrand, O. 1999 Oscillatory gamma activity in humans and its role in object representation. Trends Cogn. Sci. 3, 151-162.

Tallon-Baudry, C., Bertrand, O., Delpuech, C., Pemier, J. 1996 Stimulus specificity of phase-locked and non-phase-locked 40 Hz visual responses in human. J. Neurosci. 16, 4240-4249.

Tallon-Baudry, C., Bertrand, O., Delpuech, C., Pernier, J. 1997 Oscillatory gamma band (30-70 Hz) activity induced by visual search task in human. J. Neurosci. 17, 722-734.

Tallon-Baudry, C., Bertrand, O., Fischer, C. 2001 Oscillatory synchrony between human extrastriate areas during visual short-term memory maintenance. J. Neurosci. 21, 1-5.

Tallon-Baudry, C., Bertrand, O., Peronnet, F., Pernier, J. 1998 Induced gamma-band activity during the delay of a visual short term memory task in human. J. Neurosci. 18, 4244-4254.

Tass, P. 2003 Desynchronization by means of a coordinated reset of neural sub-populations - A novel technique for demand-controlled deep brain stimulation. Prog. Theor. Phys. Supp. 150, 281-296.

Tass, P., Rosenblum, M. G., Weule, J., Kurths, J., Pikovsky, A., Volkmann, J., Schnitzler, A., Freund, H. J. 1998 Detection of n:m phase locking from noisy data: Application to magnetoencephalography. Phys. Rev. Lett. 81, 3291-3294.

Tass, P. A. 1999 Phase Resetting in Medicine and Biology: Stochastic Modelling and Data Analysis. Springer: Berlin.

Tass, P. A., Fieseler, T., Dammers, J., Dolan, K., Morosan, P., Majtanik, M., Boers, F., Muren, A., Zilles, K., Fink, G. R. 2003 Synchronization tomography: A method for three-dimensional localization of phase synchronized neuronal populations in the human brain using magnetoencephalography. Phys. Rev. Lett. 90, 088101.

Tecchio, F., De Lucia, M., Salustri, C., Montuori, M., Bottaccio, M., Babiloni, C., Pietronero, L., Zappasodi, F., Rossini, P. M. 2004 District-related frequency specificity in hand cortical representation: dynamics of regional activation and intra-regional synchronization. Brain Res. 1014, 80-86.

Teräsvirta, T. 1998 Modeling economic relationships with smooth transition regression. In: Handbook of applies economic statistics. Eds. A. Ullah,D. E. A. Gilles. Marcel Dekker: New York

Terry, J. R., Anderson, C., Horne, J. A. 2004 Nonlinear analysis of EEG during NREM sleep reveals changes in functional connectivity due to natural aging. Hum. Brain Mapp. 23, 73-84.

Terry, J. R., Breakspear, M. 2003 An improved algorithm for the detection of dynamical interdependence in bivariate time-series. Biol. Cybern. 88, 129-136.

Thakor, N. V., Tong, S. 2004 Advances in quantitative electroencephalogram analysis methods. Annu. Rev. Biomed. Eng. 6, 453-495.

Thatcher, R. W., Krause, P. J., Hrybyk, M. 1986 Cortico-cortical associations and EEG coherence: a two-compartmental model. Electroencephalogr. Clin. Neurophysiol. 64, 123-143.

Theiler, J., Eubank, S., Longtin, A., Galdrikian, B., Farmer, J. D. 1992 Testing for nonlinearity in time series: the method of surrogate data. Physica D 58, 77-94.

Theiler, J., Rapp, P. 1996 Re-examination of the evidence for low dimensional, nonlinear structure in the human electroencephalogram. Electroencephalogr. Clin. Neurophysiol. 98, 213-222.

Theunissen, F. E., Miller, J. P. 1991 Representation of sensory information in the cricket cercal sensory system. II. Information theoretic calculation of system accuracy and optimal tuning-curve widths of four primary interneurons. J. Neurophysiol. 66, 1690-1703.

Tiesinga, P. H. 2001 Information transmission and recovery in neural communication channels revisited. Phys. Rev. E 64, 012901.

Titcombe, M., Glass, L., Guehl, D., Beuter, A. 2001 Dynamics of Parkinsonian tremor during deep brain stimulation. Chaos 11, 766-773.

Tong, H. 1990 Nonlinear Time Series: a Dynamical System Approach. Clarendon Press: Oxford.

Trabka, W., Pijn, J. P., Lopes da Silva, F. 1989 Spreading of epileptic afterdischarges between entorhinal cortex and hippocampus in acute experiments and the kindling model of epilepsy in the rat--comparing different methods of analysis. Acta Physiologica Polonica 40, 194-214.

Treisman, A. 1996 The binding problem. Current Opinion in Neurobiology 6, 171-178.

Trujillo, L., Peterson, M. A., Kasznniak, A. W., Allen, J. J. B. 2005 EEG phase synchrony differences across visual perception conditions may depend on recording and analysis methods. Clin. Neurophysiol. 116, 172-189.

TSTOOL Matlab Toolbox on Time Series Analysis. Available at http://www.physik3.gwdg.de/tstool/indexde.html.

Tucker, D. M., Roth, D. L., Bair, T. B. 1986 Functional connections among cortical regions: topography of EEG coherence. Electroencephalogr. Clin. Neurophysiol. 63, 242-250.

van den Broek, P. L. C., van Egmond, J., van Rijn, C. M., Takens, F., Coenen, A. M. L., Booij, L. 2005 Feasibility of real-time calculation of correlation integral derived statistics applied to EEG time series. Physica D 203, 198-208.

van Putten, M. J. A. M. 2003 Proposed link rates in the human brain. J. Neurosci. Methods 127, 1-10.

Varela, F. 1995 Resonant cell assemblies: a new approach to cognitive function and neuronal synchrony. Biol. Res. 28, 81-95.

Varela, F., Lachaux, J. P., Rodriguez, E., Martinerie, J. 2001 The brainweb: Phase synchronization and large-scale integration. Nature Rev. Neurosci. 2, 229-239.

Varma, N. K., Kushwaha, R., Beydoun, A., Williams, W. J., Drury, I. 1997 Mutual information analysis and detection of interictal morphological differences in interictal epileptiform discharges of patients with partial epilepsies. Electroencephalogr. Clin. Neurophysiol. 103, 426-433.

Vastano, J. A., Swinney, H. L. 1988 Information transport in spatiotemporal systems. Phys. Rev. Lett. 60, 1773-1776.

Walter, D. O., Adey, W. R. 1963 Spectral analysis of electroencephalograms recorded during learning in the cat, before and after subthalamic lesions. Exp. Neurol. 7, 481-501.

Walter, D. O., Rhodes, J. M., Brown, D., Adey, W. R. 1966 Comprehensive spectral analysis of human EEG generators in posterior cerebral regions. Electroencephalogr. Clin. Neurophysiol. 20, 224-237.

Wan, Y. H., Jian, Z., Wen, Z. H., Wang, Y. Y., Han, S., Duan, Y. B., Xing, J. L., Zhu, J. L., Hu, S. J. 2004 Synaptic transmission of chaotic spike trains between primary afferent fiber and spinal dorsal horn neuron in the rat. Neuroscience 125, 1051-1060.

Wang, G., Takigawa, M., Matsushita, T. 1992 Correlation of alpha activity between the frontal and occipital cortex. Jpn. J. Physiol. 42, 1-13.

Wang, Q., Shen, Y., Zhang, J. Q. 2005 A nonlinear correlation measure for multivariate data set. Physica D 200, 287-295.

Wang, S. Y., Tang, M. X. 2004 Exact confidence interval for magnitude-squared coherence estimates. IEEE Sign. Process. Lett. 11, 326-329.

Warne, A. 2000 Causality and Regime Inference in a Markov Switching VAR, pp. 1-41. Sveriges Riksbank Stockholm.

Wendling, F., Bartolomei, F., Bellanger, J. J., Chauvel, P. 2001 Interpretation of interdependencies in epileptic signals using a macroscopic physiological model of the EEG. Clin. Neurophysiol. 112, 1201-1218.

Wessel, R., Koch, C., Gabbiani, F. 1996 Coding of time-varying electric field amplitude modulations in a wave-type electric fish. J. Neurophysiol. 75, 2280-2293.

Widman, G., Schreiber, T., Rehberg, B., Hoeft, A., Elger, C. E. 2000 Quantification of depth of anesthesia by nonlinear time series analysis of brain electrical activity. Phys. Rev. E 62, 4898-4903.

Wiener, N. 1956 The Theory of Prediction. In: Modern mathematics for engineers *Vol. Series 1*. Ed. E. F. Beckenbach. McGraw-Hill: New York.

Wiesenfeldt, M., Parlitz, U., Lauterborn, W. 2001 Mixed state analysis of multivariate time series. Int. J. of Bifurcation and Chaos 11, 2217-2226.

Yamada, S., Nakashima, M., Matsumoto, K., Shiono, S. 1993 Information theoretic analysis of action potential trains. I. Analysis of correlation between two neurons. Biol. Cybern. 68, 215-220.

Zaveri, H. P., Williams, W. J., Iasemidis, L. D., Sackellares, J. C. 1992 Time-frequency representation of electrocorticograms in temporal lobe epilepsy. IEEE Trans. Biomed. Eng. 39, 502-509.

Zaveri, H. P., Williams, W. J., Sackellares, J. C., Beydoun, A., Duckrow, R. B., Spencer, S. S. 1999 Measuring the coherence of intracranial electroencephalograms. Clin. Neurophysiol. 110, 1717-1725.

Zheng, Z. G., Hu, G. 2000 Generalized synchronization versus phase synchronization. Phys. Rev. E 62, 7882-7885.

Zhu, L., Lai, Y. C., Hoppensteadt, F. C., He, J. 2003 Probing changes in neural interaction during adaptation. Neural Comput. 15, 2359-2377.

*Captions for figures*

*Figure 1*: (a) Cross-correlation function $C_{xy}(\tau)$ between two EEG time series recorded from the two hemispheres in a rat. (b) Same as (a) but after taking a fourth power of each data point for both time series. As $C_{xy}(\tau)$ is a measure of linear relationship between these time series, the strong correlation in (a) is decreased by a simple static nonlinear transformation in (b). The function at zero time lag is the correlation coefficient ($r_{xy}$=0.63 in (a), $r_{xy}$=0.25 in (b)).

*Figure 2*: A simulated network consisted of five systems ($X_i$, $i=1,...,5$) with causal and direct influences between several of them (Upper panel). Each system was represented by AR model and causality was introduced as time delay in the model. Corresponding PDCs are plotted in a 5x5 matrix. Frequency is expressed in arbitrary units. Off-diagonal PDC is found whenever there is any direct influence. See Baccala and Shameshima (2001b) for details.

*Figure 3:* 1000 random samples following a uniform distribution (in the [0,1] interval, *left*), a normal distribution (with zero mean and unit variance, *middle*) and a Poisson distribution (with parameter λ=10, *right*). The histograms and the corresponding probabilities are estimated taken 40 bins of a size in each case. The estimated Shannon entropy $I_X$ (Eqn. (15)) is indicated for each distribution. The distribution with the largest entropy (the uniform one) has also the largest uncertainty, since all the possible states have approximately the same probability.

*Figure 4:* Schematic representation of the use of MI to determine the efficiency of a synapse. A distal input spike-train is used to reduce the entropy of the neural output, previously transformed into a binary signal by means of a sliding window ('1': with action potential; '0': no action potential). The difference between the output and the conditioned entropy (the MI) roughly estimates the efficacy of the synapse. The study of the relationship between this efficacy and several input parameters can be tackled in this way. See London *et al.*, (2002) for details.

*Figure 5: Left:* Absolute phase difference between the *x* variables of two Rössler systems (a typical nonlinear dynamical system). A: uncoupled state; the phase difference grows and is unbounded. B: strong PS; the phase difference remains constant along time. Even in this latter case, the amplitudes remain completely uncorrelated (*right*). See Rosenblum *et al.* (1996) for details.

*Figure 6: Left:* Real part of a complex Morlet wavelet with centre frequency *f*=10 Hz and bandwidth parameters $\sigma_t$ = 0.301 (solid line) and $\sigma_t$ = 0.12 (dotted line). *Right:* their normalized power spectral densities. The frequency determines the position of the spectral peak, whereas $\sigma_t$ determines its width: a greater $\sigma_t$ gives rise to a wavelet with better frequency resolution (the sharper peak, more localized peak) but poorer localization in time.

*Figure 7:* The assessment of 1:1 PS from $\varphi'_{1,1}(t)$ for the signals shown in Fig. 5. *Left:* the values of $\varphi'_{1,1}(t)$ are shown as solid line arrows in the unit circle. For the sake of clarity, only five different times are displayed. The length of their average (dashed line arrow) is the estimated mean phase coherence, $\gamma_{1,1}$. *Right:* the distribution of $\varphi'_{1,1}(t)$, whose entropy, $I_\varphi$, is used to calculate $\rho_{1,1}$ following Eq. (25). *Top*: Unsynchronized state, (line A in Fig. 5). The relative phase is randomly scattered over the circle ($\gamma_{1,1} \sim 0$) and their values are widely distributed in the [0, 2π) interval ($I_\varphi \sim I_{MAX}$ and $\rho_{1,1} \sim 0$). *Bottom:* Strong PS (line B in Fig. 5). The relative phase is concentrated in one sector of the circle ($\gamma_{1,1} \sim 1$) and its distribution shows a sharp peak ($I_\varphi << I_{MAX}$ and $\rho_{1,1} \sim 1$).

*Figure 8:* Schematic representation of the dynamics of gamma band (30-60 Hz) EEG PS in a typical cognitive experiment. The grand-average value of the bivariate PS index is plotted as a function of the latency after the stimulus for a set of EEG channels. In all the channels, increased PS with latency of 300 ms. is followed by active desynchronization shortly afterwards. In this example, PS is greater for central electrodes than for the rest.

*Figure 9:* Basic idea of the nonlinear interdependence measures in the state space. The size of the neighborhood in one of the systems, say *X*, is compared with the size of its mapping in the other system. The example shows a Lorenz system driven by a Rössler with zero coupling (upper case) and with strong coupling (lower case). Below each attractor, the corresponding time series is shown. The *(X/Y)* interdependences are calculated in the same way, starting with a neighborhood in *Y*. See Quian Quiroga *et al.* (2002a) for details.

*Figure 10:* The concepts of nearest and mutual neighbors in terms of the time series. Two EEG time series of 5 s each (*X* and *Y*) recorded from the two hemispheres of a rat are shown. A reference vector ('*') in *Y* is a temporal pattern $\mathbf{y}_n$ spanning $(m-1)\tau$ samples. Its nearest neighbors ('o') $\mathbf{y}_{s_{n,i}}$ ($i=1,2$ in this example) are the most similar patterns in *Y*, and are entirely determined by the shape of $\mathbf{y}_n$. Its mutual neighbors ('◊') $\mathbf{y}_{r_{n,i}}$, however, are completely determined by the nearest neighbors of $\mathbf{x}_n$, as shown by the arrows. The mutual neighbors give a clue of the influence of *X* on *Y* (that of *Y* on *X* can be similarly investigated from signal *X):* in a synchronized state, they are more similar to the nearest neighbours than randomly picked vectors, whereas they are indistinguishable from these latter ones if the signals are independent. Measures described in Section 9.1 are aimed to assess this question.

*Figure 11*: Idea of event synchronization. On top, two simultaneously recorded EEG channels containing spikes. On bottom a zoom on the data. Firstly, events are detected using e.g. local maxima (markers on top of the spikes). Secondly, boxes of length $2\tau$ around each event in *X* are considered. Thirdly, the number of times that events in *X* and *Y* appear within the same box (thick boxes), the events of *X* precedes events in *Y* (solid boxes) or vice versa (dashed boxes) are counted. Event synchronization is just a quantification of the number of quasi-

synchronous events (thick boxes) and the delay asymmetry is obtained from the difference between solid and dashed boxes (equation 32).

*Figure 12*: Two EEG segments of a healthy term newborn (electrodes Fp1 (solid line) and Fp2 (dashed line)) during active sleep (*top*) and their corresponding univariate surrogates (*bottom*). *Left*: original EEG traces. *Middle*: Autocorrelation function (preserved). *Right*: Cross-correlation function (destroyed).

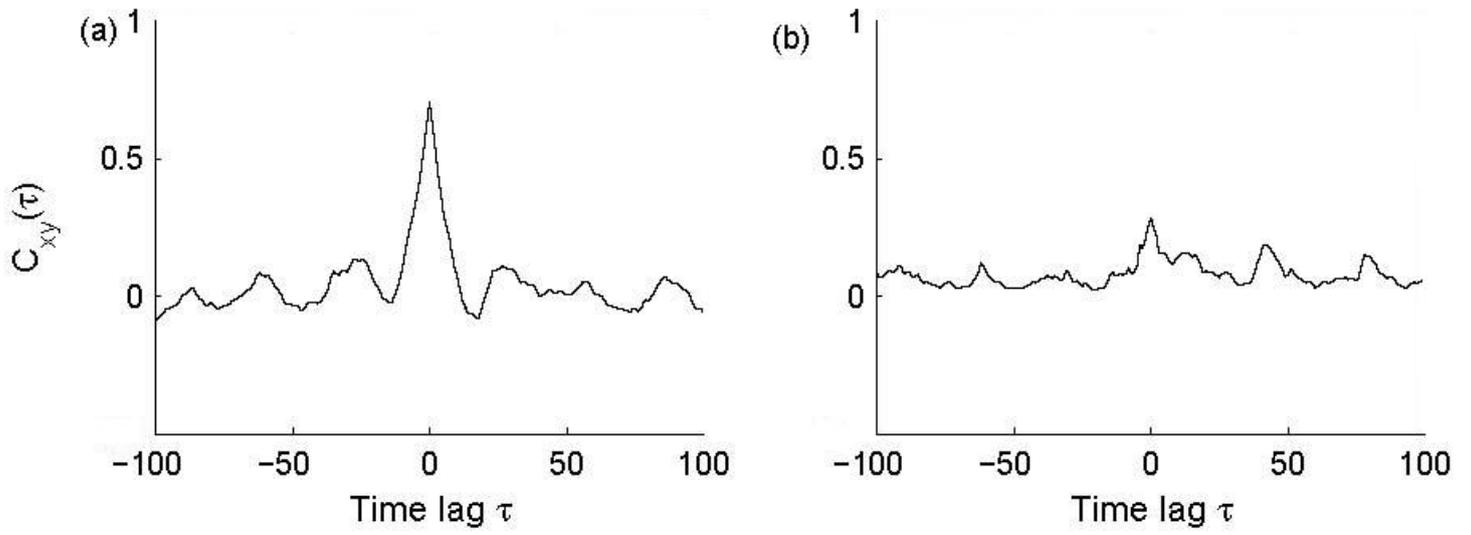

Figure 1

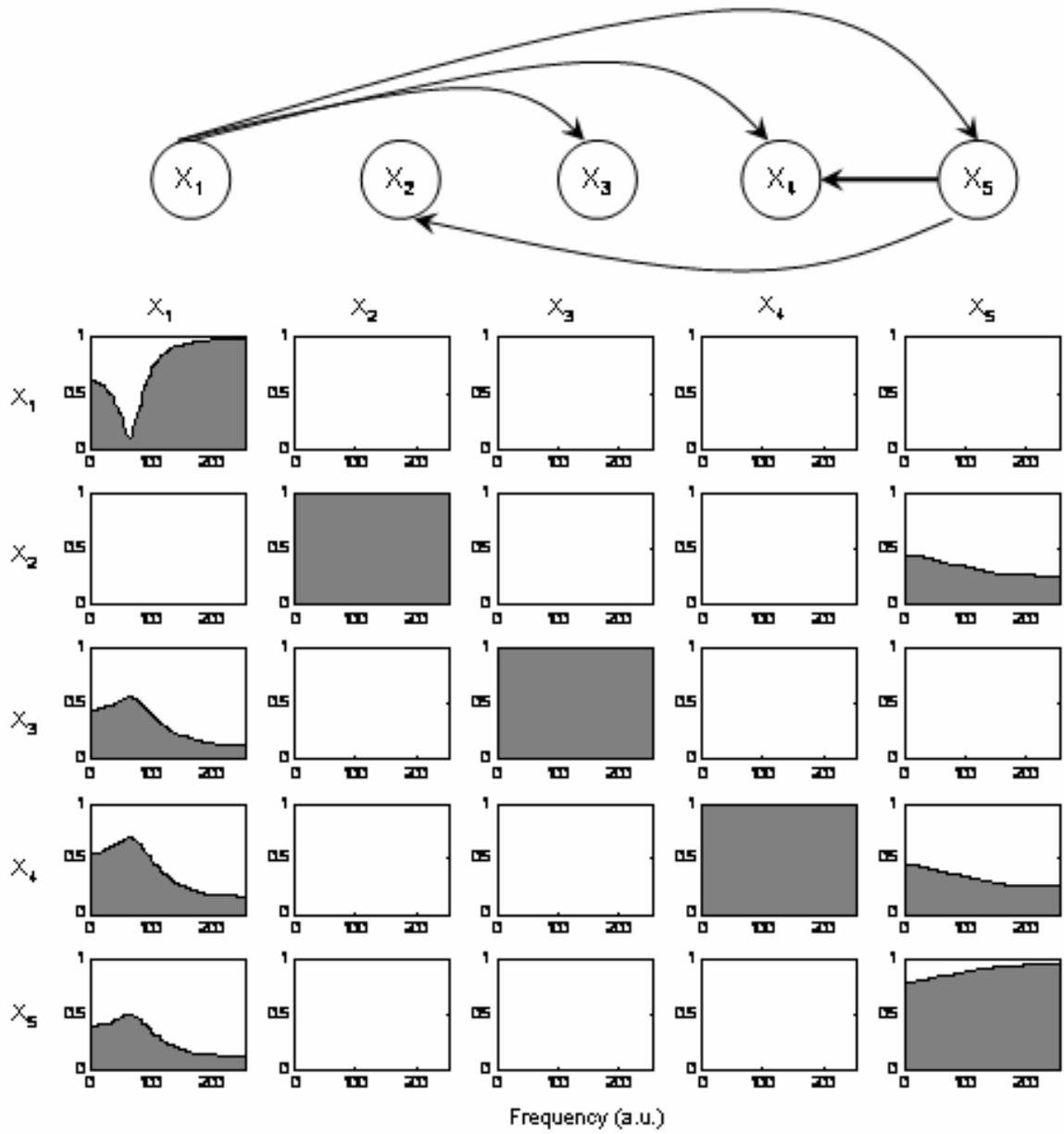

Figure 2

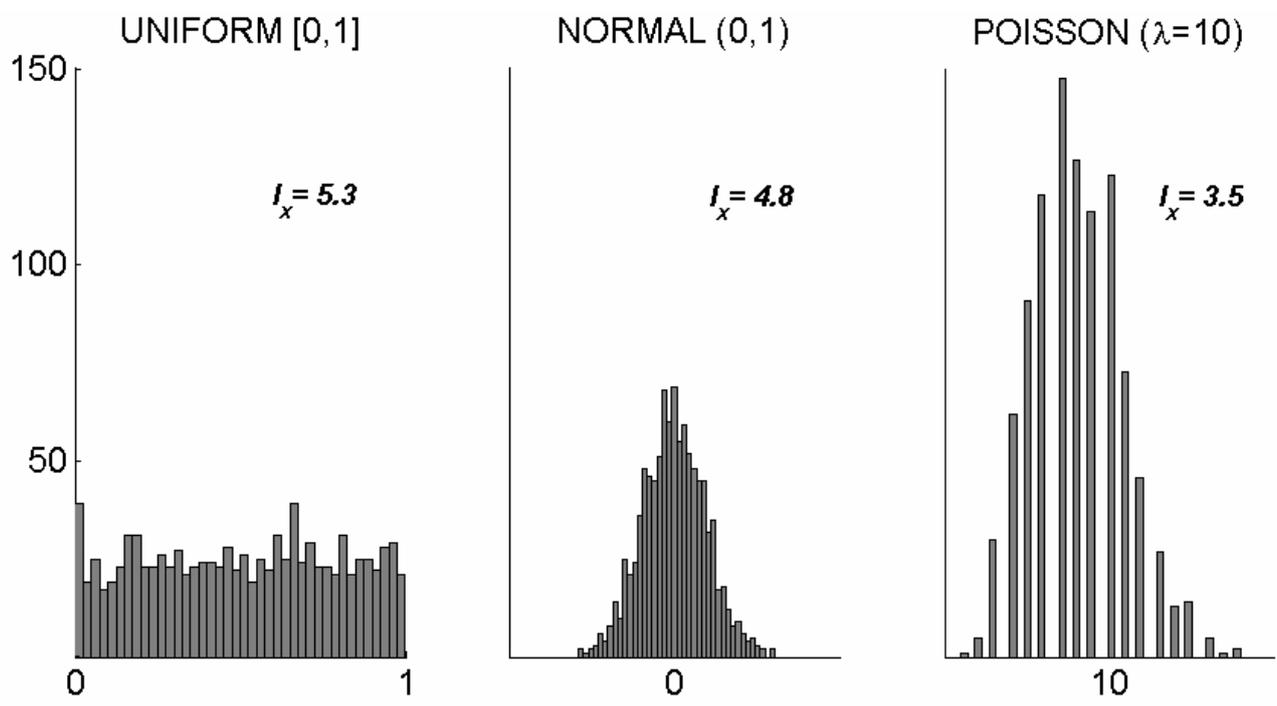

Figure 3

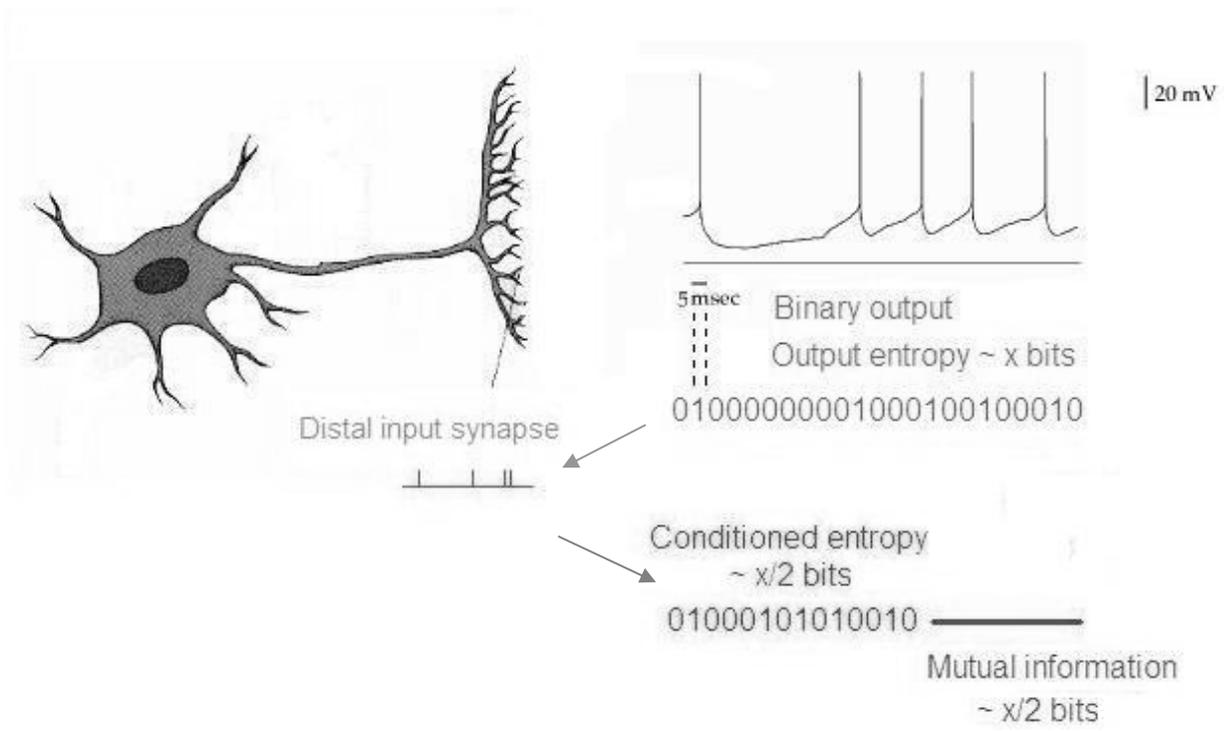

Figure 4

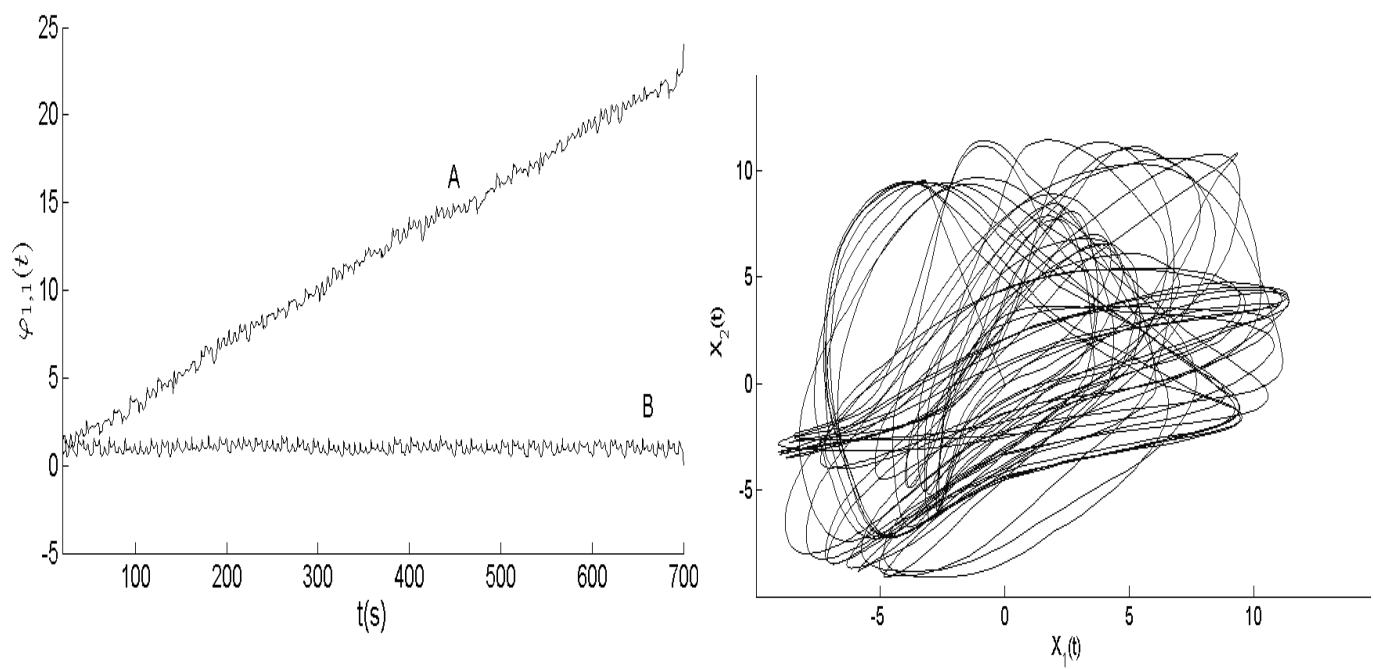

Figure 5

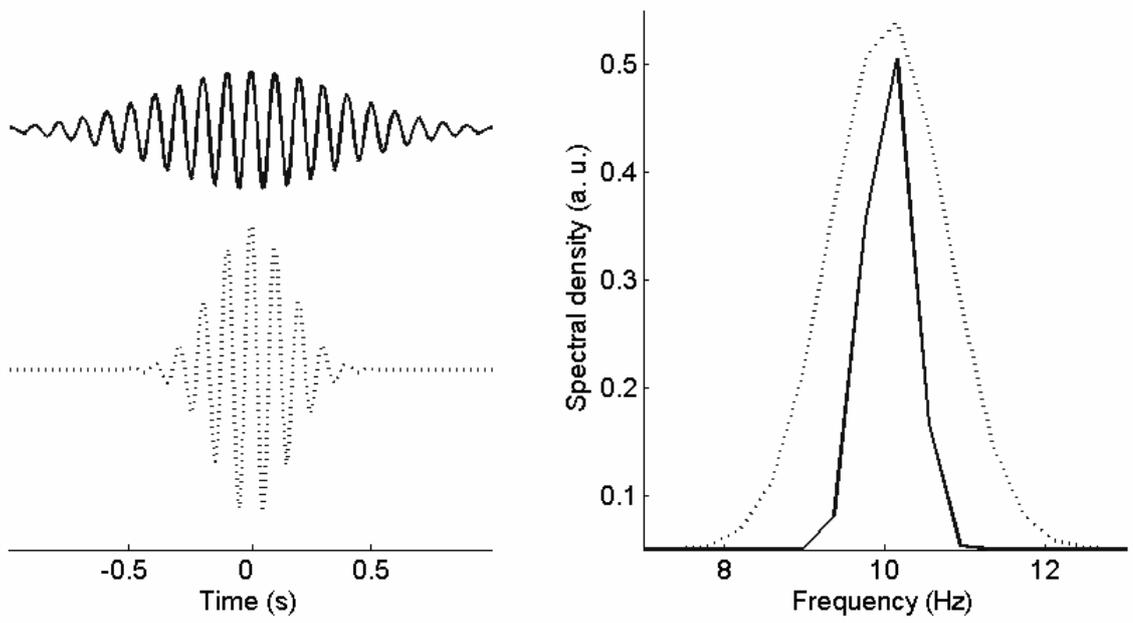

Figure 6

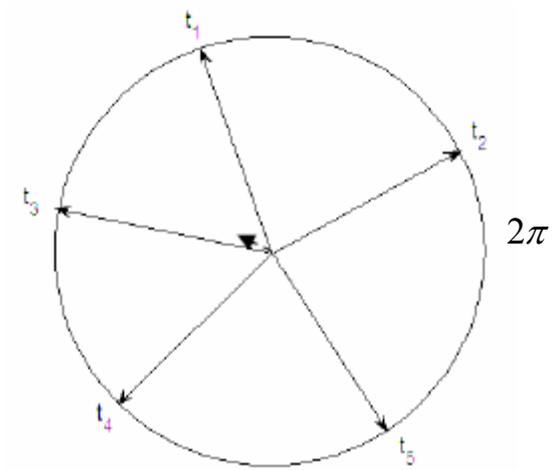
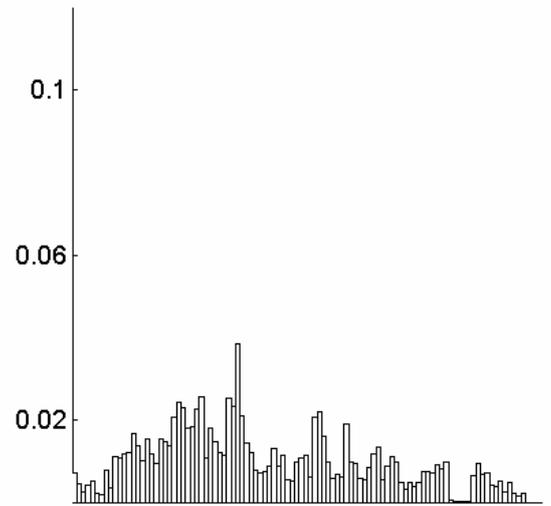
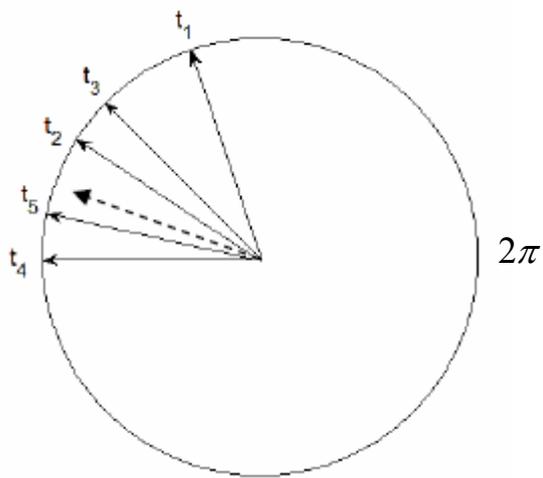
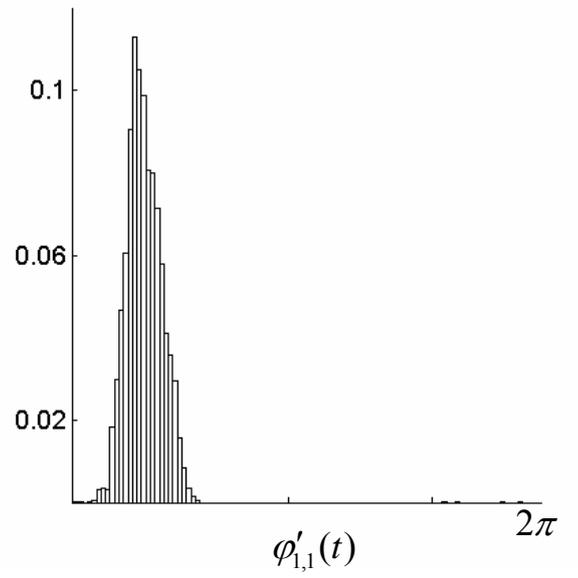

Figure 7

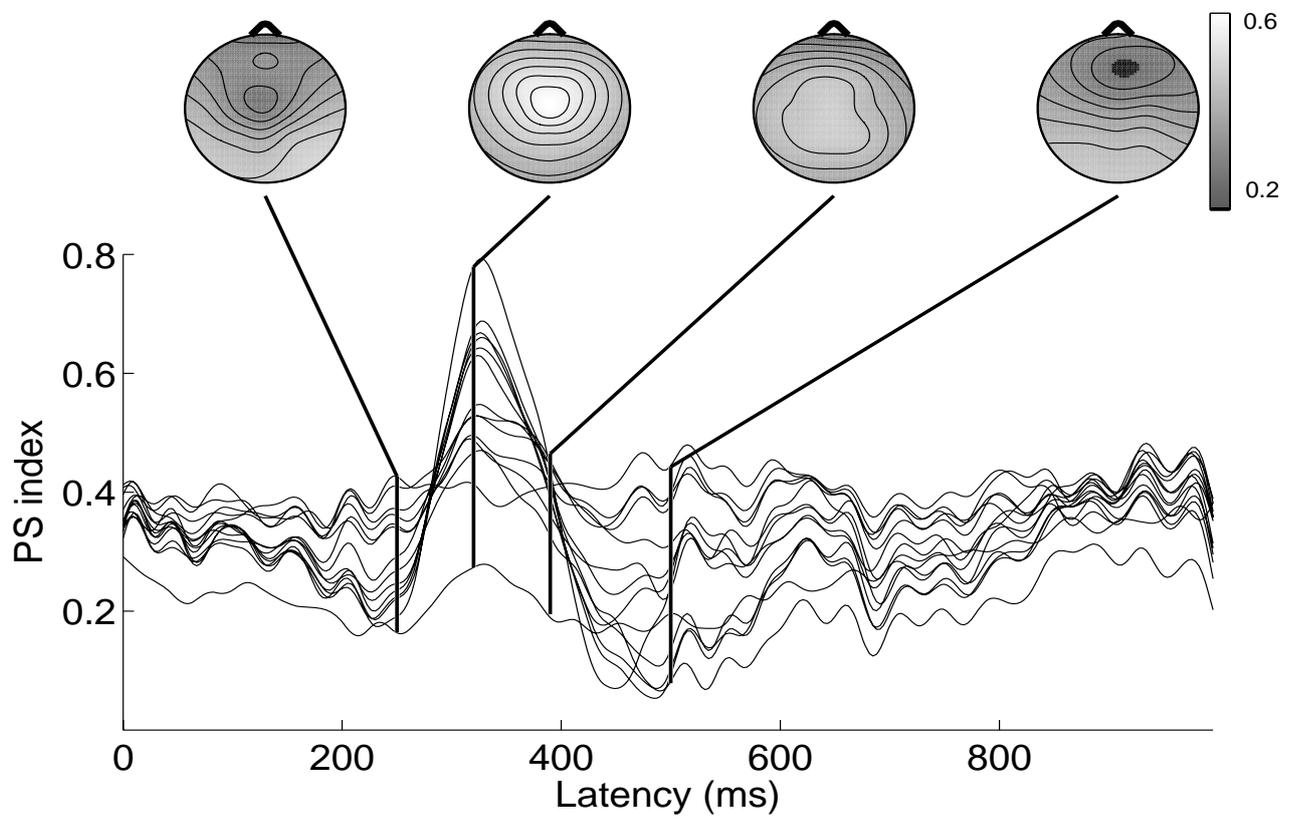

Figure 8

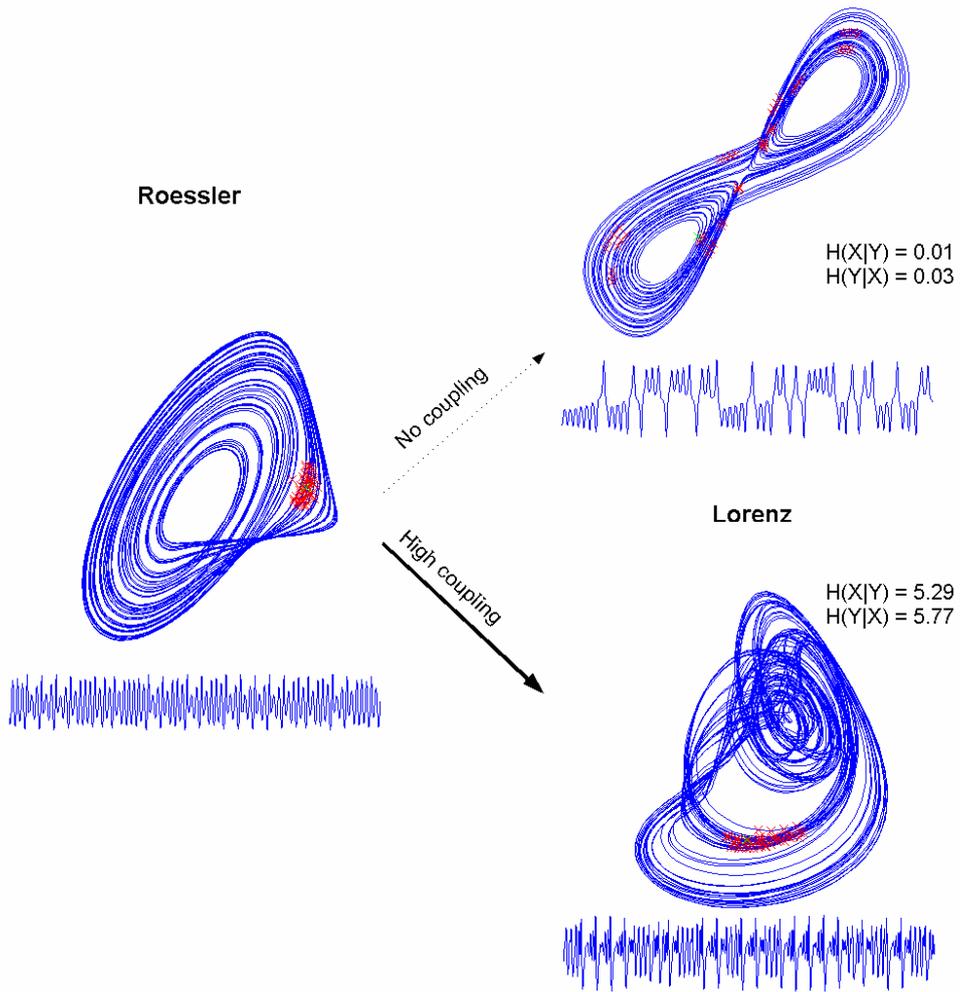

Figure 9

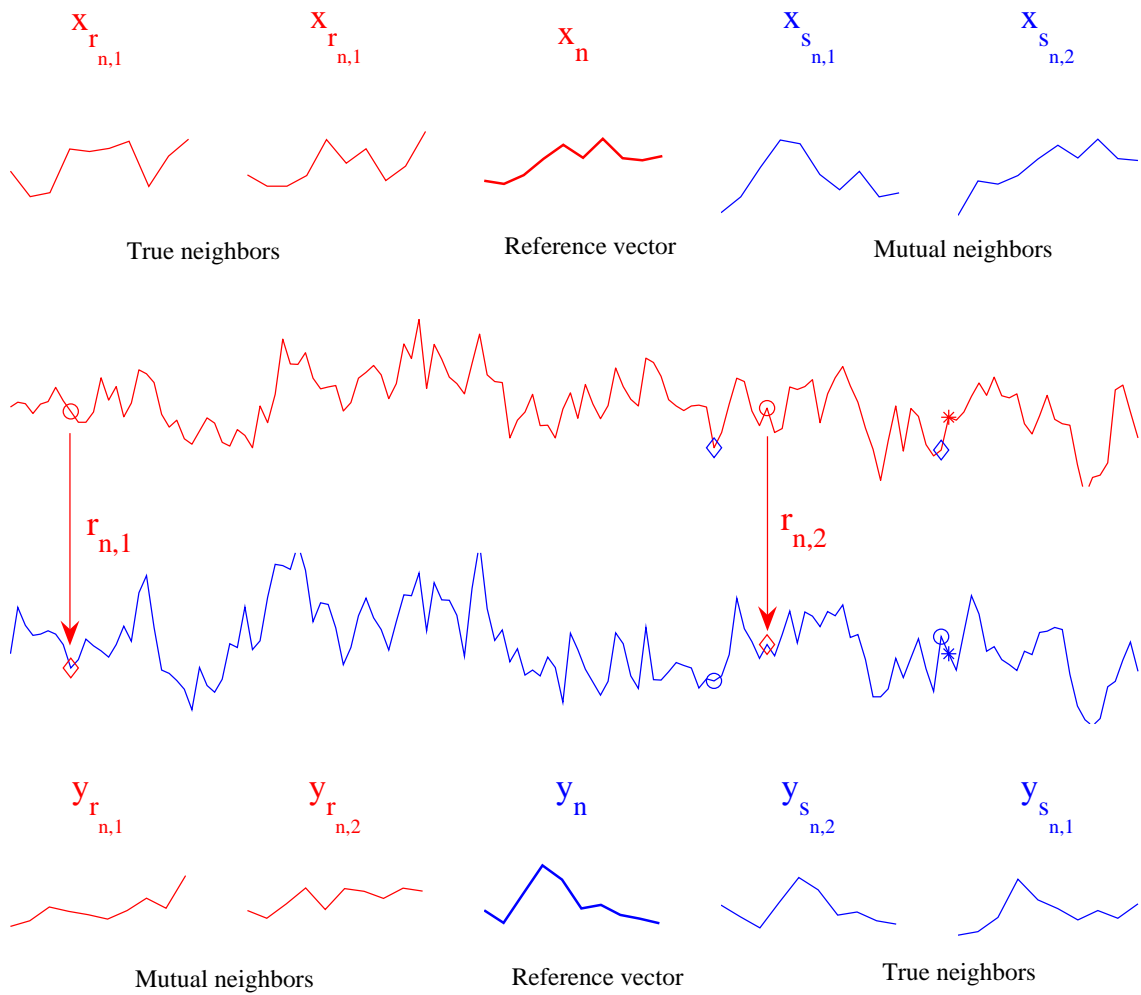

Figure 10

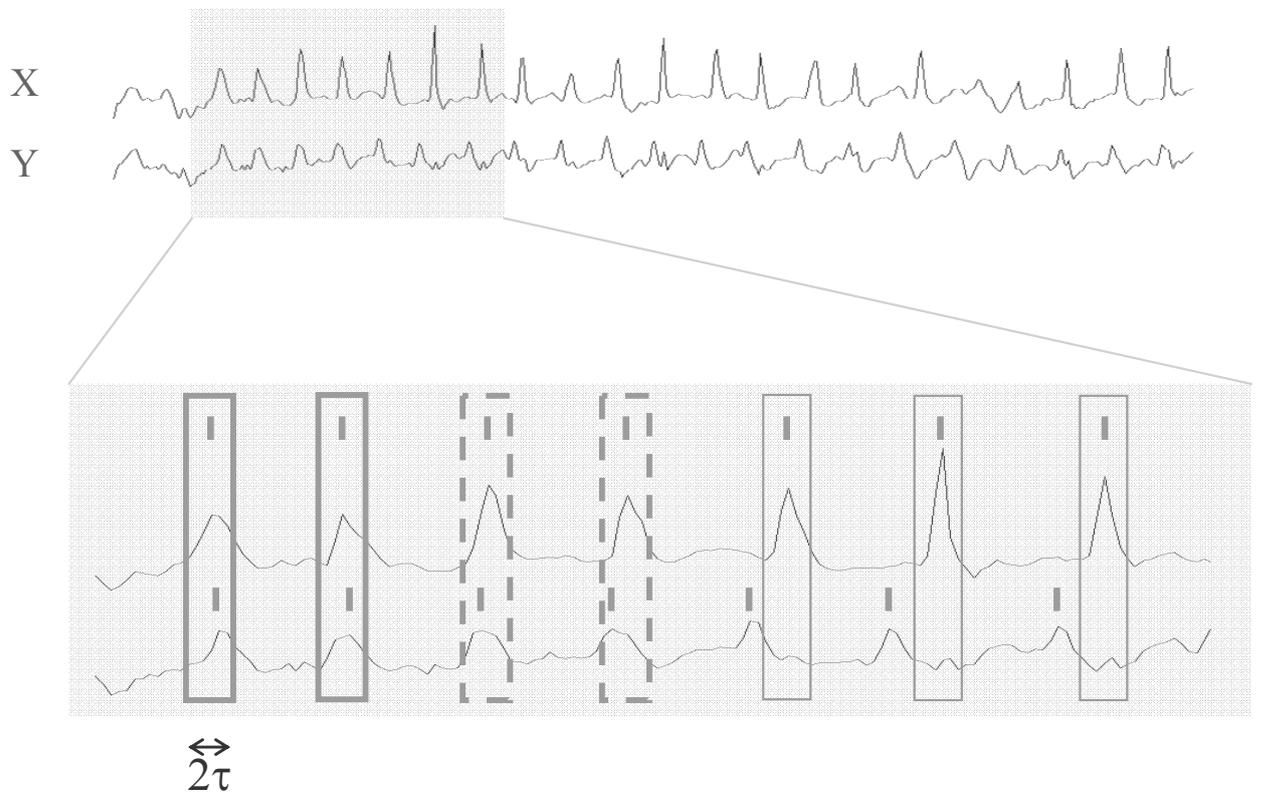

Figure 11

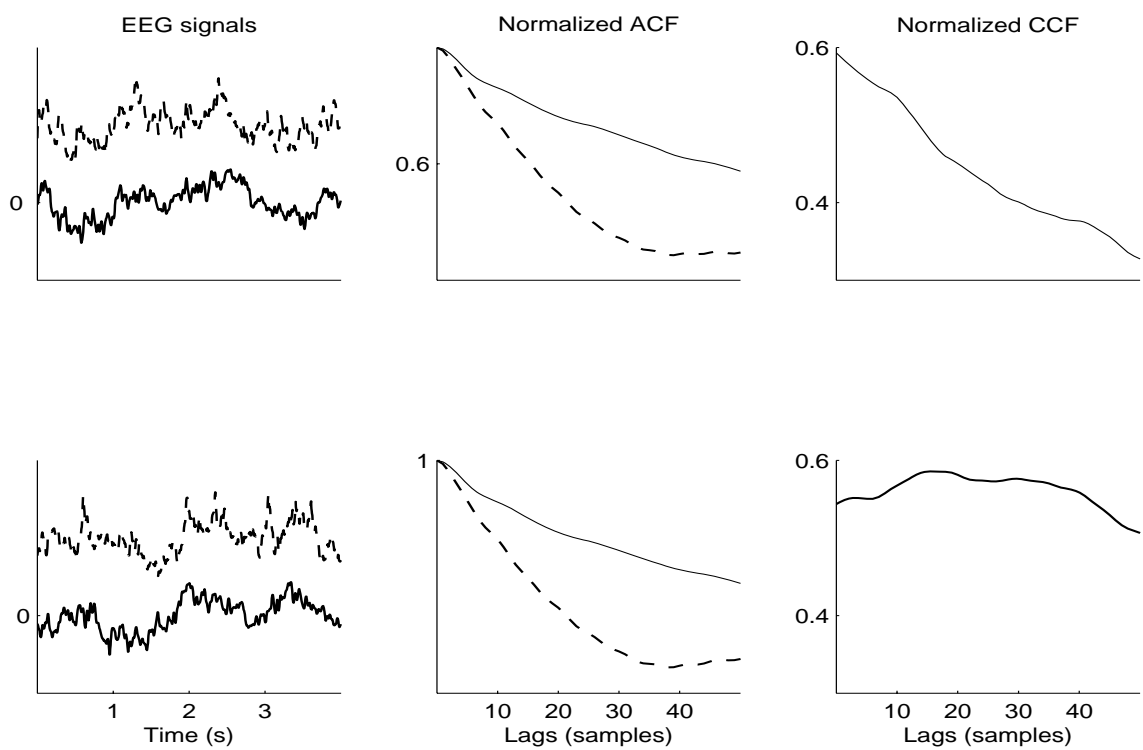

Figure 12